\documentclass[preprintnumbers, floatfix,letterpaper,aps,prd,epsfig,nofootinbib,
twocolumn
]{revtex4-1}
\pdfoutput=1
\usepackage{bm, graphicx, dcolumn, epstopdf, epsf, latexsym, mathbbol, amssymb, amsmath, color, slashed, mathrsfs, mathcomp, simplewick}
\usepackage[center]{subfigure}
\usepackage{multirow}
\usepackage{makecell}
\usepackage[colorlinks,linkcolor=blue,citecolor=blue,urlcolor=blue]{hyperref}

\begin{document}
\allowdisplaybreaks
 \newcommand{\bq}{\begin{equation}} 
 \newcommand{\eq}{\end{equation}}
 \newcommand{\bqn}{\begin{eqnarray}}
 \newcommand{\eqn}{\end{eqnarray}}
 \newcommand{\nb}{\nonumber}
 \newcommand{\lb}{\label}
 \newcommand{\f}{\frac}
 \newcommand{\p}{\partial}
\newcommand{\PRL}{Phys. Rev. Lett.}
\newcommand{\PLB}{Phys. Lett. B}
\newcommand{\PRD}{Phys. Rev. D}
\newcommand{\CQG}{Class. Quantum Grav.}
\newcommand{\JCAP}{J. Cosmol. Astropart. Phys.}
\newcommand{\JHEP}{J. High. Energy. Phys.}
\newcommand{\red}{\textcolor{red}}

\title{Periodic orbits and their gravitational wave radiations in a polymer black hole in loop quantum gravity}

\author{Ze-Yi Tu${}^{a, b}$}
\email{zeyi$\_$tu@zjut.edu.cn}

\author{Tao Zhu${}^{a, b}$}
\email{zhut05@zjut.edu.cn; Corresponding author}

\author{Anzhong Wang${}^{c}$}
\email{anzhong$\_$wang@baylor.edu}

\affiliation{${}^{a}$Institute for Theoretical Physics \& Cosmology, Zhejiang University of Technology, Hangzhou, 310023, China\\
		${}^{b}$ United Center for Gravitational Wave Physics (UCGWP),  Zhejiang University of Technology, Hangzhou, 310023, China\\
		${}^{c}$ GCAP-CASPER, Physics Department, Baylor University, Waco, Texas 76798-7316, USA}

\date{\today}

\begin{abstract}

This article provides a detailed investigation into the motion of the surrounding particles around a polymer black hole in loop quantum gravity (LQG). Using effective potential, the critical bound orbits and innermost stable circular orbits (ISCO) are analyzed. The study finds that the radii and angular momentum of the critical bound orbits decrease with an increase in the parameter $A_\lambda$ which labels the LQG effects, while the energy and angular momentum of the ISCO also decreases with an increase in $A_\lambda$. Based on these findings, we then explore the periodic orbits of the polymer black hole in LQG using rational numbers composed of three integers. Our results show that the rational numbers increase with the energy of particles and decrease with the increase of angular momentum based on a classification scheme. Moreover, compared to a Schwarzschild black hole, the periodic orbits in a polymer black hole in LQG consistently have lower energy, providing a potential method for distinguishing a polymer black hole in LQG from a Schwarzschild black hole. Finally, we also examine the gravitational wave radiations of the periodic orbits of a test object which orbits a supermassive polymer black hole in LQG, which generates intricate GW waveforms that can aid in exhibiting the gravitational structure of the system.

\end{abstract}


\maketitle

\section{Introduction}
\renewcommand{\theequation}{1.\arabic{equation}} \setcounter{equation}{0}

Black holes are a unique and immensely powerful force of gravity that gives rise to a range of fascinating astronomical phenomena in their vicinity, including gravitational waves \cite{gws}, gravitational lensing \cite{VLBI_deflection}, shadows \cite{Akiyama:2019fyp, Akiyama:2019eap}, etc. Through the study of the geodesics of test particles in the vicinity of a black hole, we are able to explore these phenomena and potentially tackle some of the most challenging problems in the universe. This approach allows us to delve deeply into the nature of gravity and gain a more profound understanding of Einstein's theory of general relativity (GR).
One type of orbit for the test particle, the periodic orbit around a black hole, is an important phenomenon of GR. Periodic orbits are special because they capture fundamental information about orbits around a black hole and all generic black hole orbits are small deviations from periodic orbits \cite{Levin:2008mq}. In particular, periodic orbits are crucial in solving some of the most difficult problems in astrodynamics, such as understanding the motion of planetary satellites, the long-term stability of the solar system, and the motion of galactic potentials. However, while periodic orbits have been extensively studied in these contexts, their behavior in relativistic astrophysical systems, such as compact binary stars and the gravitational radiation of an extreme-mass-ratio inspiral (EMRI) system, remains a topic of active research.

Stellar-mass black holes are commonly known to be tightly bound in orbit around a significantly larger black hole, which can be approximated as a timelike test particle orbiting a supermassive black hole. Such binary systems, known as the EMRI systems, are one of the main targets of future space-based gravitational detectors, such as LISA \cite{Danzmann:1997hm,LISA:2017pwj}, Taiji \cite{CMS:2017asf}, Tianqin \cite{TianQin:2015yph, Gong:2021gvw}, etc. Possible detections of these systems provide a positive to explore the nature of gravity and cosmology\cite{LISA:2022kgy,LISACosmologyWorkingGroup:2022jok}. The bound orbits of stellar-mass black holes around a supermassive black hole may exhibit peculiar behavior during the inspiral stage of gravitational wave detection. As a result of gravitational wave radiation, two black holes with an extreme mass ratio move closer to each other. During this process, periodic orbits act as continuous transitions and play an important role in studying gravitational wave radiation \cite{Glampedakis:2002ya}. 

Given this, Levin et al. proposed a classification of periodic orbits for mass particles, which is highly useful for understanding the dynamics of black hole mergers. Their classification scheme follows $Poincar\acute{e}$ paradigm, which states that the behavior of a dynamical system can be understood by studying its periodic trajectories. In the zoom-whirl classification  \cite{Levin:2008mq}, each periodic orbit is characterized by three topological integers: $z$, $w$, and $v$, which represent scaling, rotation, and vertex behaviors of the orbit, respectively. The tracing order of leaves is also demonstrated. The rational number $q$ explicitly measures the extent of periapsis precession beyond the ellipse and the orbit's topology. With this taxonomy, the study of the periodic orbits has been conducted for a lot of black hole spacetimes, such as Schwarzschild black holes, Kerr black holes, charged black holes, Kerr-Sen black holes, naked singularities, etc, see refs. \cite{Levin:2008mq, Levin:2008ci, Levin:2009sk, Misra:2010pu, Babar:2017gsg, Liu:2018vea,Lin:2023rmo,Wang:2022tfo,Zhang:2022zox,Mummery:2022ana,Habibina:2022ztd,Zhang:2022psr,Lin:2022wda,Gao:2021arw,Lin:2021noq,Deng:2020yfm,Zhou:2020zys,Gao:2020wjz,Deng:2020hxw,Azreg-Ainou:2020bfl,Wei:2019zdf,Pugliese:2013xfa,Healy:2009zm,Zhang:2022zox,Lin:2022llz,Bambhaniya:2020zno,Azreg-Ainou:2020bfl,Wei:2019zdf,Rana:2019bsn} and references therein.

Recently, a quantum extension of the Schwarzschild black hole was constructed based on polymer quantization in the context of LQG \cite{LQG_BH1, LQG_BH2}, called polymer black hole. This is an effective quantum spacetime that arises from a specific $\bar \mu$-scheme based on the polymerlike quantization inspired by LQG, in which the quantum theory of black hole is achieved by replacing the canonical variables $(b, c)$ in the phase space of the black hole spacetime with their regularized counterparts, $b \to \frac{\sin(\delta_b b)}{\delta_b}$ and $c \to \frac{\sin(\delta_c c)}{\delta_c}$, where $\delta_b$ and $\delta_c$ are two quantum polymeric parameters that control the relevant scales of the quantum effects of LQG \cite{Ashtekar20}. In this picture, the quantum effect is controlled by the parameter $A_\lambda$ which sensitively depends on $\delta_b$ and $\delta_c$ and its exact value in LQG has not been determined yet. In this effective quantum spacetime, similar to the case in loop quantum cosmology, the spacetime singularity of the classical Schwarzschild black hole can be replaced by a quantum bounce that connects the black hole region and the white hole region. Based on this quantum-extended Schwarzschild black hole, a rotating spacetime with LQG effects has been constructed using the Newman-Janis algorithm \cite{Brahma:2020eos}. Several phenomenological implications of this black hole have been studied. For example, in \cite{Papanikolaou:2023crz}, how quantum effects can influence primordial black hole formation within a quantum gravity framework has been discussed in detail. In addition, people have also tested the LQG black holes with the Event Horizon Telescope observations \cite{Islam:2022wck, Afrin:2022ztr} and constrain the parameter arises in LQG black hole with the observational data of M87* and Sgr A* \cite{KumarWalia:2022ddq, Yan:2022fkr, Vagnozzi:2022moj}. Some other phenomenological studies on testing LQG with black holes observations and GWs can be found in \cite{Brahma:2020eos, Pugliese:2020ivz, Pawlowski:2014nfa, Vaid:2012pr, Barrau:2011md, Modesto:2009ve, Alesci:2011wn, Chen:2011zzi, Dasgupta:2012nk, Barrau:2014yka, Hossenfelder:2012tc, Sahu:2015dea, Cruz:2015bcj, Fu:2023drp, Yang:2023gas,Addazi:2021xuf,Garcia-Chung:2020zyq,Garcia-Chung:2022pdy,Garcia-Chung:2023oul} and references therein.

\begin{figure*}
\includegraphics[width=8cm]{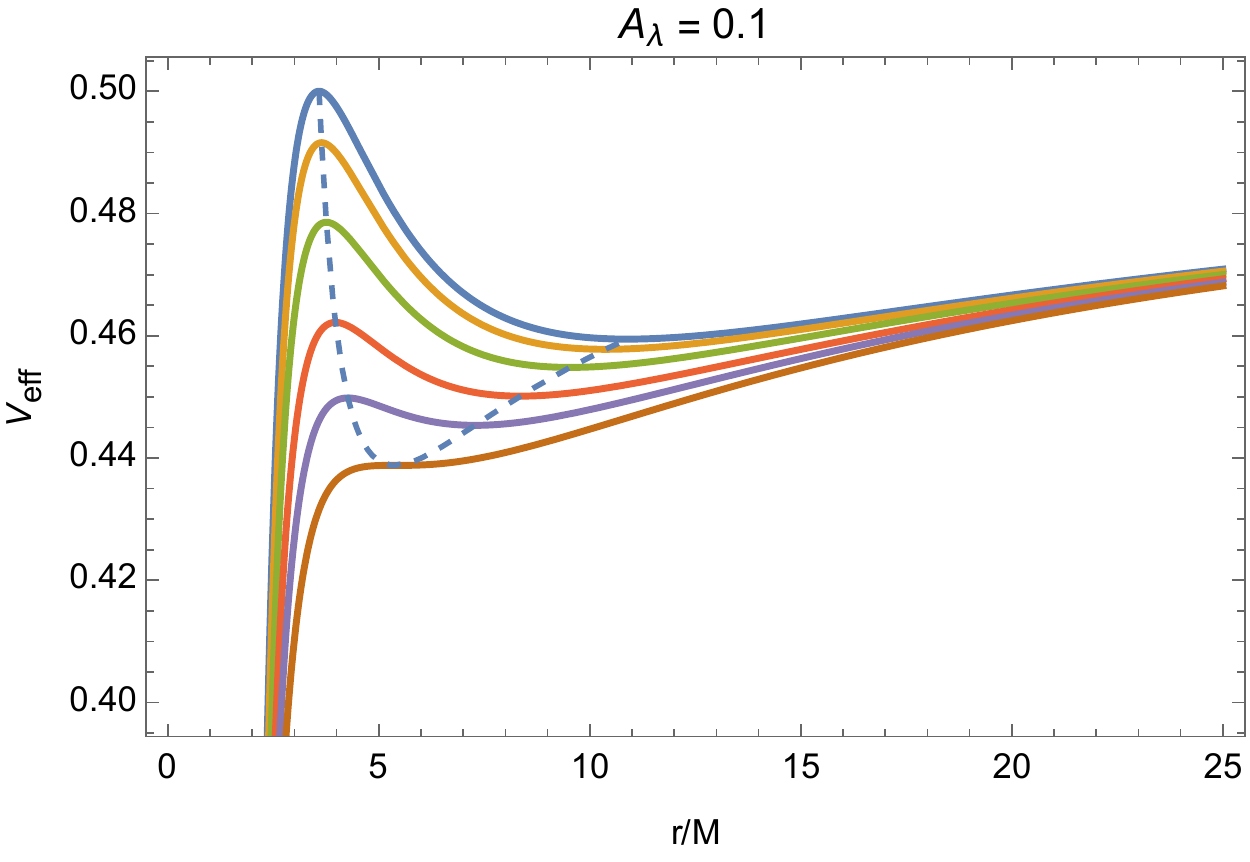}
\includegraphics[width=8cm]{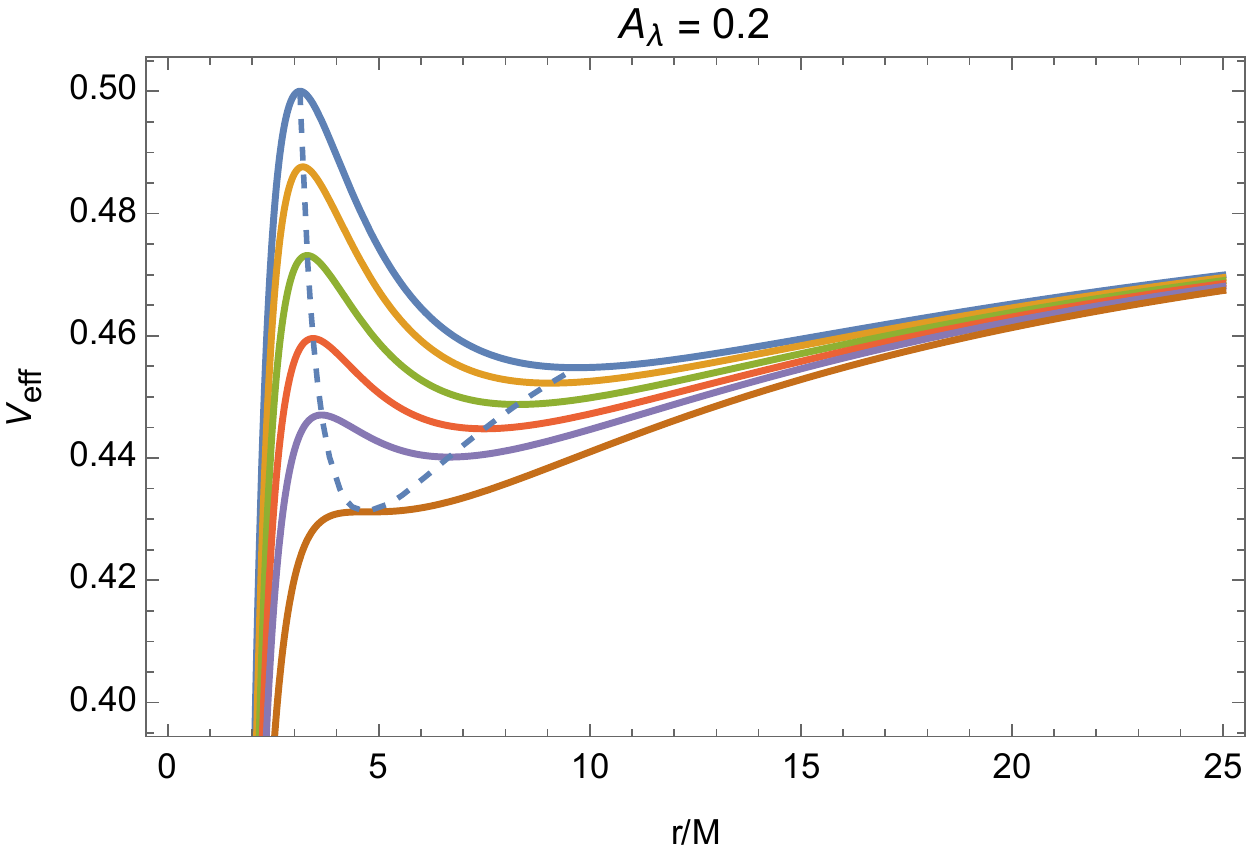} 
\caption{$V_{\rm eff}$ as a function of polymer black holes in LQG. The angular momentum varies from MBO to ISCO from top to bottom. The extremal points of $V_{\rm eff}$ are represented by the dashed line. In the left panel, from top to bottom, each curve with different colors corresponds to $L=3.761823,\;3.7,\;3.6,\;2\sqrt{3},\;3.35,\;3.229269$, respectively, while in the right panel,  each curve with different colors correspond to $L=3.491068,\;3.41,\;3.31,\;3.21,\;3.11,\;2.960660$, respectively.}
\label{veff}
\end{figure*}

The main purpose of this paper is devoted to the study of the periodic orbital behaviors of the surrounding particles around the polymer black hole in LQG. We explore in detail how the LQG effect affects the behaviors of orbits. In addition to studying the periodic orbits of the polymer black hole in LQG, we have also delved into the gravitational wave radiation of the periodic orbits. In gravitational wave astronomy, one generally uses effective potentials to understand "zoom-whirl" behaviors of the periodic orbits, which are commonly studied in the context of classical mechanics and scattering theory. These types of orbits are predicted to be prevalent in EMRI systems \cite{Cutler:1994pb}, where small compact objects are absorbed by supermassive black holes. They are a key source for space-based laser interferometer space antennae such as LISA \cite{Danzmann:1997hm, LISA:2017pwj}, Taiji \cite{CMS:2017asf}, and Tianqin \cite{TianQin:2015yph, Gong:2021gvw}. An EMRI system composed of the polymer black hole in LQG and a stellar-mass compact object which experiences a periodic orbit may provide an unprecedented opportunity to explore the properties of a polymer black hole in LQG.

This article focuses on the periodic orbits of a massive test particle around a polymer black hole in LQG. The paper is structured as follows. In Sec.~\ref{Polymer}, we present a brief review of the polymer black hole solution in LQG, and in Sec.~\ref{Geodesics} we explore the effective potential for a test massive particles around the polymer Black Hole in LQG and study the marginally bound orbits (MBO) and the innermost stable circular orbits (ISCO) using the effective potential. Sec.~\ref{Periodic} is dedicated to the study of periodic orbits characterized by three rational numbers in polymer black holes in LQG, by taking into account the zoom-whirl structure \cite{Levin:2008mq} and the classification of bound orbits. In Sect.~\ref{gravitational}, we delve into the gravitational wave radiation of the periodic orbits around the polymer black holes in LQG. Finally, Sect.~\ref{Conclusion} presents the conclusions and discussions. Through the paper, we use a geometrized unit system with $G=c=1$, and adopt the metric convention $(-,+,+,+)$. 

\section{polymer black hole in loop quantum gravity}
\label{Polymer}
\renewcommand{\theequation}{2.\arabic{equation}} \setcounter{equation}{0}

In this section, we present a brief review of the polymer black hole in LQG.  This black hole is a quantum extension of the static and spherically symmetric metric by solving the LQG effective equations. The metric of this polymer black hole is given by \cite{LQG_BH1, LQG_BH2, Brahma:2020eos}
\bqn
ds^2 = - 8 A_\lambda M_{\rm b}^2 {\cal A}(r) dt^2 + \frac{dr^2}{8 A_\lambda M_{\rm b}^2 {\cal A}(r)}  + {\cal B}(r) d\Omega^2, \nb\\
\eqn
where the metric functions ${\cal A}(r)$ and ${\cal B}(r)$ are defined in terms of radial variable $r$ as 
\bqn
{\cal A}(r) &=& \frac{1}{{\cal B}(r)} \left(1+ \frac{r^2}{8 A_\lambda M_{\rm b}^2}\right) \left(1- \frac{2 M_{\rm b}}{\sqrt{8 A_\lambda M_{\rm b}^2 + r^2}}\right),\nb\\
&&\;\; \\
{\cal B}(r) &=& \frac{512 A_\lambda^3 M_{\rm b}^4 M_{\rm w}^2 + (r+\sqrt{8 A_\lambda M_{\rm b}^2+r^2})^6}{8 \sqrt{8 A_\lambda M_{\rm b}^2+r^2}(\sqrt{8 A_\lambda M_{\rm b}^2+r^2} + r^2)^3},
\eqn
with $M_{\rm b}$ and $M_{\rm w}$ being two Dirac observables of the loop quantum model of this black hole. The parameter $A_{\lambda}$ is defined as $A_\lambda \equiv (\lambda_k/(M_{\rm b}/M_{\rm w}))^{2/3}/2$, where $\lambda_k$ denotes a quantum parameter related to holonomy modifications in LQG \cite{LQG_BH1, LQG_BH2}. It is worth mentioning here that parameter $\lambda_k$ can be eliminated after fixing the integration constants and introducing the two Dirac observables $M_{\rm b}$ and $M_{\rm w}$ for solving the effective equations in LQG \cite{LQG_BH1, LQG_BH2}.

One important feature of this polymer black hole in LQG is that it is free of any singularity in its interior.  When the radial variable $r \to 0$, the areal radius ${\cal B}(r)$ reaches a minimum which smoothly connects an asymptotically Schwarzschild black hole to a white hole with mass $M_{\rm b}$ and $M_{\rm w}$, respectively \cite{LQG_BH1, LQG_BH2}. This is similar to the quantum bounce in LQC. If the bounce is symmetric, then one has $M_{\rm b} = M_{\rm w}$. In this paper, similar to \cite{LQG_BH1, LQG_BH2, Brahma:2020eos}, we consider such interesting and meaningful symmetric bounce scheme and set $M=M_{\rm b}=M_{\rm w}$. Then the metric functions ${\cal A}(r)$ and ${\cal B}(r)$ can be rewritten in the form of
\bqn
{\cal A}(r) &=& \frac{1}{{\cal B}(r)} \left(1+ \frac{r^2}{8 A_\lambda M^2}\right)\left(1- \frac{2M}{\sqrt{8 A_\lambda M^2 + r^2}}\right), \nb\\
&&\;\; \\
{\cal B}(r) &=& 2 A_\lambda M^2 + r^2.
\eqn

It is easy to obtain the location of the horizon of this polymer black hole in LQG by solving ${\cal A}(r) =0$, which gives
\bqn
r_{\rm h} = 2 M \sqrt{1- 2 A_\lambda}. 
\eqn
Obviously, the horizon does not exist if $A_\lambda > 1/2$. Here we also need to mention that when all the effects of LQG are absent, i.e., $A_\lambda =0$,  the above metric reduces to the Schwarzschild spacetime precisely. 

For later convenience, let us introduce a new metric function, $\tilde{\cal A}(r) = 8 A_\lambda M^2 {\cal A}(r)$, then the metric of this polymer black hole in LQG can be cast into the form of
\bqn
ds^2 = - \tilde{\cal A}(r) dt^2 + \frac{dr^2}{\tilde{\cal A}(r)} + {\cal B}(r) (d\theta^2 + \sin^2\theta d\phi^2).
\eqn

\section{Geodesics, marginally bound orbits, and the innermost stable circular orbits}
\label{Geodesics}
\renewcommand{\theequation}{3.\arabic{equation}} \setcounter{equation}{0}

\subsection{Geodesics and effective potential}

Let us first consider the evolution of a particle in the black hole spacetime. We start with the Lagrangian of the particle,
\bqn
\mathcal{L} = \frac{1}{2}g_{\mu \nu} \frac{d x^\mu} {d \lambda } \frac{d x^\nu}{d \lambda},
\eqn
where $\lambda$ denotes the affine parameter of the world line of the particle. For massless particles, we have $\mathcal{L}=0$ and for massive ones $\mathcal{L}<0$. Then the generalized momentum $p_\mu$ of the particle can be obtained via
\bqn
p_{\mu} = \frac{\partial \mathcal{L}}{\partial \dot x^{\mu}} = g_{\mu\nu} \dot x^\nu,
\eqn
which leads to four equations of motions for a particle with energy $E$ and angular momentum $L$,
\bqn
p_t &=& g_{tt} \dot t  = - E,\\
p_\phi &=& g_{\phi \phi} \dot \phi = L, \\
p_r &=& g_{rr} \dot r,\\
p_\theta &=& g_{\theta \theta} \dot \theta.
\eqn
Here a dot denotes the derivative with respect to the affine parameter $\lambda$ of the geodesics. From these expressions we obtain 
\bqn
\dot t = - \frac{ E  }{ g_{tt} } = \frac{E}{\tilde{\cal A}(r)},\\
\dot \phi = \frac{L}{g_{\phi\phi}} = \frac{L}{ {\cal B}(r) \sin^2\theta}.
\eqn
For timelike geodesics, we have $ g_{\mu \nu} \dot x^\mu \dot x^\nu = -1$. Substituting $\dot t$ and $\dot \phi$ we can get
\bqn
g_{rr} \dot r^2 + g_{\theta \theta} \dot \theta^2 &=& -1 - g_{tt} \dot t^2  - g_{\phi\phi} \dot \phi^2\nb\\
&=& -1 +\frac{E^2}{\tilde{\cal A}(r)}- \frac{L^2}{ {\cal B}(r)\sin^2\theta}.
\eqn

We are interested in the evolution of the particle in the equatorial circular orbits. For this reason, we can consider $\theta=\pi/2$ and $\dot \theta=0$ for simplicity. Then the above expression can be simplified into the form
\bqn
\frac{1}{2}\dot r ^2 = \varepsilon_{\rm eff} - V_{\rm eff}(r),
\eqn
where $V_{\rm eff}(r)$ denotes the effective potential which is given by

\begin{figure}[htbp]
\centering
{\includegraphics[width=8cm]{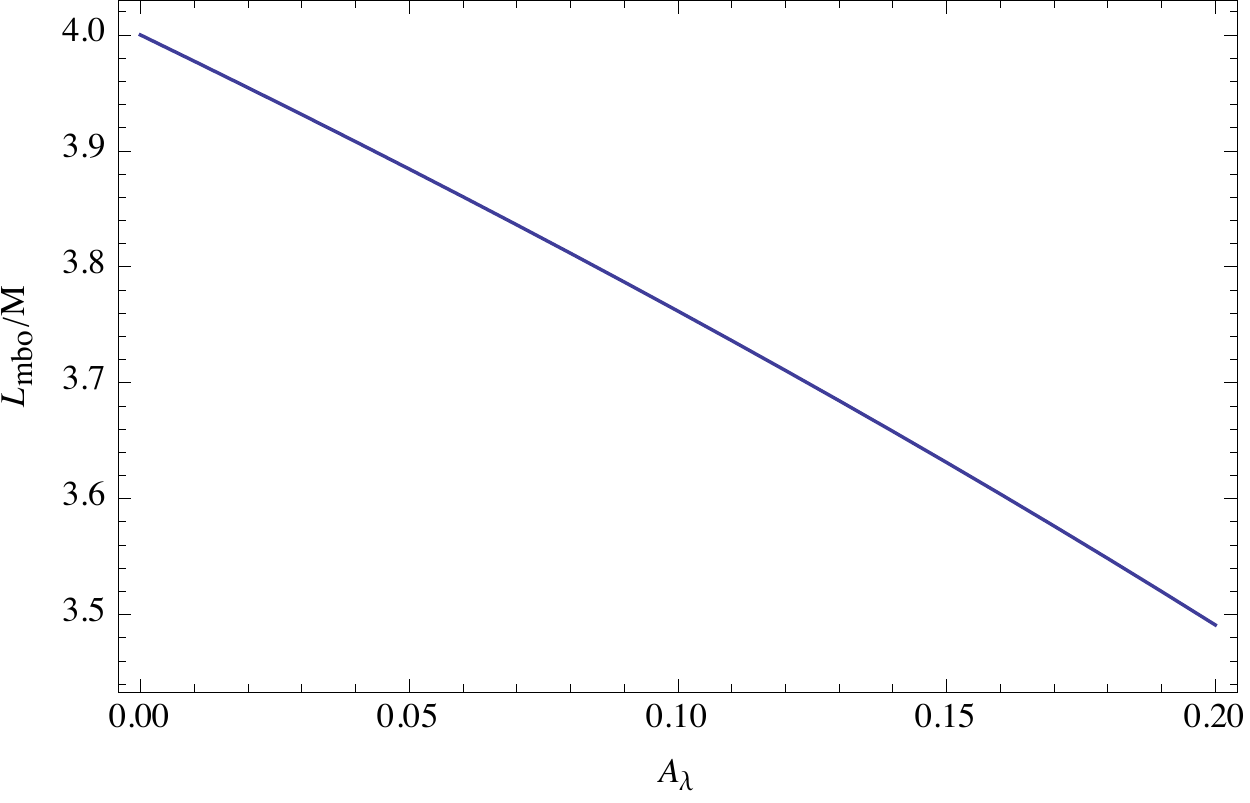}}
{\includegraphics[width=8cm]{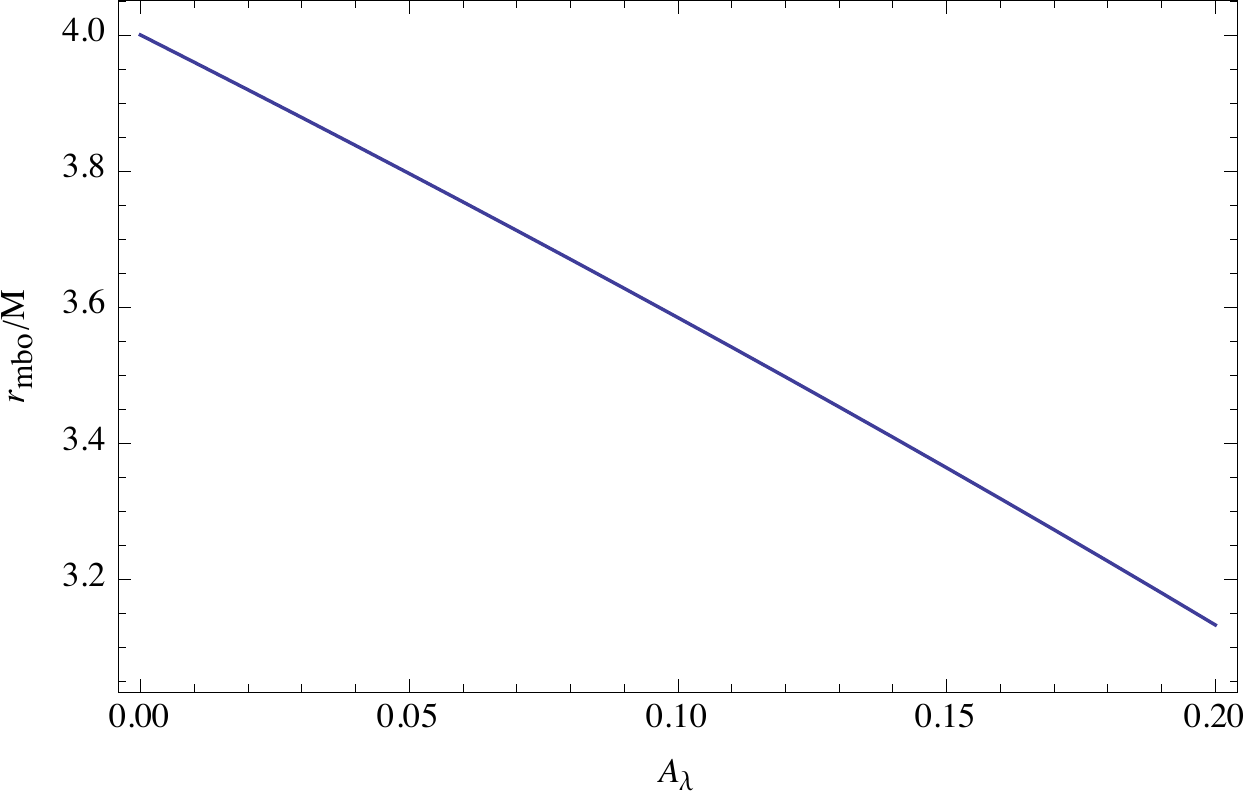}}
\caption{The angular momentum $L_{\rm mbo}$ (upper panel) and radius $r_{\rm mbo}$ (bottom panel) for the MBOs in the polymer black hole in LQG.}
\label{mbo}
\end{figure}

\bqn \lb{Veff}
V_{\rm eff}(r)= \frac{1}{2}\left(1+\frac{L^2}{ {\cal B}(r)}\right)\tilde{\cal A}(r),
\eqn
and
\bqn
\varepsilon_{\rm eff} =\frac{1}{2}E^2.
\eqn

One immediately observes that $V_{\rm eff}(r) \to \frac{1}{2}$  as $r \to +\infty$, as expected for asymptotically flat spacetime. In this case, the particles with energy $E >1$ are able to escape to infinity, and $E = 1$ is the critical case between bound and unbound orbits. In this sense, the maximum energy for the bound orbits is $E=1$. We also plot the effective potential for two different values of LQG parameter $A_\lambda$ respectively in Fig.~\ref{veff}. Different curves in each figure of Fig.~\ref{veff} correspond to different values of angular momentum $L$ and energy $E$ of the geodesics.

\subsection{Marginally bound orbits}

\begin{figure}
\centering
\subfigure{\includegraphics[height=4.5cm,width=8cm]{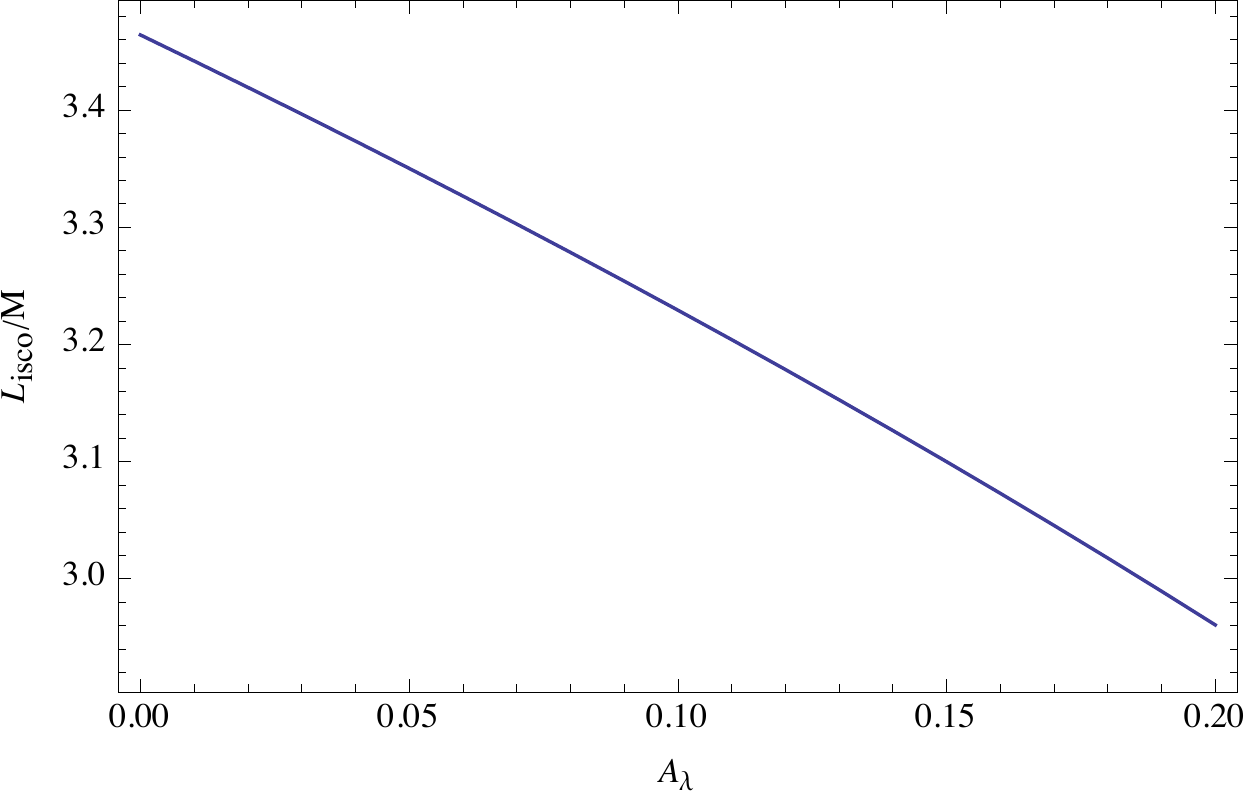}}
\subfigure{\includegraphics[height=4.5cm,width=8cm]{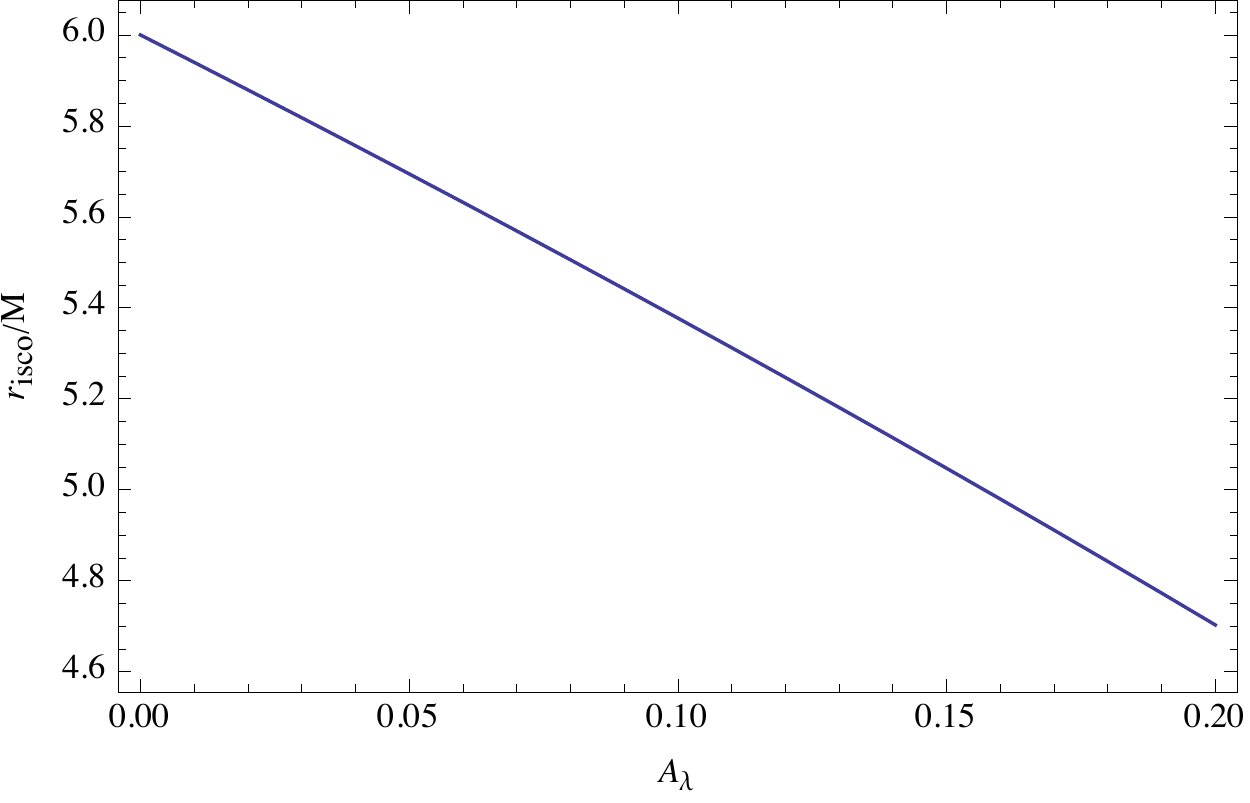}}
\subfigure{\includegraphics[height=4.5cm,width=8cm]{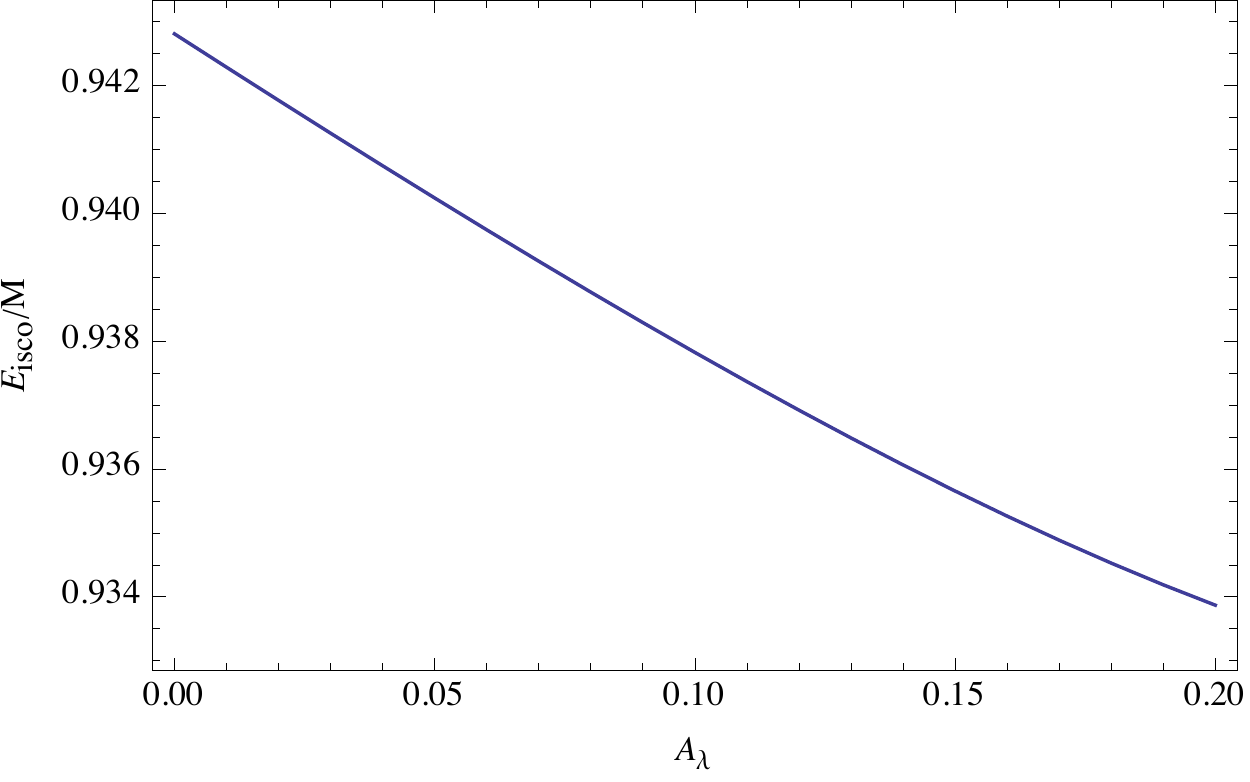}}
\caption{The angular momentum $L_{\rm isco}$ (upper panel), radius $r_{\rm isco}$ (middle panel), and energy $E_{\rm isco}$ (bottom panel) for the innermost stable circular orbits.}
\label{picisco}
\end{figure}

Now let us turn to consider the MBO in a polymer black hole in LQG, which is one of the important circular orbits and has the maximum energy. This bound orbit is defined by the following conditions,
\bqn\lb{mar_condition}
\dot r^2=E^2- 2V_{\rm eff}=0,\;\; \frac{dV_{\rm eff}(r)}{dr}=0,
\eqn
with $E=1$. Solving these conditions one can determine the radius $r_{\rm mbo}$ of the MBO and the angular momentum $L_{\rm mbo}$ for this orbit. For a polymer black hole in LQG, the above equations do not have an exact solution and one can find an approximate solution by treating $A_\lambda$ as a small quantity, i.e., 
\bqn
r_{\rm mbo} \simeq 4M \left(1-\frac{9}{4}A_\lambda\right),
\eqn
and
\bqn
L_{\rm mbo} \simeq 4M (1-A_\lambda).
\eqn
where we use $( L_{\rm mbo},  r_{\rm mbo})$ denotes $(L_{\rm mbo}, r_{\rm mbo})$ for  polymer black holes in LQG. From the above equations, we can plot $r_{\rm mbo}$ and $L_{\rm mbo}$ vs  $A_\lambda$ for MBO, whose behaviors with respect to $A_\lambda$ are displayed in Fig.~\ref{mbo}.

\begin{figure}
\includegraphics[width=1\hsize]{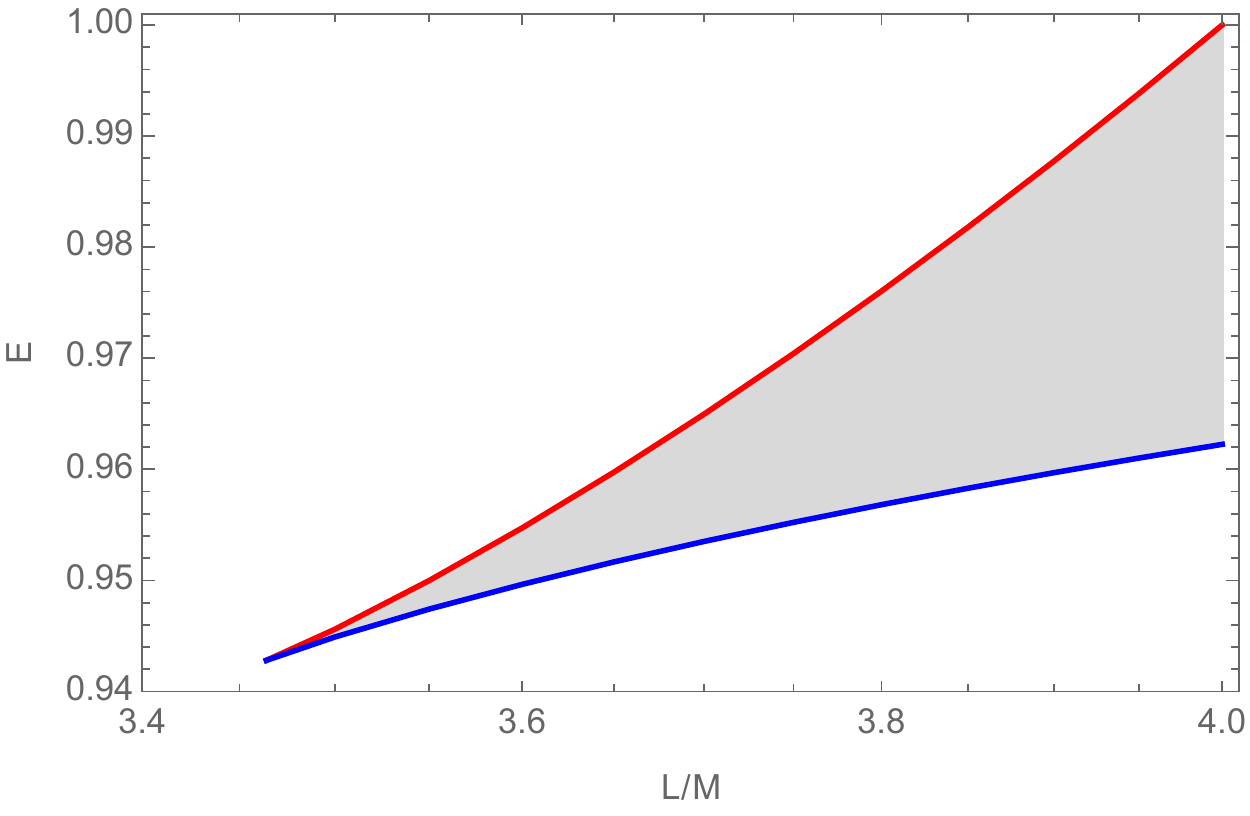}\\
\includegraphics[width=1\hsize]{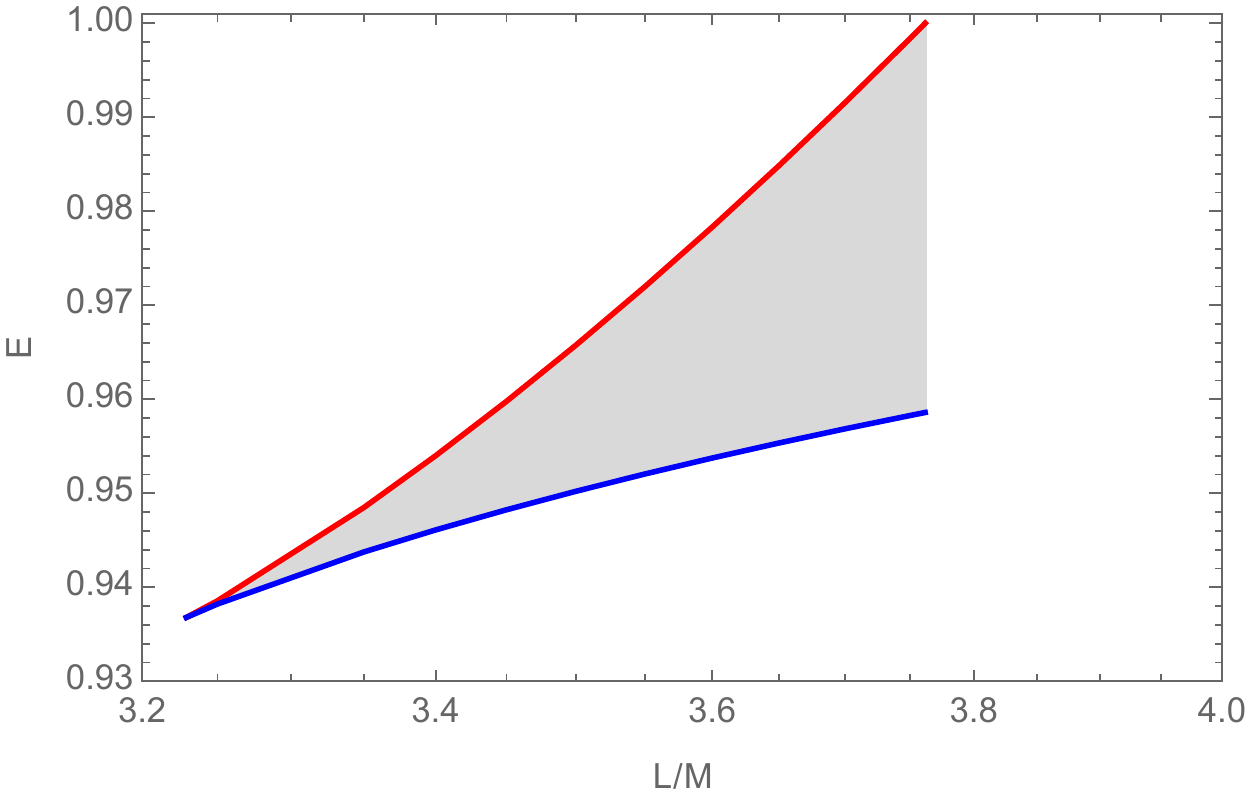} \\
\includegraphics[width=1\hsize]{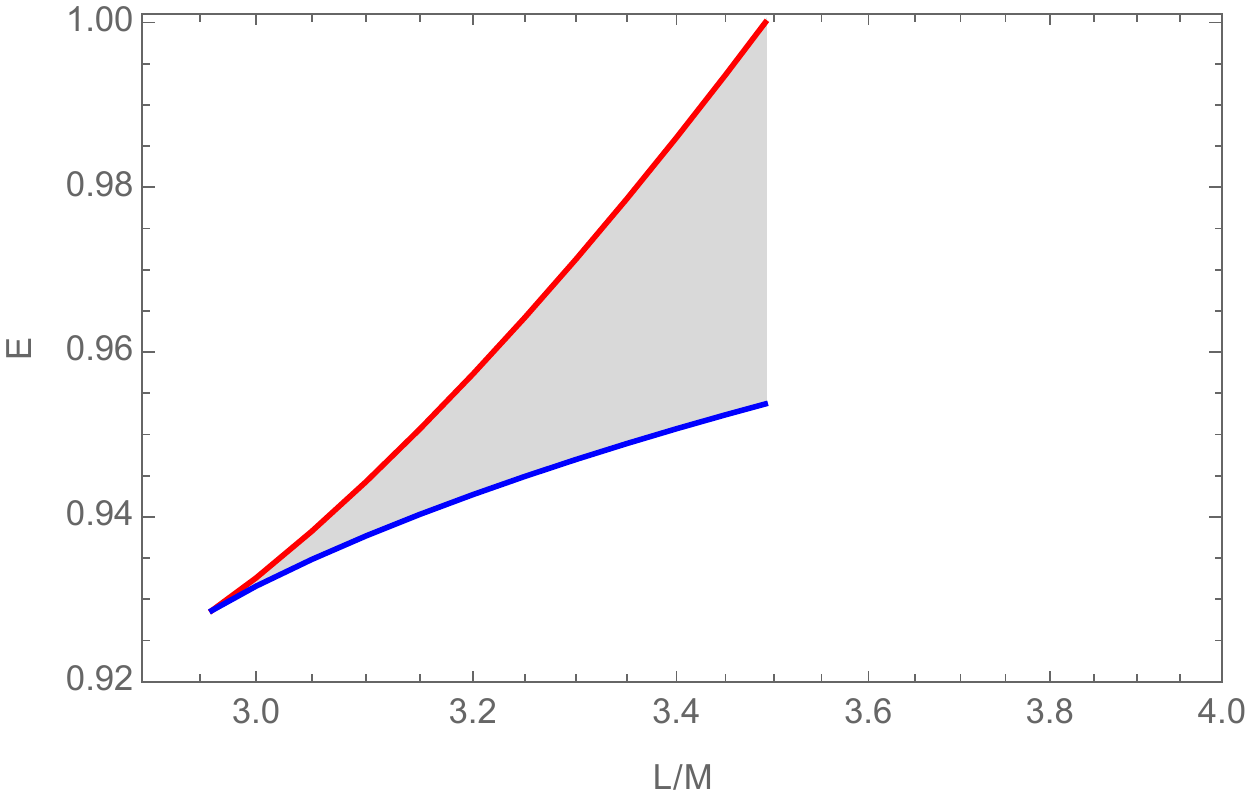} 
\caption{The allowed regions (in shadow) of the ($L$, $E$) for the timelike particle’s bound orbits around polymer black holes in LQG. The values of $A_\lambda$ from top to bottom are 0, 0.1, 0.2. The shadow area moves to the left as $A_\lambda$ becomes larger. In these figures, the red and blue boundaries of the shadow regions represent the outermost and innermost stable circular orbits for different angular momenta $L$, respectively.}
\label{shadow}
\end{figure}

\subsection{Innermost stable circular orbits}

\begin{table*}
\caption{The energy $E$ for the orbits with different $(z, w, v)$ and different black hole parameter $A_\lambda$. The angular momentum parameter $\epsilon=0.3$. Note that $A_\lambda=0$ denotes the Schwarzschild black hole case.}\label{tab1}
 \centering 
\begin{ruledtabular}
    \begin{tabular}{ccccccccc} 
  $A_\lambda$ & $E_{(1,1,0)}$ & $E_{(1,2,0)}$ & $E_{(2,1,1)}$ & $E_{(2,2,1)}$
  &$E_{(3,1,2)}$ & $E_{(3,2,2)}$& $E_{(4,1,3)}$ & $E_{(4,2,3)}$\\
  \hline
  0  & 0.953628 & 0.957086 & 0.956607 & 0.957170
     & 0.956864 & 0.957178 & 0.956946 & 0.957181\\
     \hline
  0.02& 0.957761 & 0.955270 & 0.954769 & 0.955360
     & 0.955036 & 0.955369 & 0.955122 & 0.955371\\
     \hline
  0.04& 0.949823 & 0.953483 & 0.952960 & 0.953579
     & 0.953239 & 0.953589 & 0.953328 & 0.953592\\
     \hline
  0.06& 0.947972 & 0.951731 & 0.951186 & 0.951833
     & 0.951475 & 0.951843 & 0.951568 & 0.951847\\
     \hline
  0.08& 0.946162 & 0.950018 & 0.949452 & 0.950126
     & 0.949751 & 0.950137 & 0.949848 & 0.950140\\
     \hline
  0.1& 0.944402 & 0.948351 & 0.947765 & 0.948464
     & 0.948073 & 0.948475 & 0.948173 & 0.948479
\end{tabular}
   \end{ruledtabular}
\end{table*}

\begin{table*}
\caption{The energy $E$ for the orbits with different $(z, w, v)$ and different black hole parameter $A_\lambda$. The angular momentum parameter $\epsilon=0.5$.}\label{tab2}
 \centering 
\begin{ruledtabular}
\begin{tabular}{ccccccccc}
  $A_\lambda$ & $E_{(1,1,0)}$ & $E_{(1,2,0)}$ & $E_{(2,1,1)}$ & $E_{(2,2,1)}$
  &$E_{(3,1,2)}$ & $E_{(3,2,2)}$& $E_{(4,1,3)}$ & $E_{(4,2,3)}$\\\hline
  0  & 0.965425 & 0.968383 & 0.968026 & 0.968434
     & 0.968225 & 0.968438 & 0.968285 & 0.96844\\\hline
  0.02& 0.962852 & 0.965915 & 0.965538 & 0.965971
     & 0.965747 & 0.965975 & 0.965810 & 0.965977\\\hline
  0.04& 0.960303 & 0.963474 & 0.963075 & 0.963535
     & 0.963295 & 0.963540 & 0.963362 & 0.963541\\\hline
  0.06& 0.957782& 0.961063 & 0.960642 & 0.961129
     & 0.960873 & 0.961135 & 0.960944 & 0.961137\\\hline
  0.08& 0.955293 & 0.958687 & 0.958243 & 0.958758
     & 0.958485 & 0.958765 & 0.958560 & 0.958767\\\hline
  0.1& 0.952843 & 0.956350 & 0.955883 & 0.956427
     & 0.956136 & 0.956434 & 0.956216 & 0.956436
\end{tabular}
   \end{ruledtabular}
\end{table*}
As we mentioned in above, the marginally bound orbit corresponds to the bound orbit that has the maximum energy $E=1$. All the bound orbits which have energy $E<1$ can only exist beyond $r_{\rm mbo}$, i.e., $r>r_{\rm mbo}$. The stabilities of these orbits are determined by the sign of $d^2 V_{\rm eff}(r)/dr^2$. Stable orbits correspond to $d^2 V_{\rm eff}(r)/dr^2>0$, and unstable ones have $d^2 V_{\rm eff}(r)/dr^2<0$. The critical condition,
\bqn\lb{third}
\frac{d^2V_{\rm eff}(r)}{dr^2}=0,
\eqn
together with the conditions in (\ref{mar_condition}) for $E<1$ determine the radius of the ISCOs. 
As we mentioned previously, the bound orbits can only exist when $E<1$. For this reason, in order to find the ISCO, which represents the innermost and stable bounded orbit of the polymer black hole, one has to ensure $E<1$ here. 

For a polymer black hole in LQG, these conditions yield,
\bqn
E_{\rm isco} &=&\sqrt{(1+\frac{L_{z}^2}{{\cal B}(r)})\tilde{\cal A}(r)},
\eqn
\bqn
L_{\rm isco} &=&\sqrt{\frac{\tilde{\cal A}^\prime(r){\cal B}^2(r)}{{\cal B}(r)\tilde{\cal A}^\prime(r)-\tilde{\cal A}(r){\cal B}^\prime(r)}},
\eqn
with the radius $r_{\rm isco}$ of the ISCO satisfying the relation
\bqn
\label{risco}
&&\frac{L^2 \tilde{\mathcal{A}}(r) \mathcal{B}^{\prime}(r)^2}{\mathcal{B}(r)^3}-\frac{L^2 \mathcal{B}^{\prime}(r) \tilde{\mathcal{A}}^{\prime}(r)}{\mathcal{B}(r)^2} -\frac{L^2 \tilde{\mathcal{A}}(r) \mathcal{B}^{\prime \prime}(r)}{2 \mathcal{B}(r)^2}\nb\\ 
&&+\frac{1}{2}\left(1+\frac{L^2}{\mathcal{B}(r)}\right) \tilde{\mathcal{A}}^{\prime \prime}(r)-r_{\rm isco}=0.
\eqn
Again, with the LQG effects, the above equations in general do not admit exact solutions and one can find approximate solutions by treating the LQG parameter $A_\lambda \ll 1$, which gives 
\bqn
L_{\rm isco}\simeq2\sqrt{3}M \left(1-\frac{23}{36}A_\lambda \right),
\eqn
and
\bqn
r_{\rm isco} \simeq6M(1-A_\lambda),
\eqn
where $L_{\rm isco}$ and $r_{\rm isco}$ represent the angular momentum and radius of the ISCO in the polymer black hole in LQG. This approximate result clearly shows that the positive LQG parameter $A_\lambda$ tends to decrease the radius of the ISCO. In Fig.~\ref{picisco}, we plot the results of $r_{\rm isco}$, $E_{\rm isco}$, and $L_{\rm isco}$ with respect to the LQG parameter $A_\lambda$ of the polymer black hole in LQG. It is shown that the radius, energy, and angular momentum for the ISCO all decrease with  $A_\lambda$. When the  $A_\lambda$ are absent (i.e. $A_\lambda=0$), all these quantities reduce to those of the Schwarschild black hole.

Fig. \ref{veff} shows the behaviors of the effective potential with LQG parameter $A_\lambda=0.1$ (left panel) and $A_\lambda=0.2$ (right panel) respectively. The corresponding angular momentum $L$ for each curve in both figures vary from $L_{\rm isco}$ to $L_{\rm mbo}$ from top to bottom. The extremal points of the effective potential $V_{\rm eff}$ for each curve are represented by the dashed line. In addition, the effective potential of ISCO has only one extreme value, and in other cases, there are two extreme values. To support a bound orbit, the energy $E$ has to be restricted to be $E_{\rm isco}^2 \leq E  \leq E_{\rm mbo}^2=1$ for a given particle. It indicates that $E$ cannot be too high, otherwise, there is no solution, and $E$ cannot be too small, otherwise, the particle will fall into the black hole. The allowed range of $E$ also depends on the angular momentum $L$ of the particle. In  Fig. \ref{shadow}, we plot the allowed regions of $E-L$ in the $E-L$ diagram for the bound orbits in the polymer black hole in LQG for different values of $A_\lambda$. 


\section{Periodic orbits}
\label{Periodic}
\renewcommand{\theequation}{4.\arabic{equation}} \setcounter{equation}{0}

In this section, we shall seek the periodic timelike orbits around the polymer black holes in LQG. This is a spherically symmetric black hole. We adopt the taxonomy introduced in \cite{Levin:2008mq} for indexing different periodic orbits around the polymer black holes in LQG with a triplet of integers $(z, w, v)$, which denote the zoom, whirl, and vertex behaviors. Normally, the periodic orbits are those orbits that can return exactly to their initial conditions after a finite time, which requires that the ratio between the two frequencies of oscillations in the $r$-motion and $\phi$-motion has to be a rational number. And a generic orbit around the black hole can be approximated by a nearby periodic orbit since any irrational number can be approximated by a nearby rational number. Therefore, the exploration of the periodic orbits would be very helpful for understanding the structure of any generic orbits and the corresponding radiation of the gravitational waves. It plays an important role in the study of gravitational wave radiation.

According to the taxonomy of ref.~\cite{Levin:2008mq}, we introduce the ratio $q$ between the two frequencies, $\omega_r$ and $\omega_\phi$ of oscillations in the $r$-motion and $\phi$-motion respectively, in terms of three integers $(z, w, v)$ as
\bqn
q\equiv \frac{\omega_\phi}{\omega_r}-1 =\frac{\bigtriangleup\phi}{2\pi}-1= w + \frac{v}{z}.
\eqn
Since $\frac{\omega_\phi}{\omega_r}=\Delta \phi/(2\pi)$ with $\Delta \phi \equiv \oint d\phi$ being the equatorial angle during one period in $r$, which is required to be an integer multiple of $2\pi$. Using the geodesic equations of the  polymer black holes in LQG, $q$ can be calculated via
\bqn
q &=& \frac{1}{\pi} \int_{r_1}^{r_2} \frac{\dot \phi}{\dot r} dr -1\nb\\
&=& \frac{1}{\pi} \int_{r_1}^{r_2} \frac{L}{{\cal B}(r)\sqrt{E^2- \left(1+\frac{L_{z}^2}{ {\cal B}(r)}\right)\tilde{\cal A}(r)}}-1,
\eqn

\begin{figure*}[htbp]
\centering
\subfigure[]{\includegraphics[width=8cm]{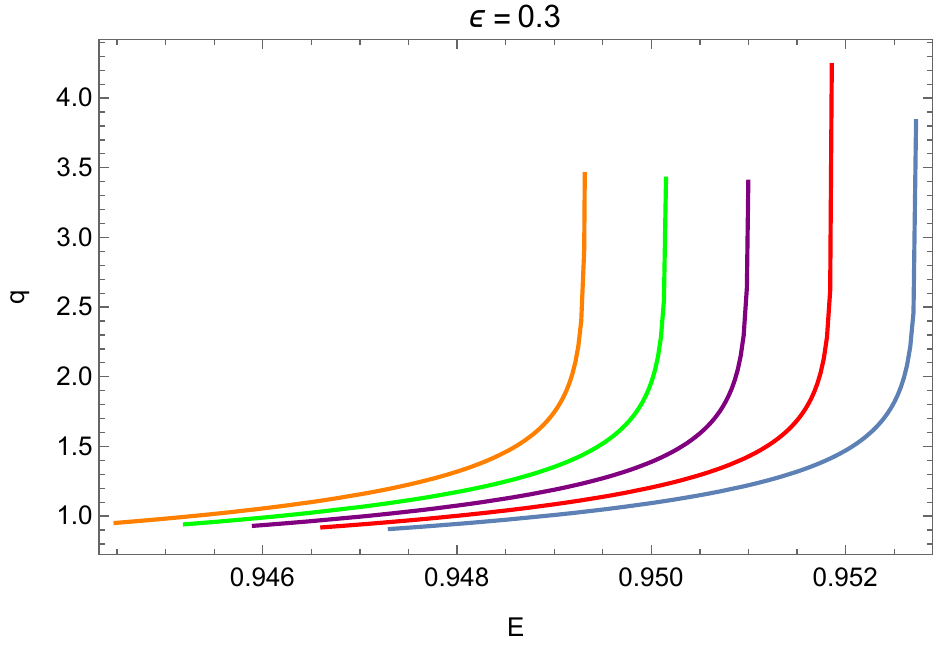}}
\subfigure[]{\includegraphics[width=8cm]{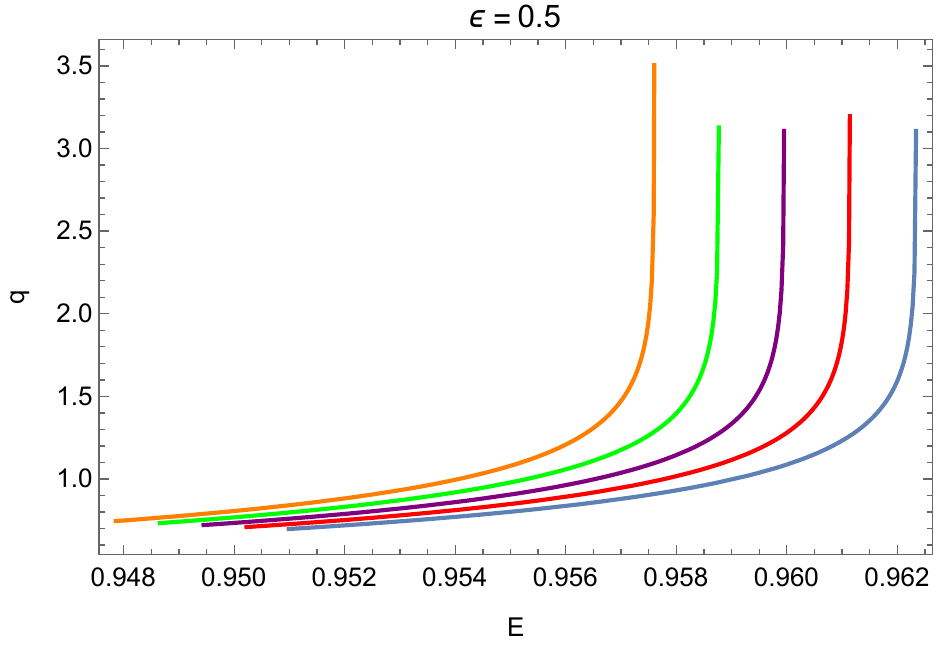}}
\subfigure[]{\includegraphics[width=8cm]{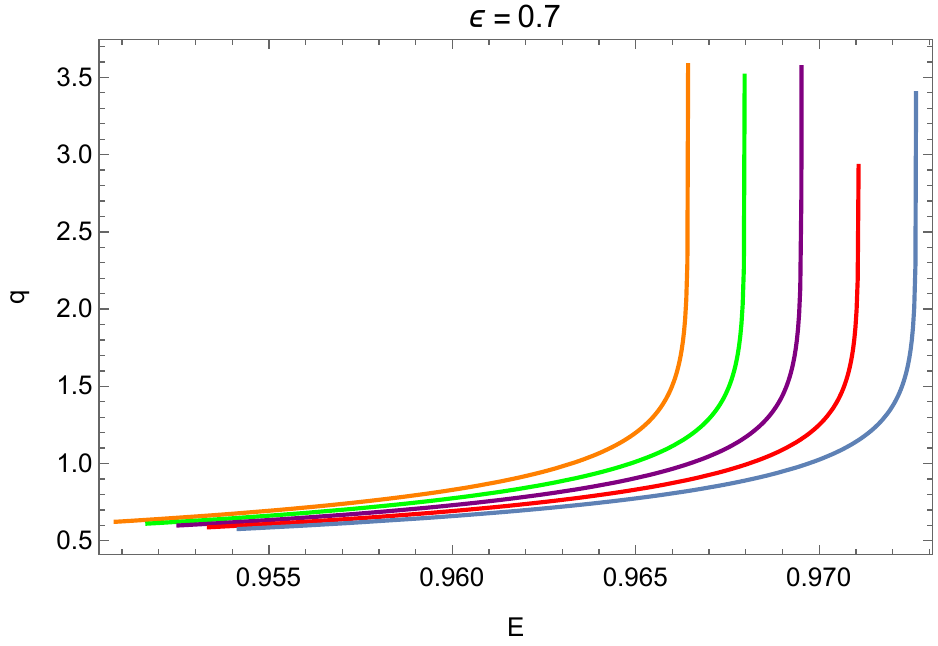}}
\subfigure[]{\includegraphics[width=8cm]{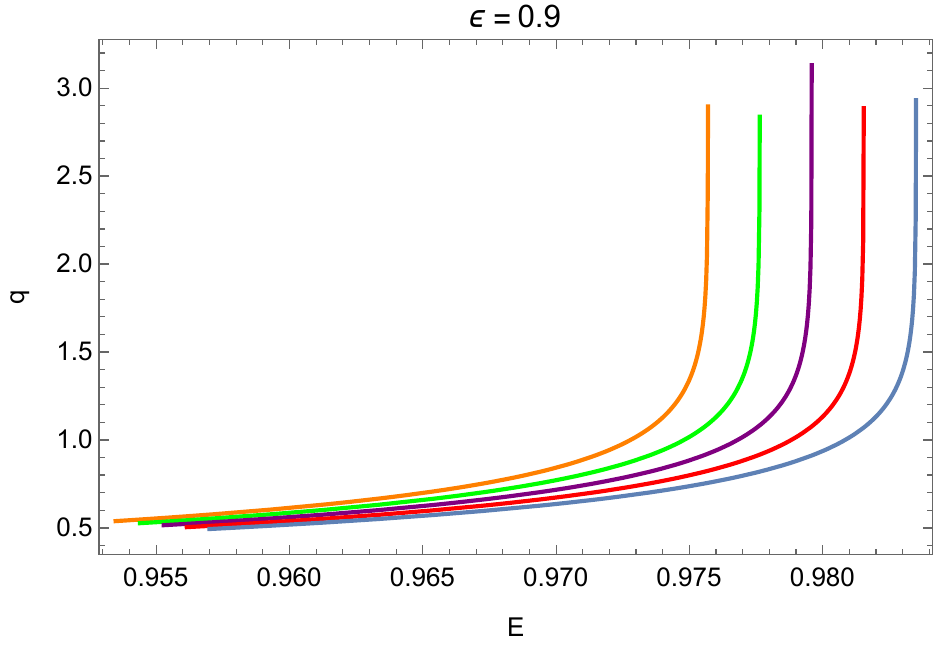}}
\caption{$q$ vs $E$. The parameter $A_\lambda=0.05, 0.06, 0.07, 0.08, 0.09$, and from right to left. (a)\;$\epsilon=0.3$. (b)\;$\epsilon=0.5$. (c)\;$\epsilon=0.7$. (d)\;$\epsilon=0.9$.}
\label{veffa}
\end{figure*}

where $r_1$ and $r_2$ are two turning points. For different periods, the azimuth through which particles pass $\bigtriangleup\phi$ can be expressed as
\bqn
\bigtriangleup\phi&=& 2\oint d\phi,
\eqn
Similarly, using the geodesic equations of the polymer black holes in LQG, $\bigtriangleup\phi$ can be calculated via
\bqn
\bigtriangleup\phi&=& 2\int_{\phi_1}^{\phi_2} d\phi\nb\\
&=&2\int_{r_1}^{r_2}\frac{\dot \phi}{\dot r} dr\nb\\
&=&2 \int_{r_1}^{r_2} \frac{L}{{\cal B}(r)\sqrt{E^2- \left(1+\frac{L^2}{ {\cal B}(r)}\right)\tilde{\cal A}(r)}}dr.
\eqn
For a bound orbit, the value of its angular momentum only changes from ISCO to MBO. In order to facilitate our analysis and calculation, we write the angular momentum $L$ for a given bound orbit in the following form, 
\bqn
L=L_{\rm isco}+\epsilon(L_{\rm mbo}-L_{\rm isco}),
\eqn
where $\epsilon$=0 and  $\epsilon$=1 represent the angular momentum of ISCO and MBO respectively and will be limited to the range of $(0,1)$, because when the parameter  $\epsilon$ is greater than 1, there is no bound orbit. Therefore, the angular momentum can be determined by taking different values of the parameter $\epsilon$. In the Fig. \ref{veffa}, when we take different values $\epsilon$, the rational number $q$ for bound orbits are displayed in four subfigures by varying energy $E$, which correspond to $\epsilon=0.3$, $\epsilon=0.5$, $\epsilon=0.7$, and $\epsilon=0.9$, respectively. We find from Fig. \ref{veffa}  that the rational number $q$ increases slowly with the increase of energy $E$. When it approaches the maximum value of energy, $q$ suddenly explodes. When $\epsilon$ are the same, the maximum energy decreases with the increase of the LQG parameter $A_\lambda$. By comparing different $\epsilon$, we can also find that the maximum energy increases with the increase of $E$.

\begin{figure*}[htbp]
\centering
\subfigure[]
{\includegraphics[width=8cm]{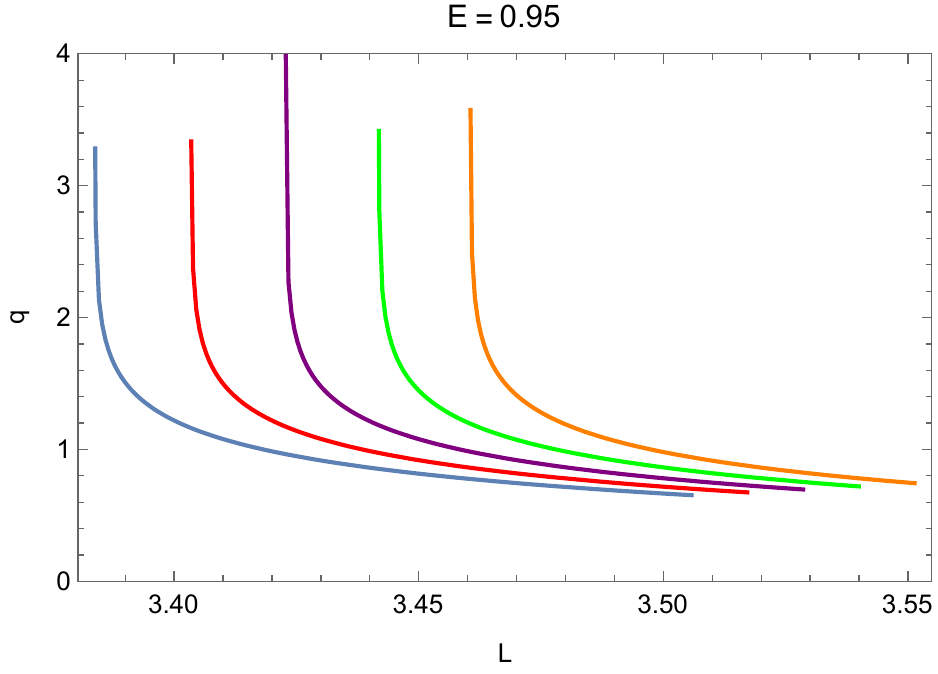}}
\subfigure[]
{\includegraphics[width=8cm]{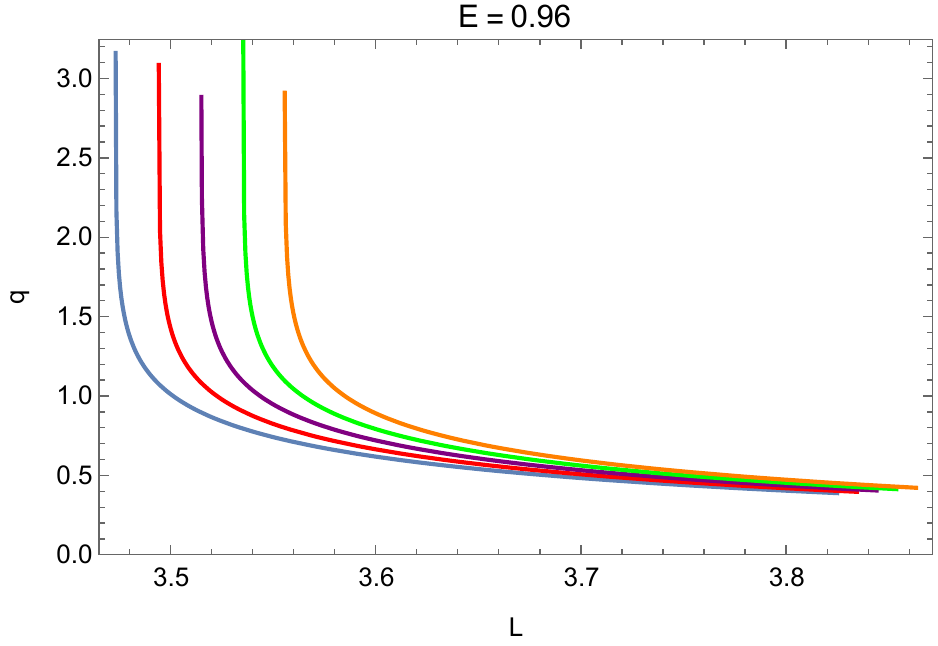}}
\subfigure[]
{\includegraphics[width=8cm]{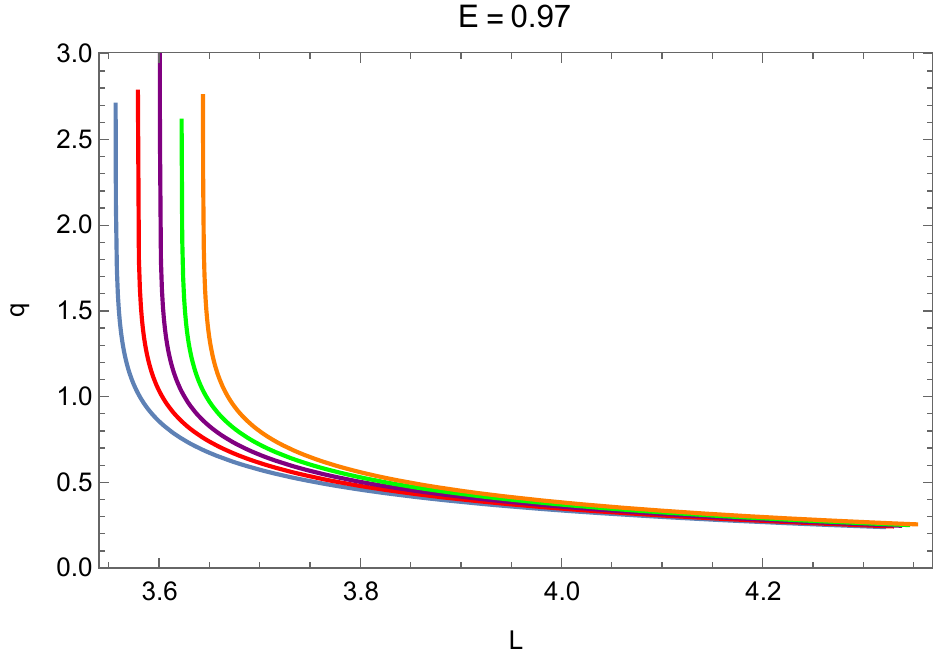}}
\subfigure[]
{\includegraphics[width=8cm]{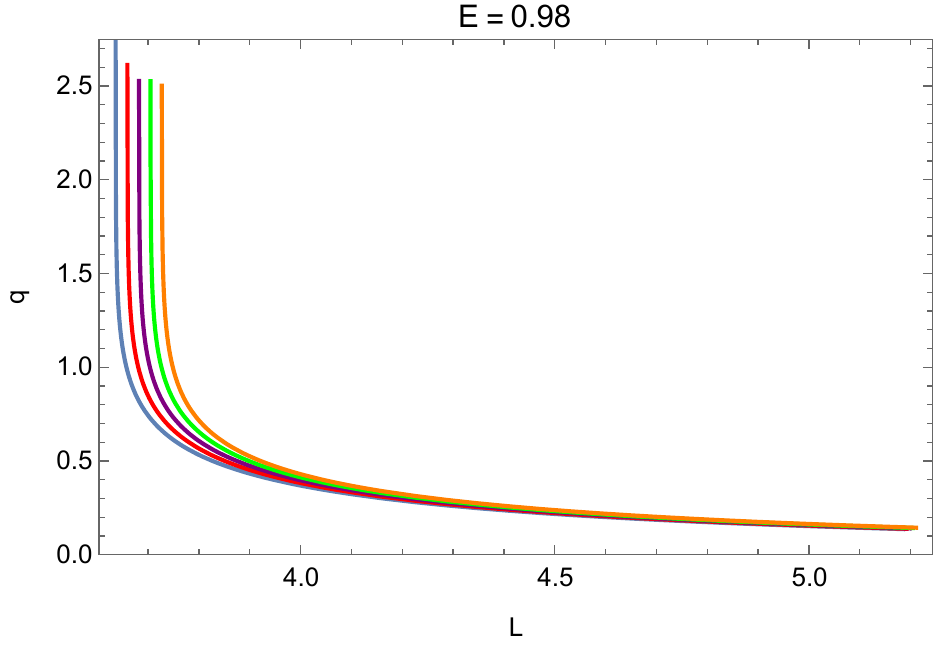}}
\caption{$q$ vs $L$. The parameter $A_\lambda=0.05, 0.06, 0.07, 0.08, 0.09$, and from right to left. (a)\;$E=0.95$. (b)\;$E=0.96$. (c)\;$E=0.97$. (d)\;$E=0.98$.}
\label{veffb}
\end{figure*}

\begin{figure*}[htbp]
\centering
\subfigure[\;$L=3.58494$]{\includegraphics[height=4.5cm,width=4.5cm]{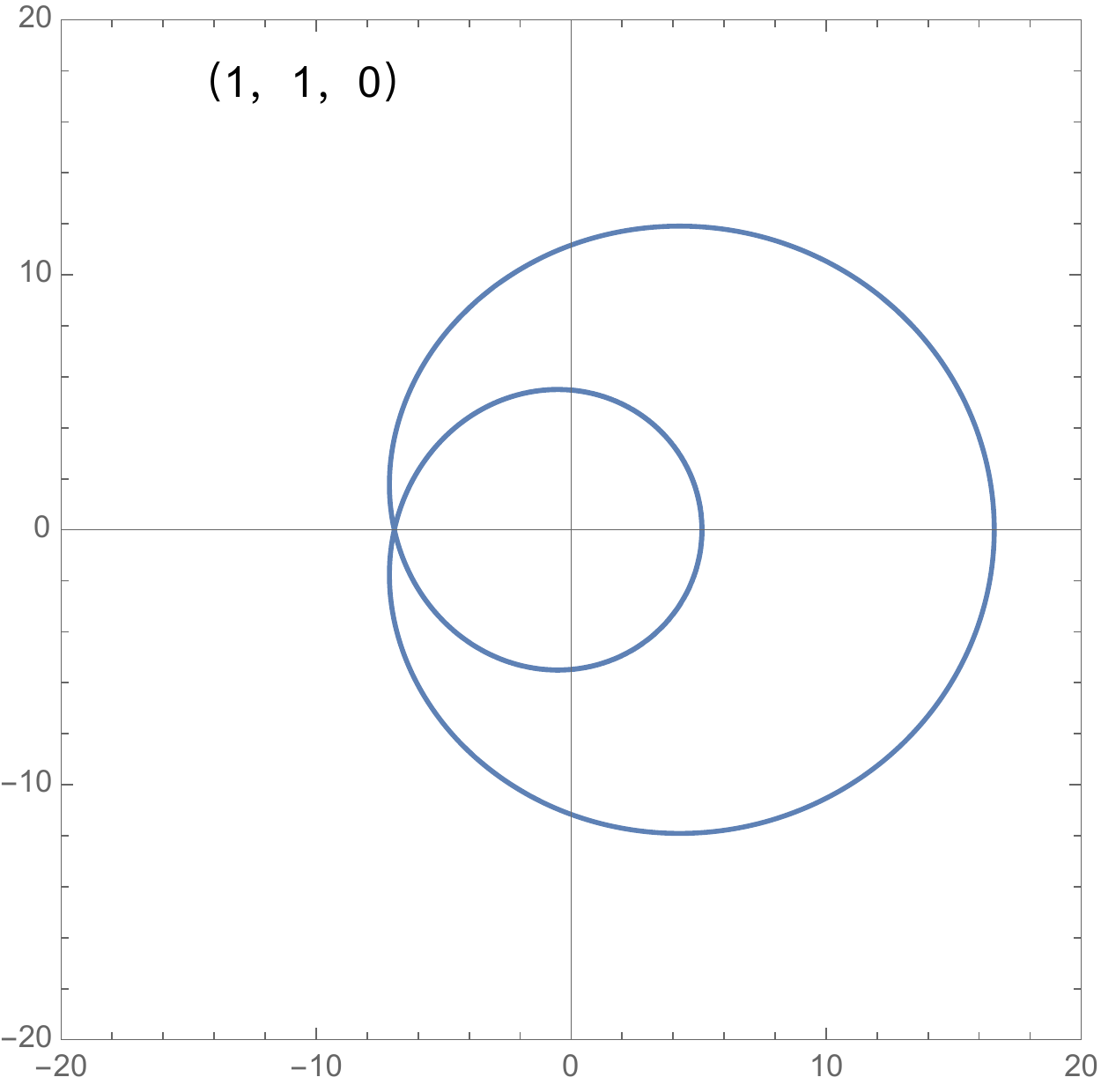}}
\subfigure[\;$L=3.55638$]{\includegraphics[height=4.5cm,width=4.5cm]{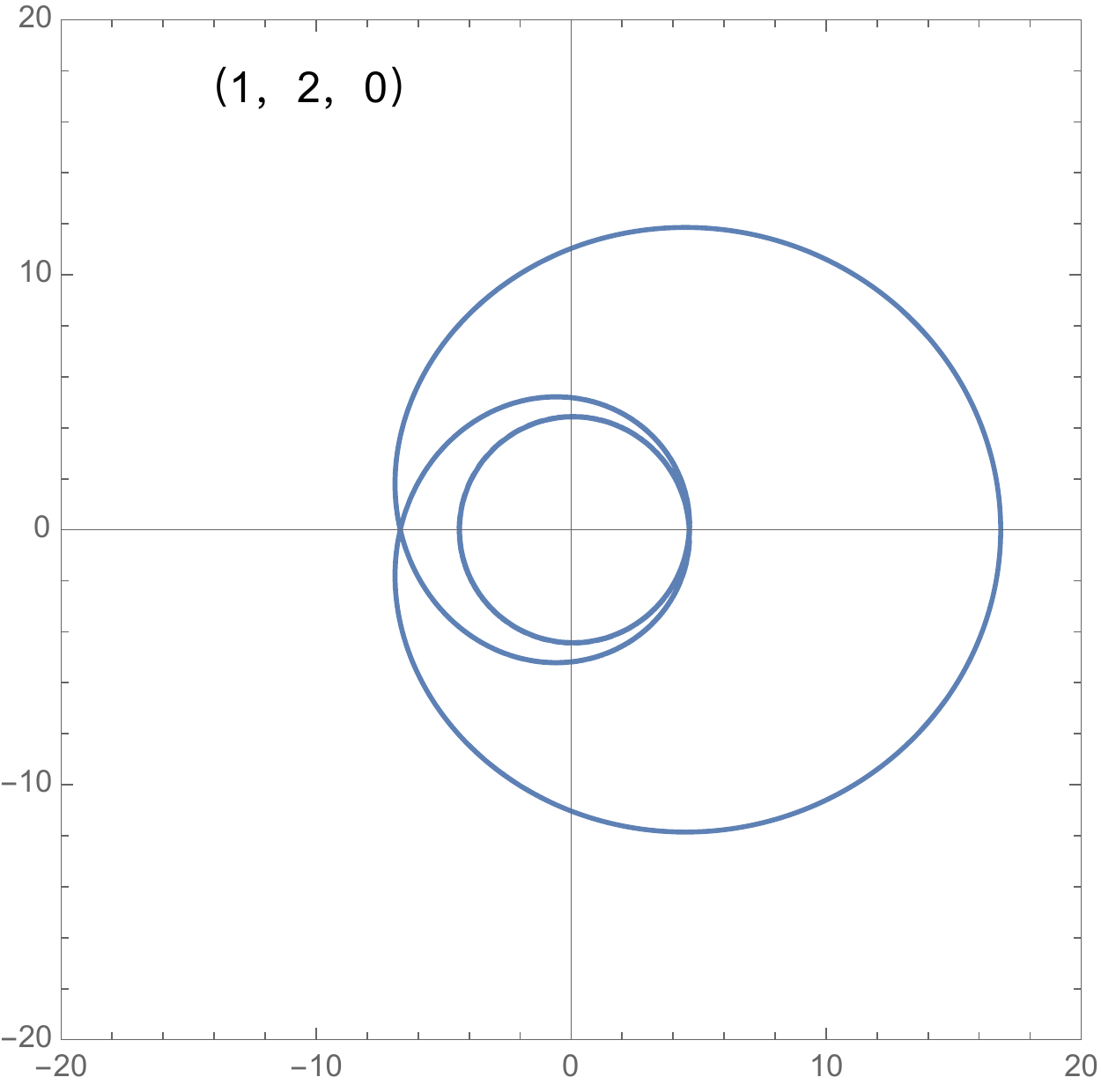}}
\subfigure[\;$L=3.55564$]{\includegraphics[height=4.5cm,width=4.5cm]{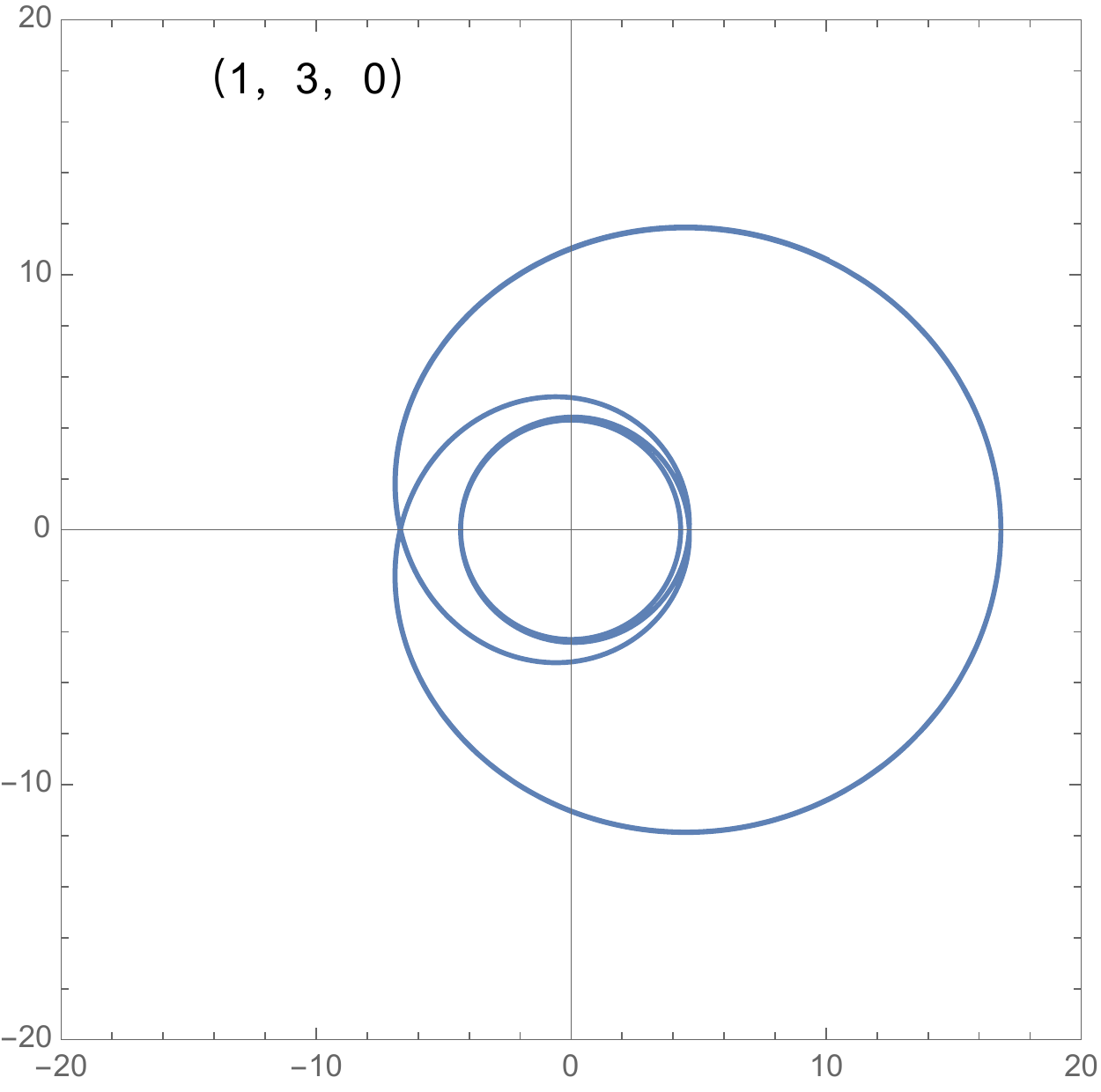}}\\
\subfigure[\;$L=3.56025$]{\includegraphics[height=4.5cm,width=4.5cm]{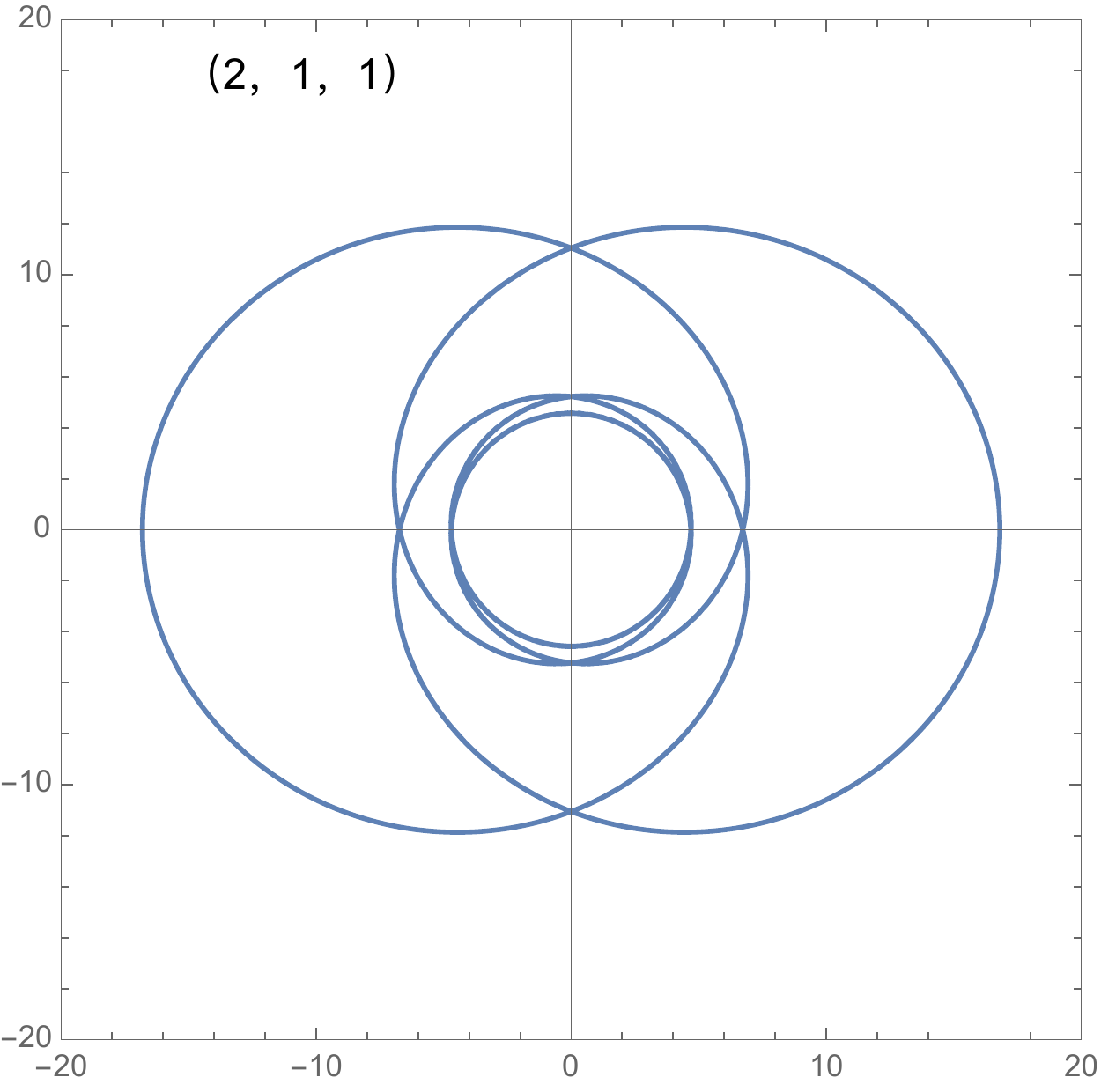}}
\subfigure[\;$L=3.55575$]{\includegraphics[height=4.5cm,width=4.5cm]{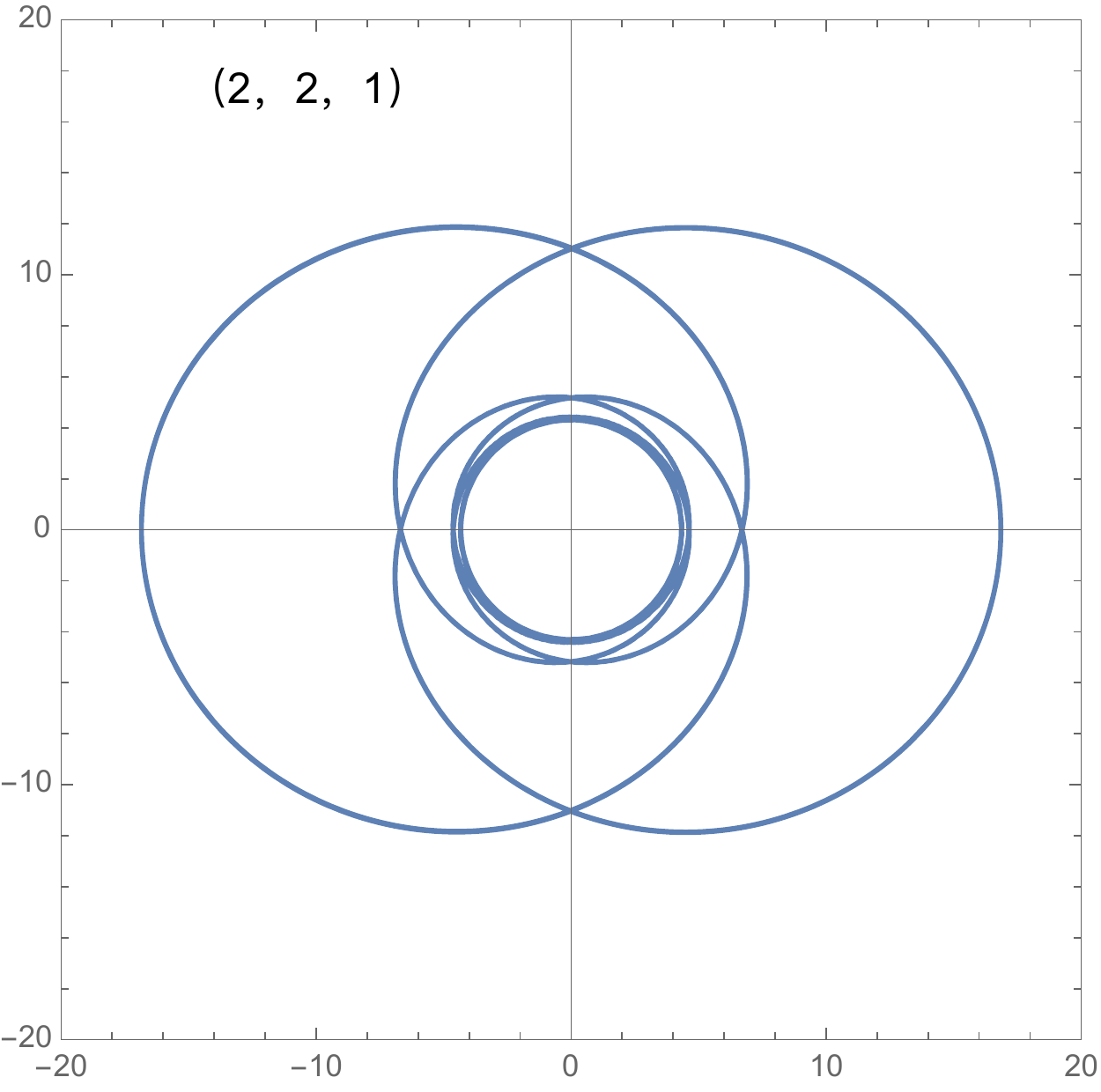}}
\subfigure[\;$L=3.55562$]{\includegraphics[height=4.5cm,width=4.5cm]{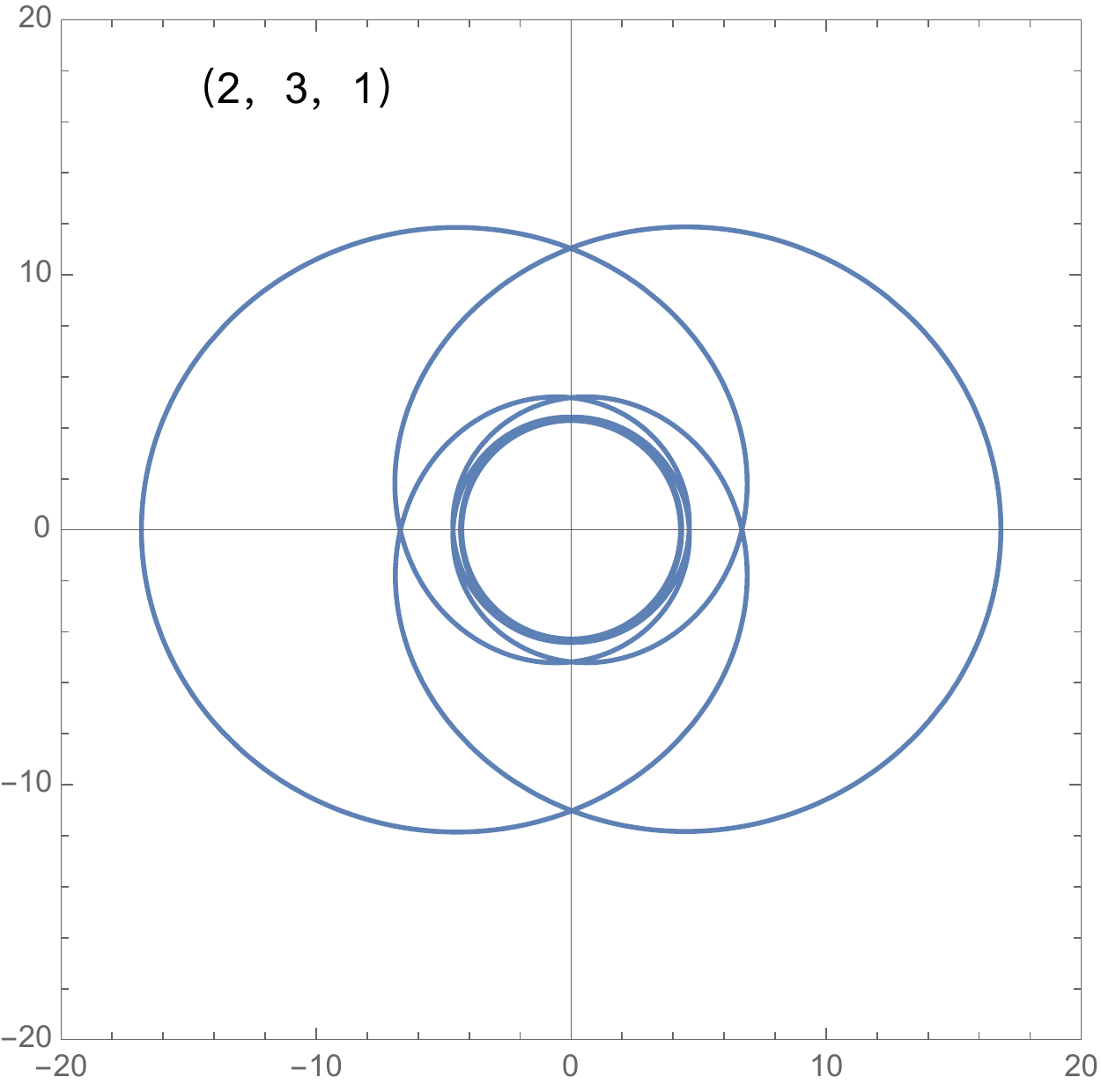}}\\
\subfigure[\;$L=3.55814$]{\includegraphics[height=4.5cm,width=4.5cm]{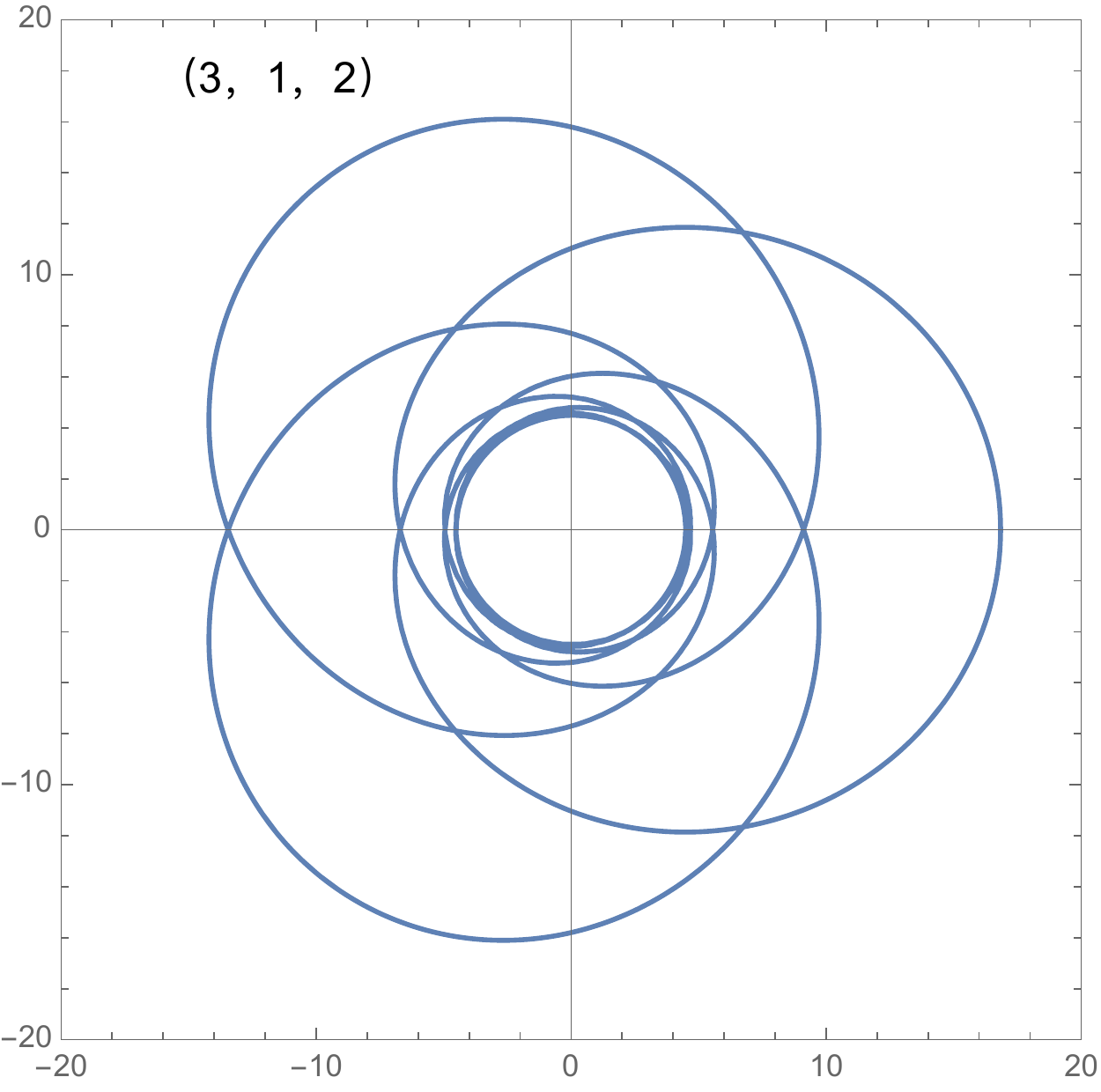}}
\subfigure[\;$L=3.55569$]{\includegraphics[height=4.5cm,width=4.5cm]{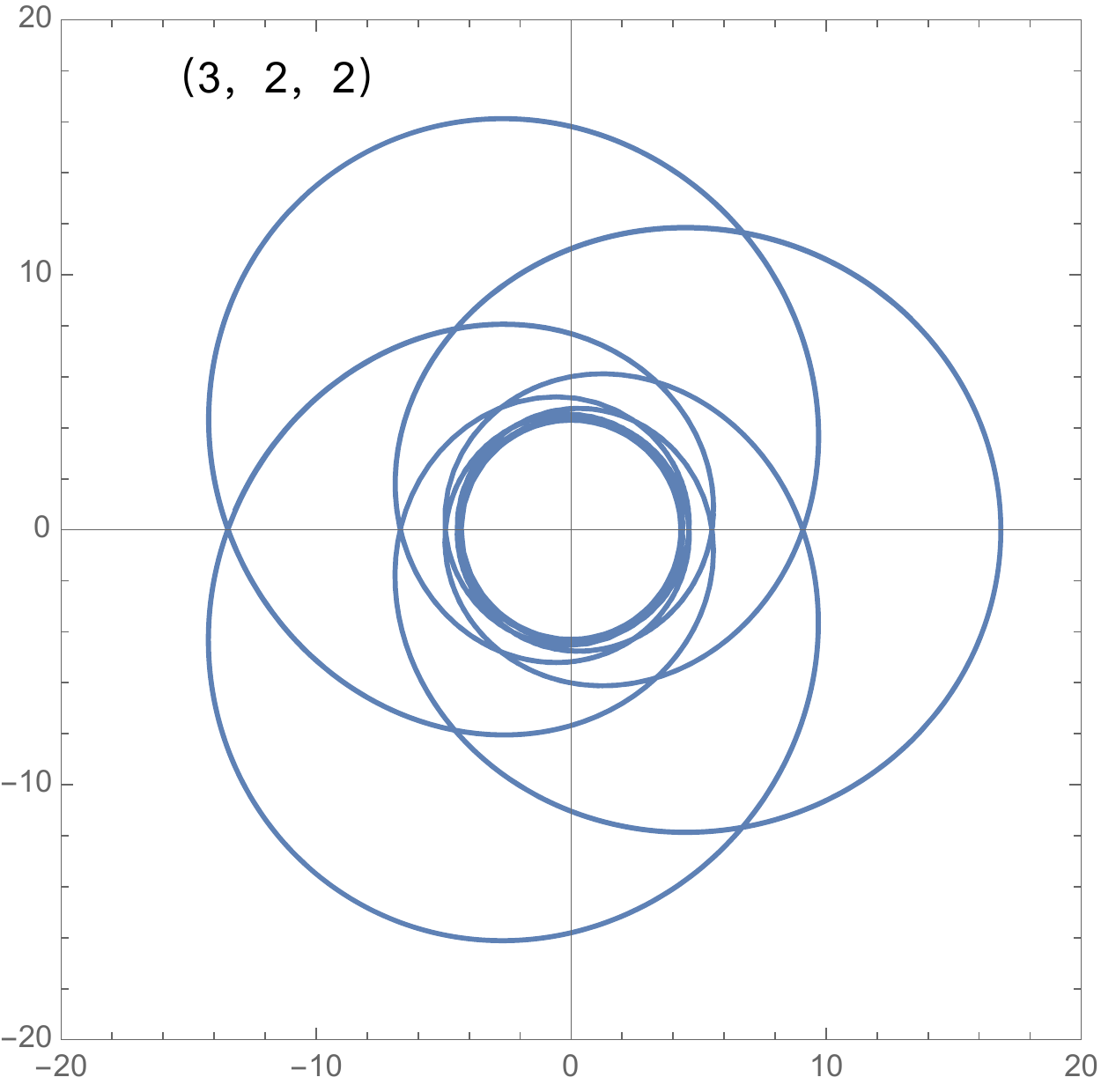}}
\subfigure[\;$L=3.55562$]{\includegraphics[height=4.5cm,width=4.5cm]{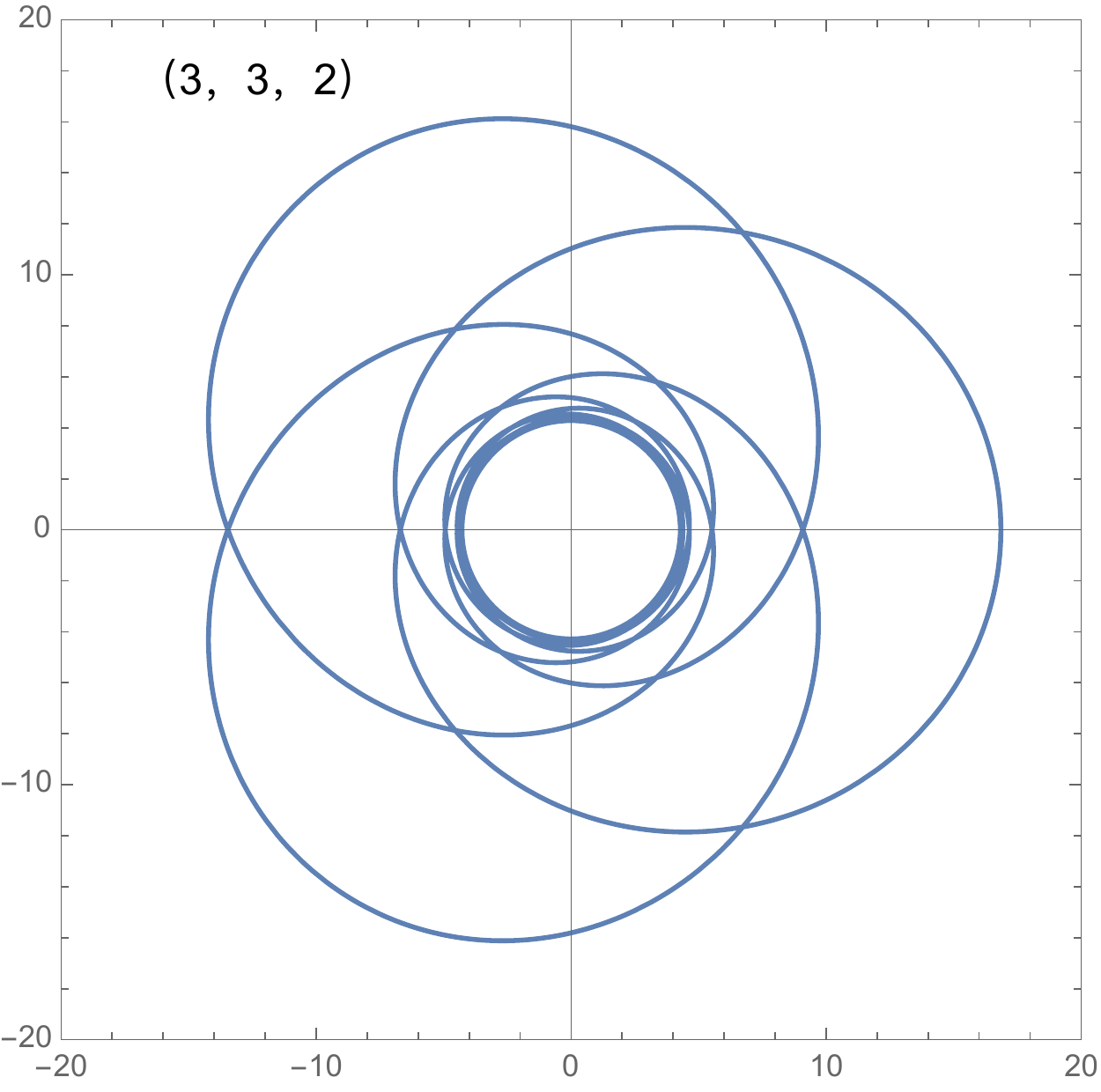}}\\
\subfigure[\;$L=3.63246$]{\includegraphics[height=4.5cm,width=4.5cm]{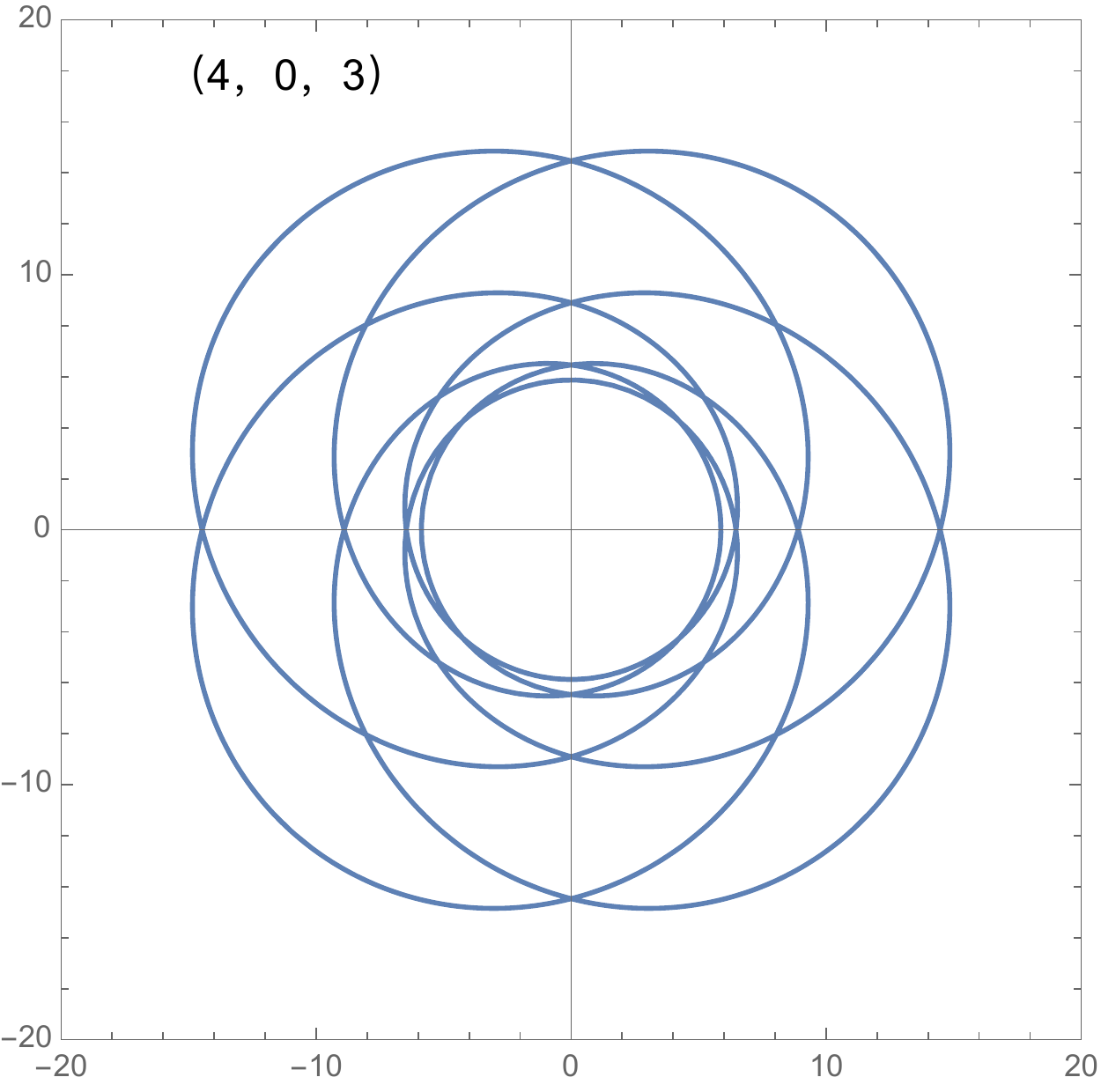}}
\subfigure[\;$L=3.55749$]{\includegraphics[height=4.5cm,width=4.5cm]{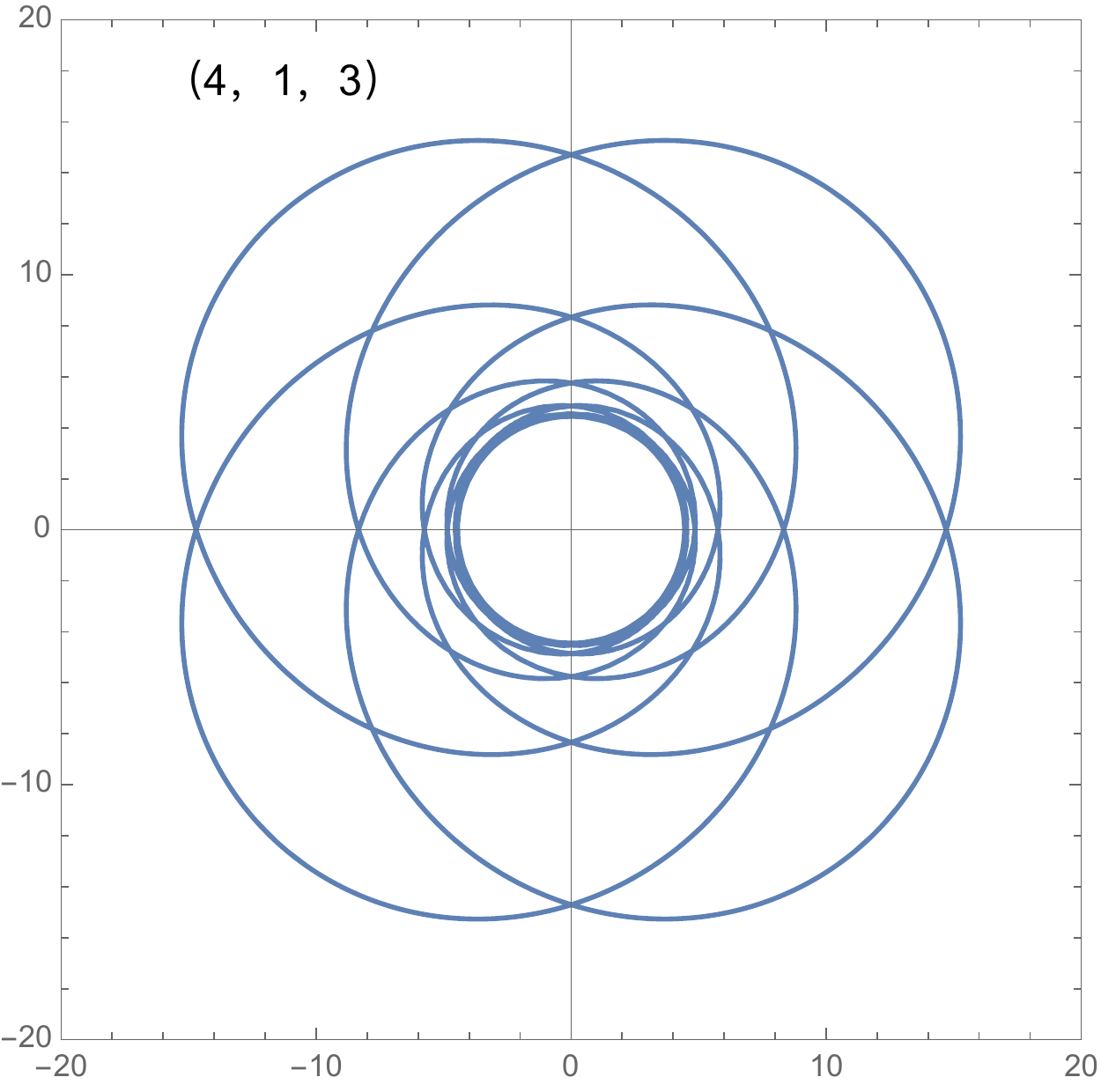}}
\subfigure[\;$L=3.55567$]{\includegraphics[height=4.5cm,width=4.5cm]{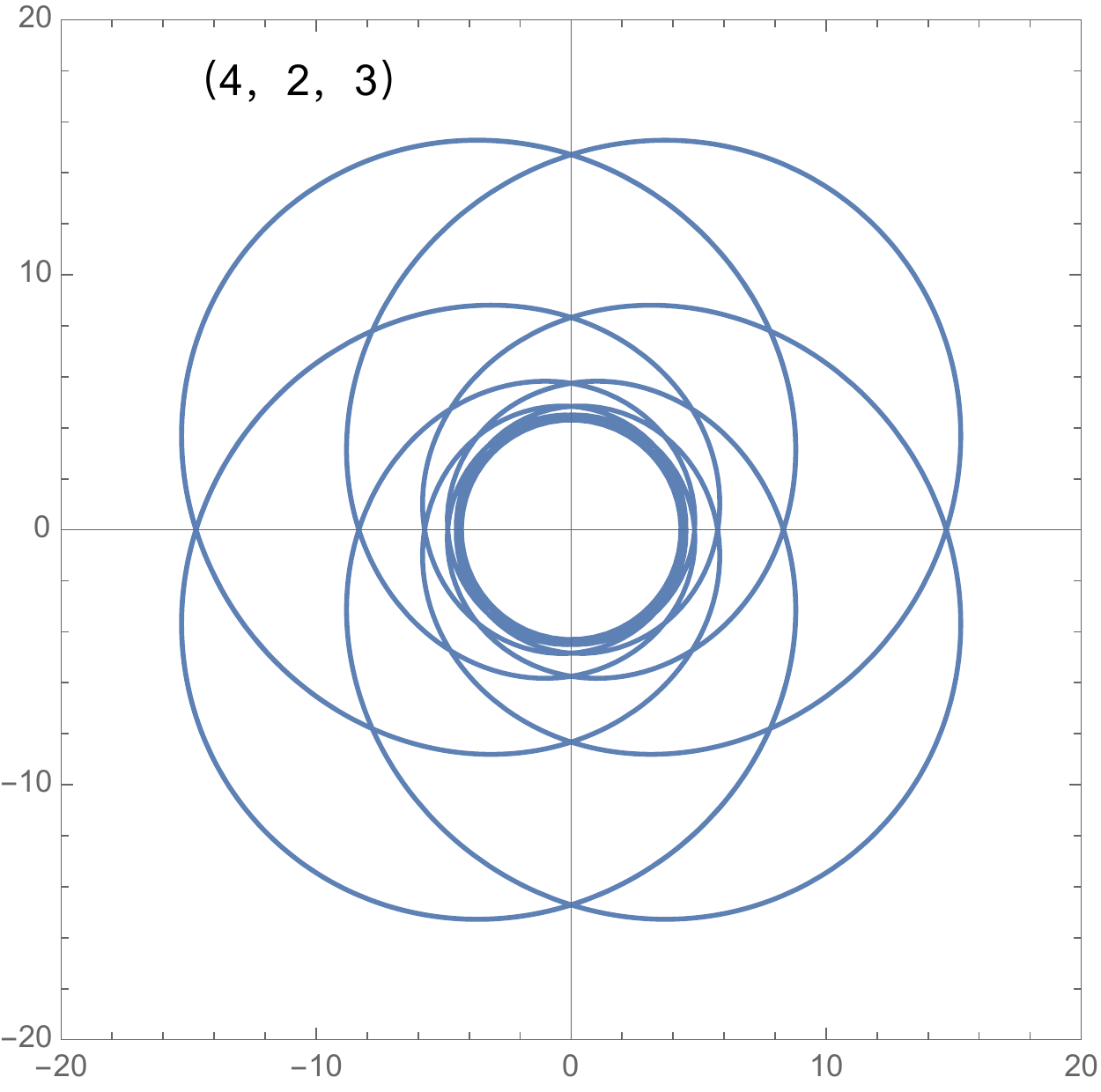}}
\caption{ Periodic orbits of different($z$, $w$, $v$) around the polymer black holes in LQG with $A_\lambda=0.05$ and $E=0.96$.}
\label{orbitse}
\end{figure*}

\begin{figure*}[htbp]
\centering
\subfigure[\;$E=0.959039$]{\includegraphics[height=4.5cm,width=4.5cm]{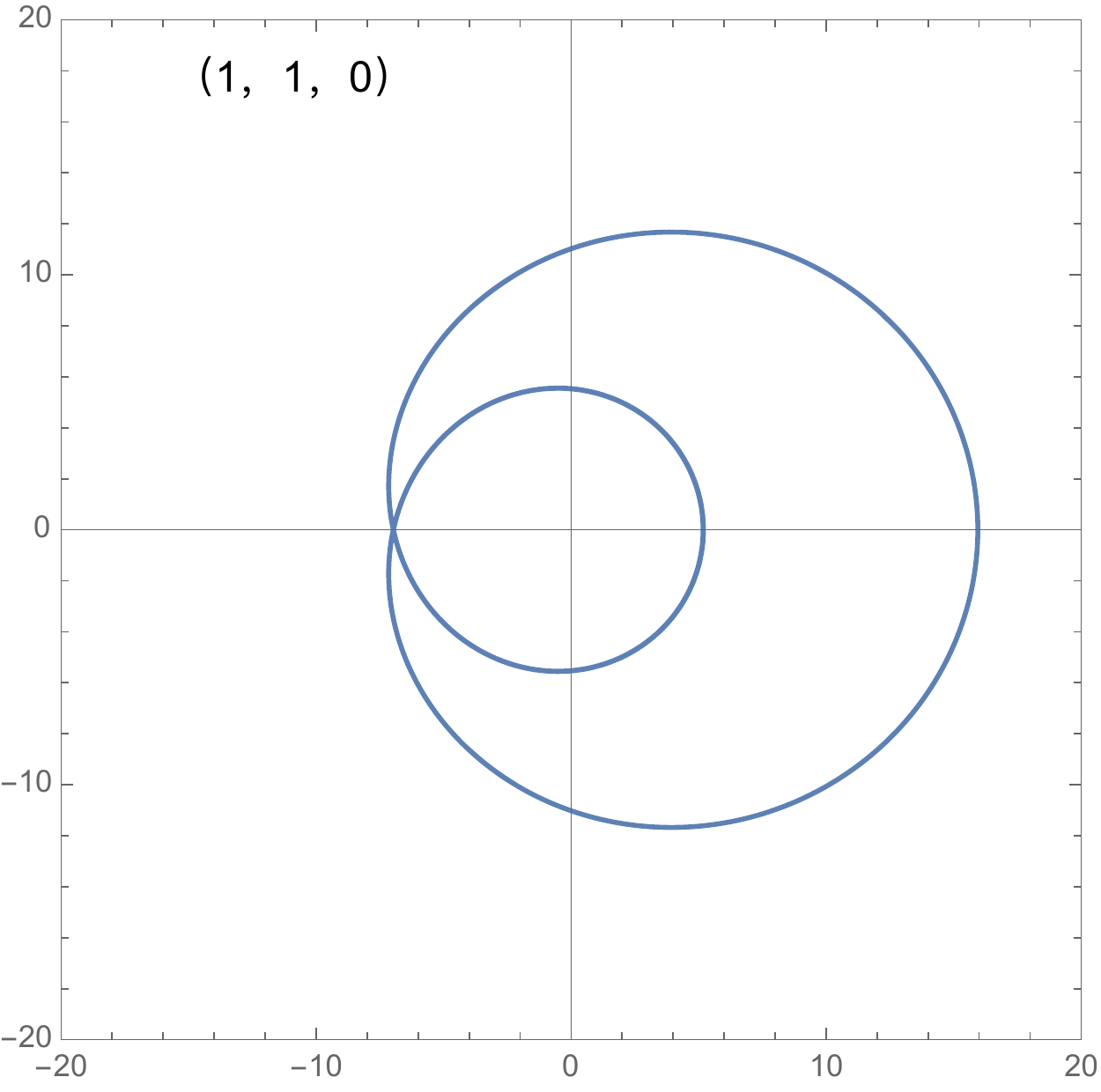}}
\subfigure[\;$E=0.962264$]{\includegraphics[height=4.5cm,width=4.5cm]{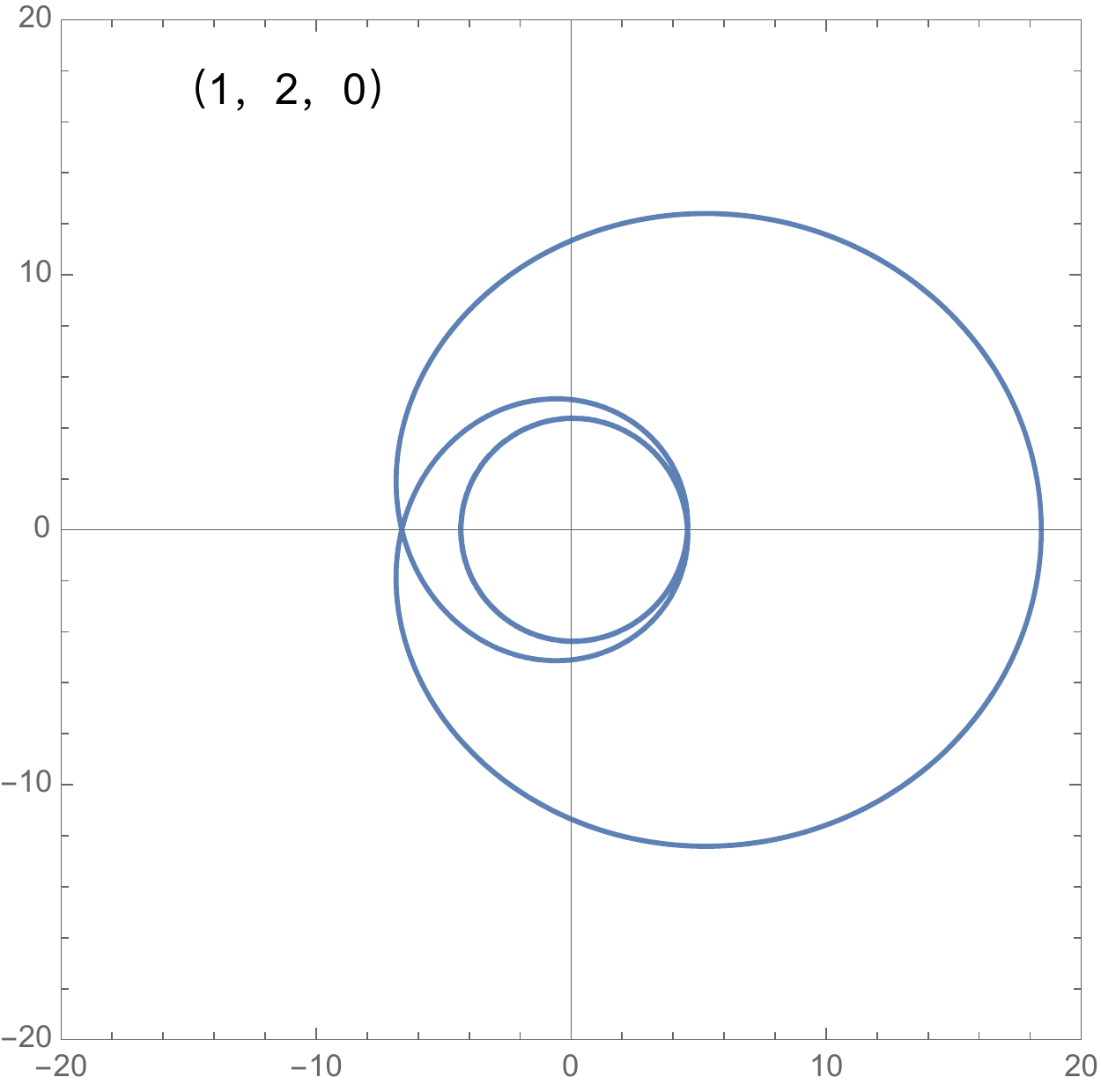}}
\subfigure[\;$E=0.962338$]{\includegraphics[height=4.5cm,width=4.5cm]{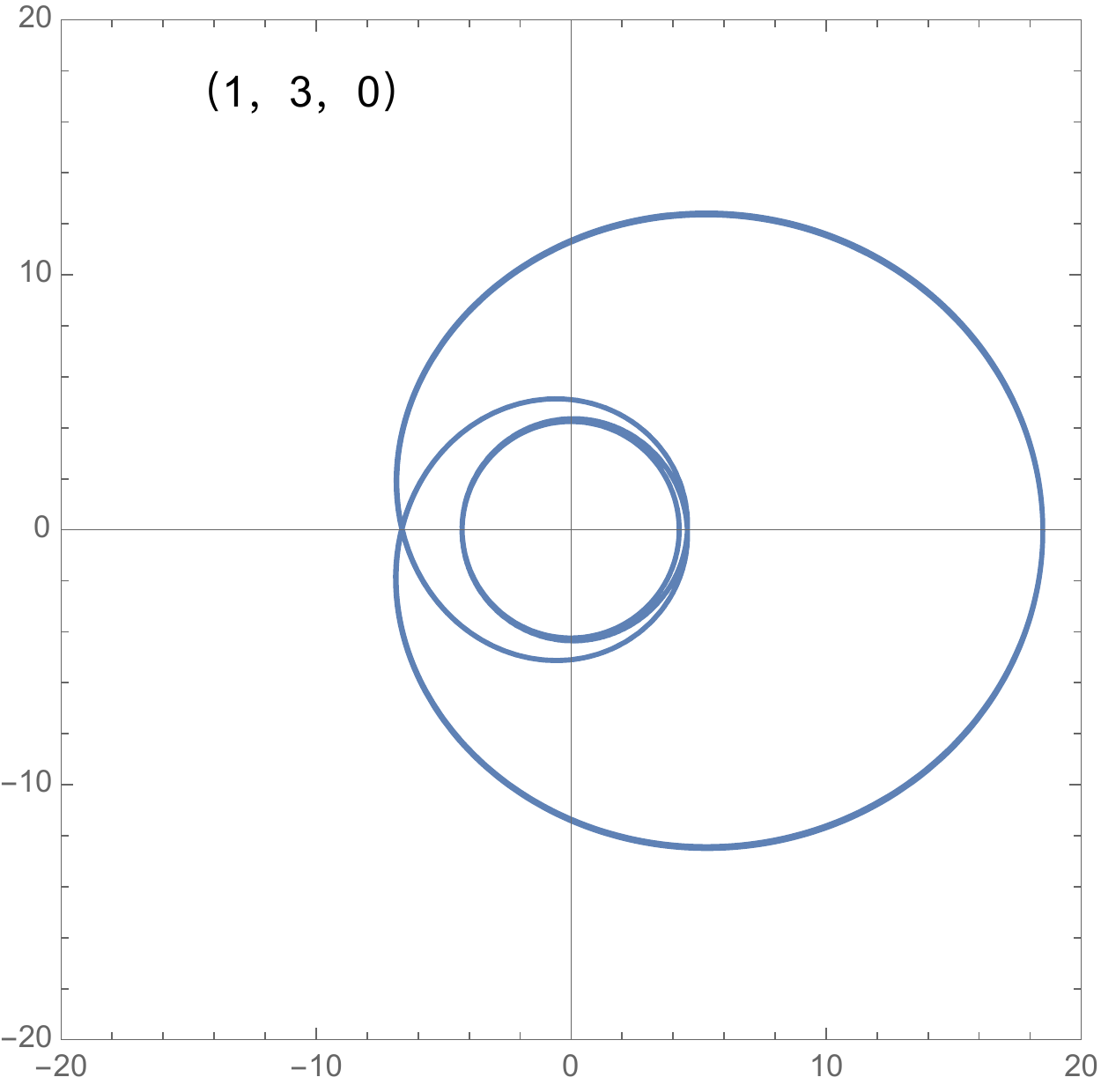}}\\
\subfigure[\;$E=0.961855$]{\includegraphics[height=4.5cm,width=4.5cm]{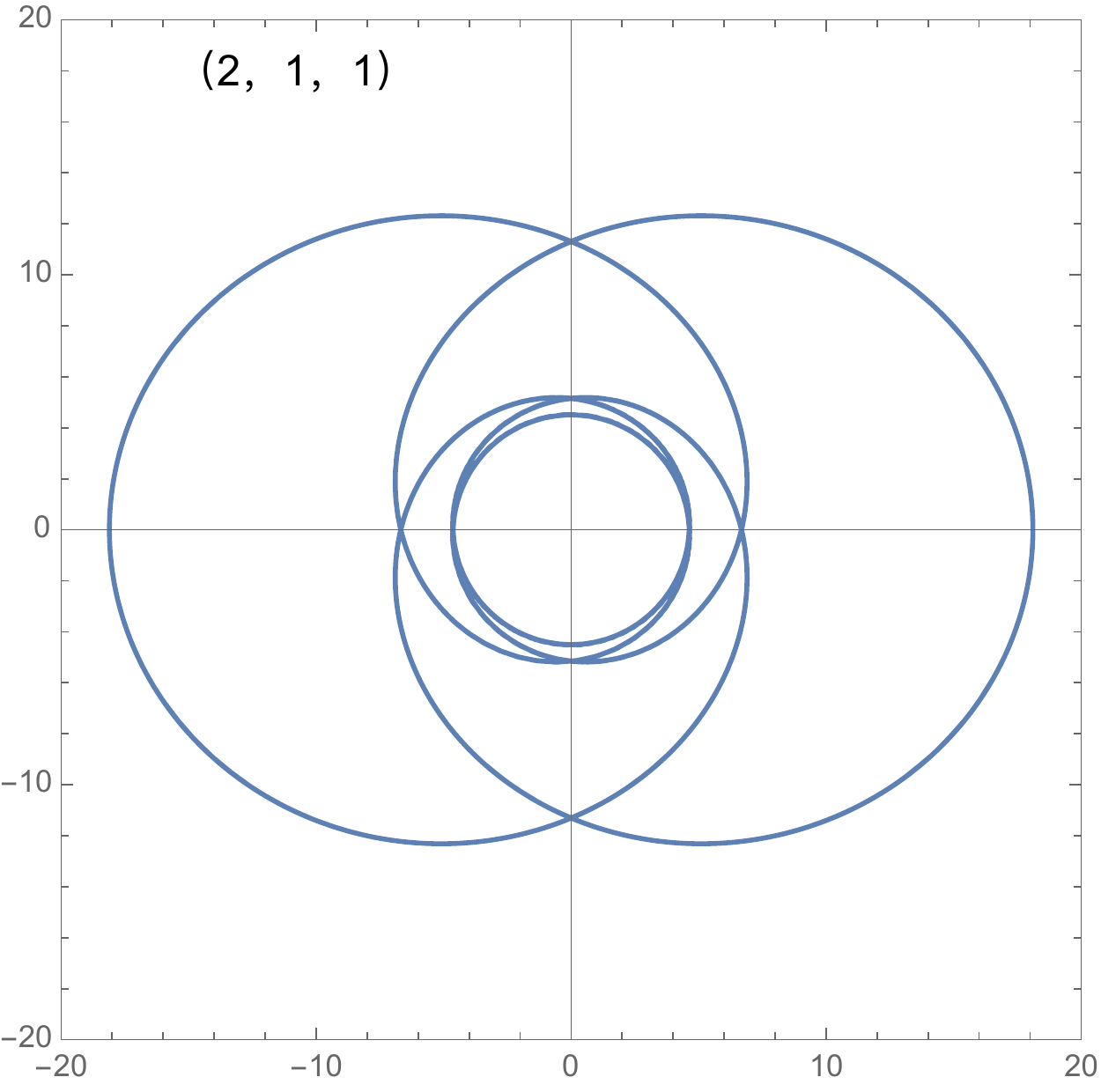}}
\subfigure[\;$E=0.962328$]{\includegraphics[height=4.5cm,width=4.5cm]{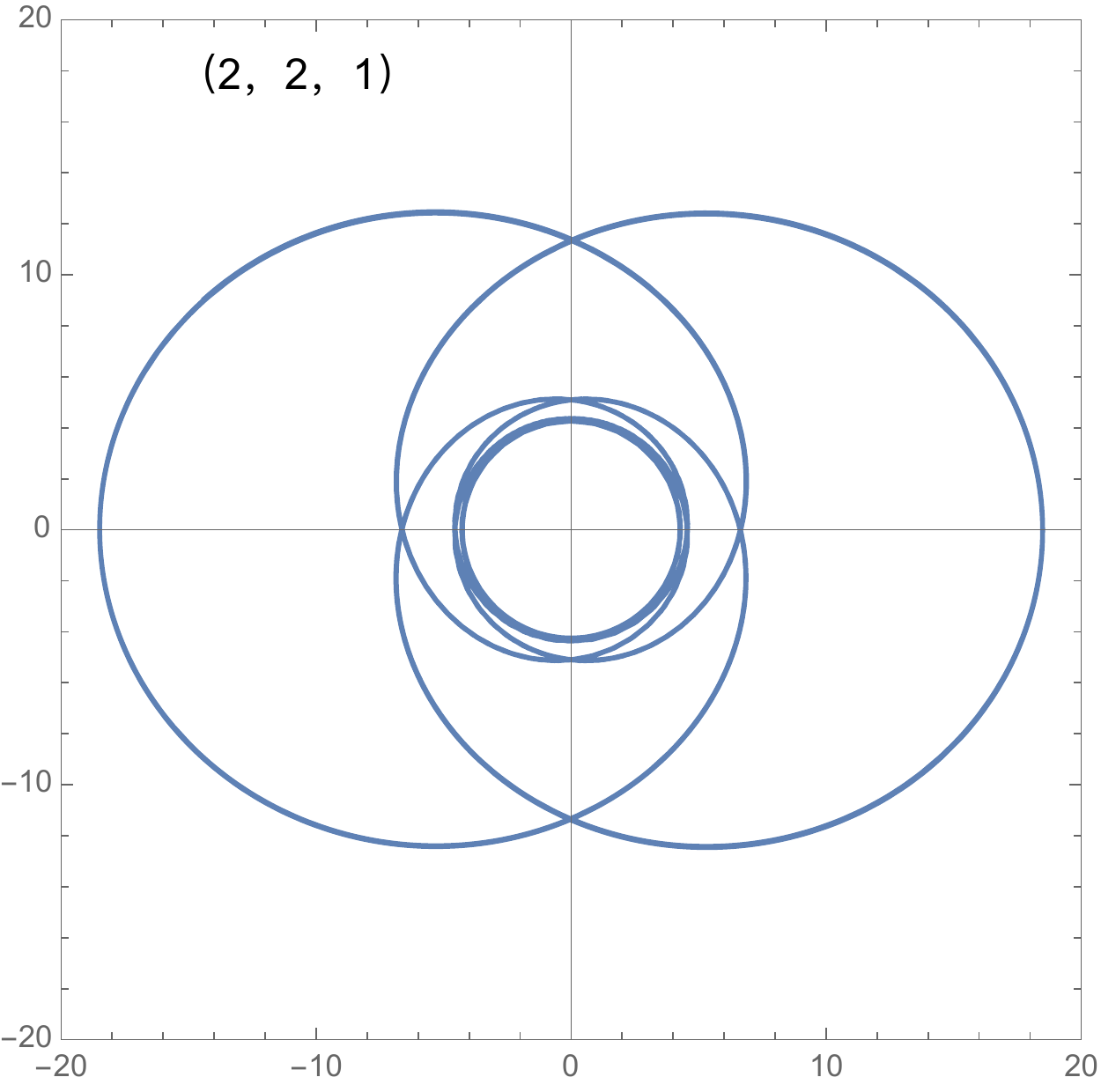}}
\subfigure[\;$E=0.962339$]{\includegraphics[height=4.5cm,width=4.5cm]{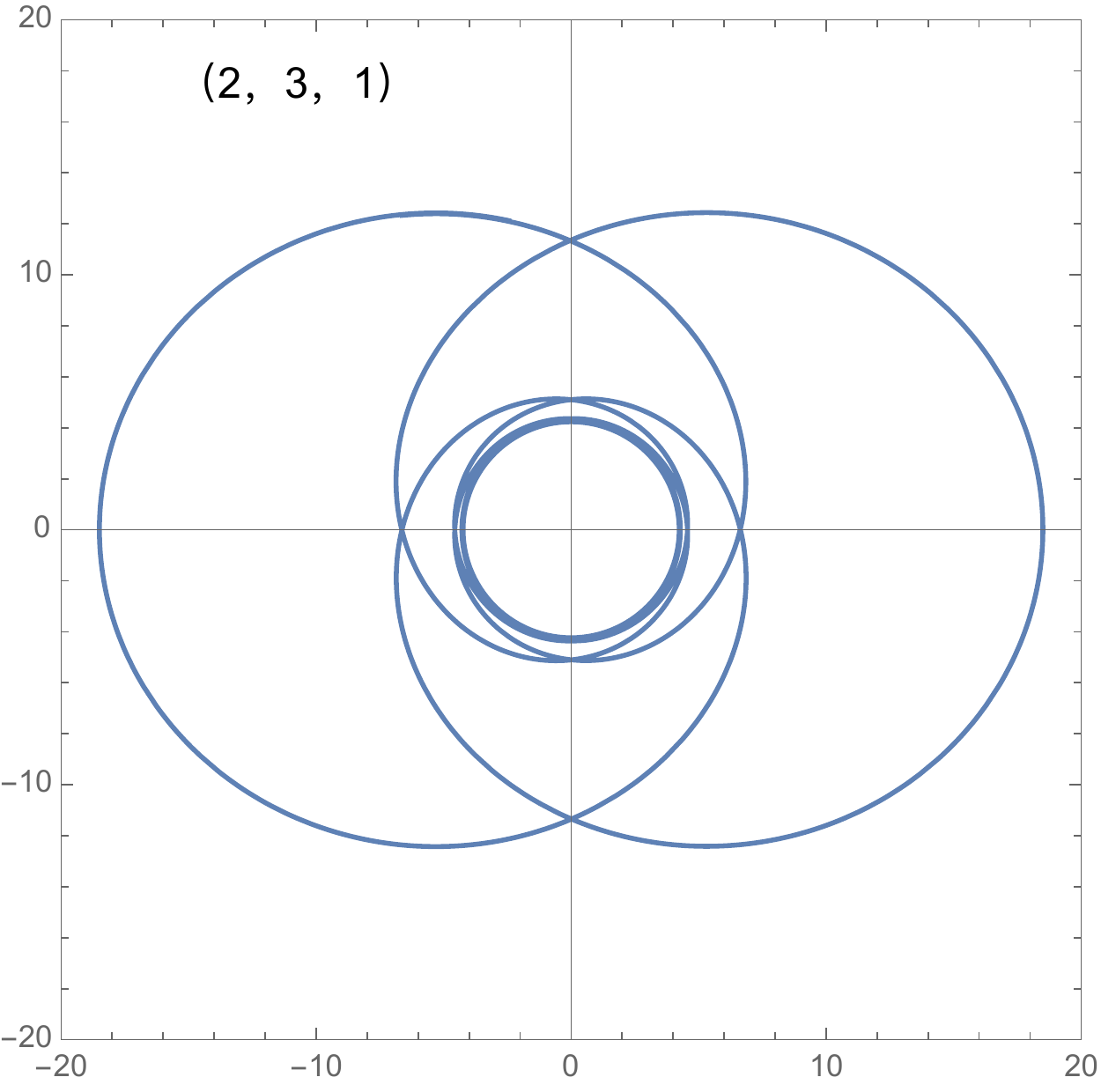}}\\
\subfigure[\;$E=0.962080$]{\includegraphics[height=4.5cm,width=4.5cm]{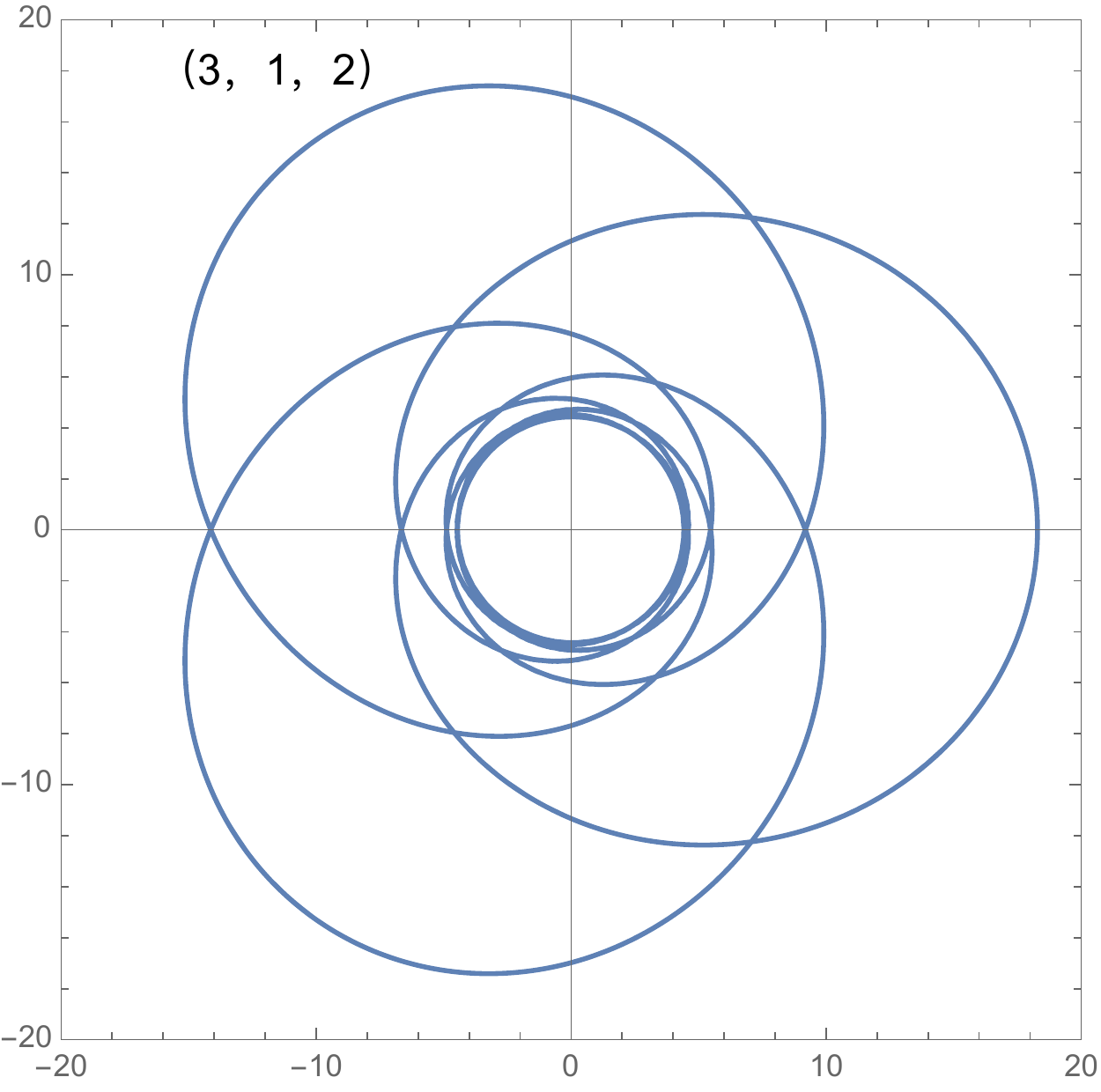}}
\subfigure[\;$E=0.962333$]{\includegraphics[height=4.5cm,width=4.5cm]{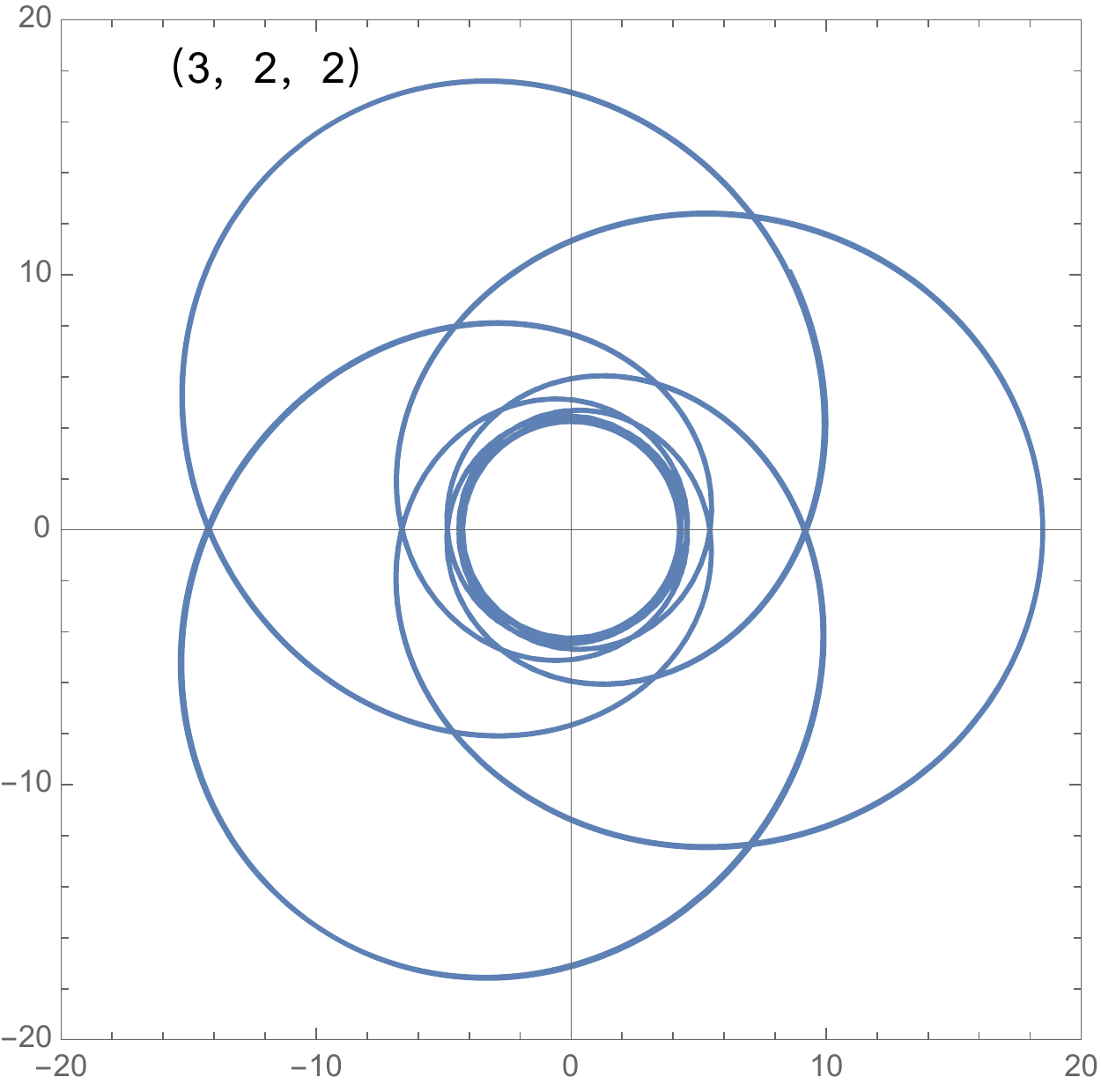}}
\subfigure[\;$E=0.962339$]{\includegraphics[height=4.5cm,width=4.5cm]{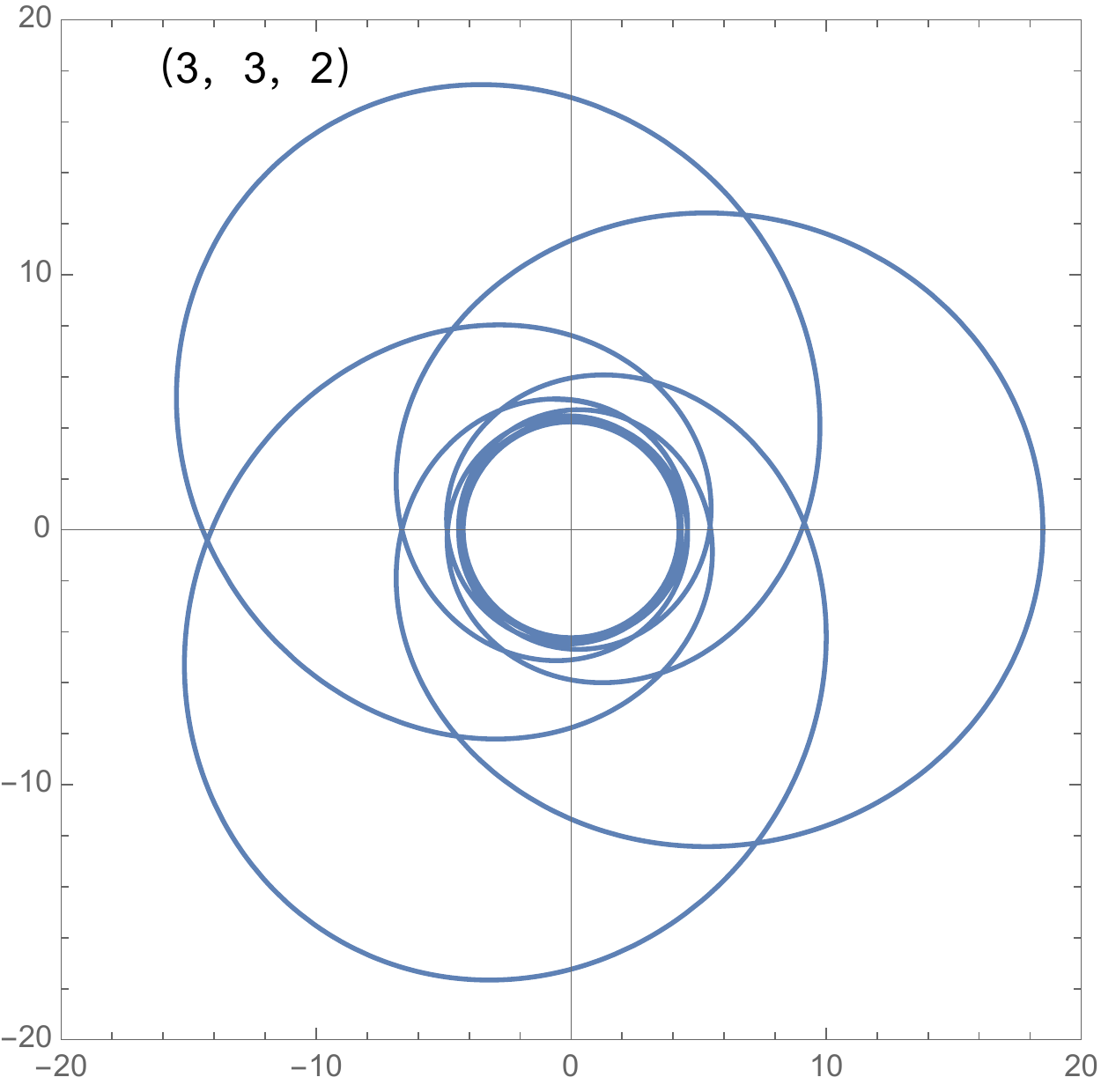}}\\
\subfigure[\;$E=0.953240$]{\includegraphics[height=4.5cm,width=4.5cm]{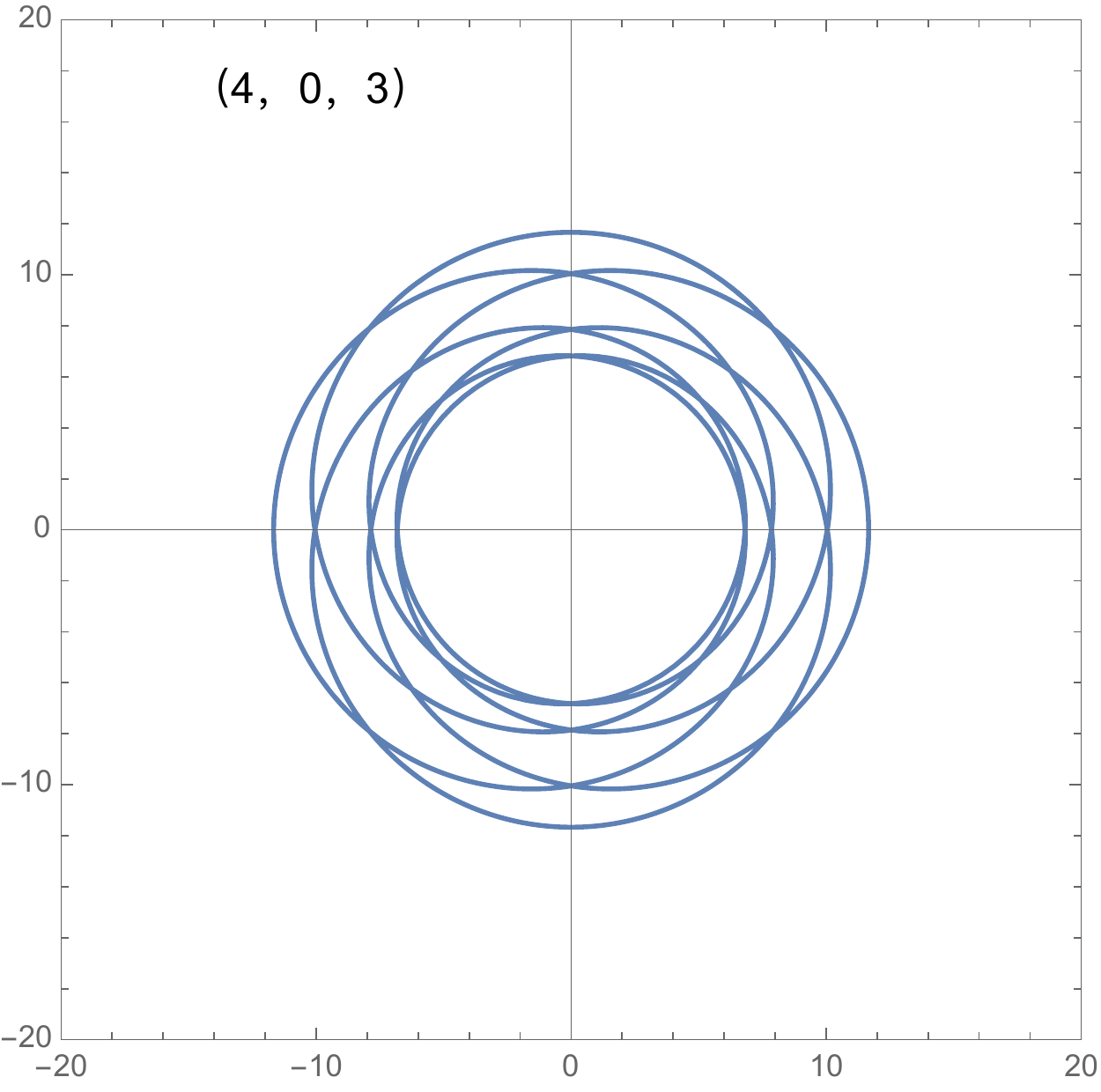}}
\subfigure[\;$E=0.962149$]{\includegraphics[height=4.5cm,width=4.5cm]{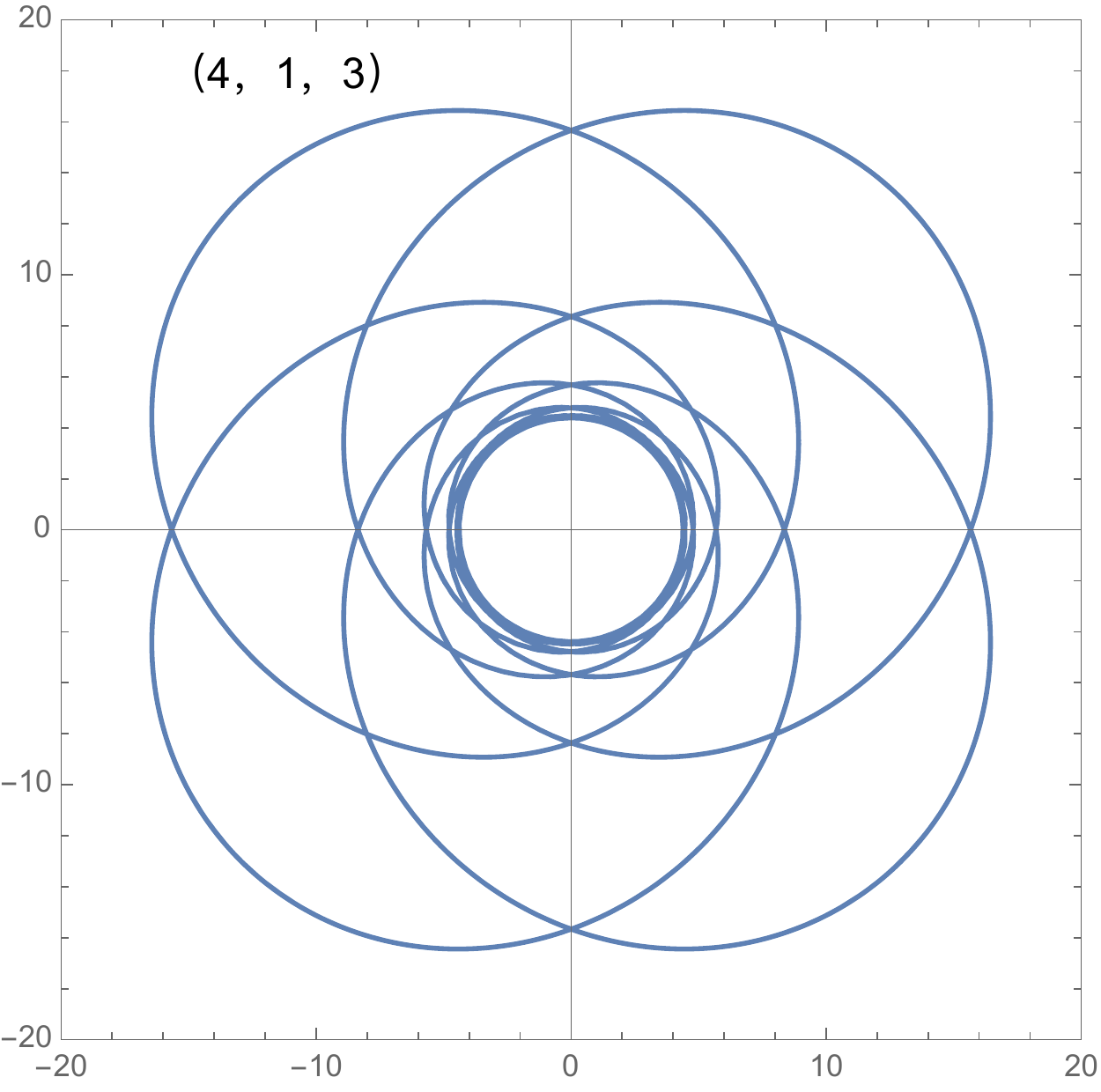}}
\subfigure[\;$E=0.962335$]{\includegraphics[height=4.5cm,width=4.5cm]{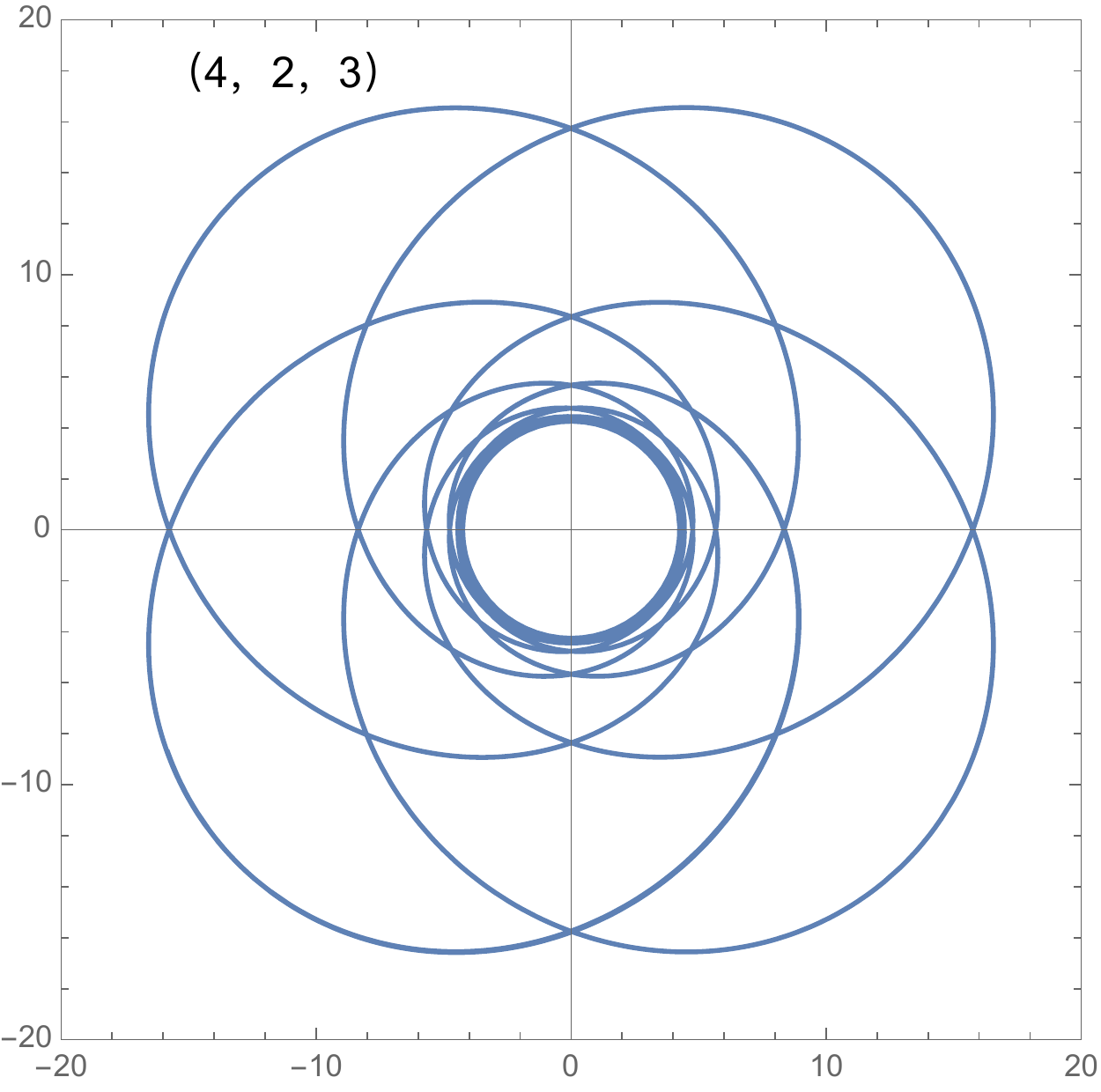}}
\caption{ Periodic orbits of different ($z$, $w$, $v$) around the polymer black holes in LQG with $A_\lambda=0.05$ and $\epsilon=0.5$.}
\label{orbitsl}
\end{figure*}


\begin{figure*}
  \centering
  \begin{minipage}[b]{0.4\textwidth}
    \centering
    \includegraphics[width=\textwidth]{Orbitsa312.pdf}
  \end{minipage}
  \hfill
  \begin{minipage}[b]{0.5\textwidth}
    \centering
    \includegraphics[width=1\textwidth]{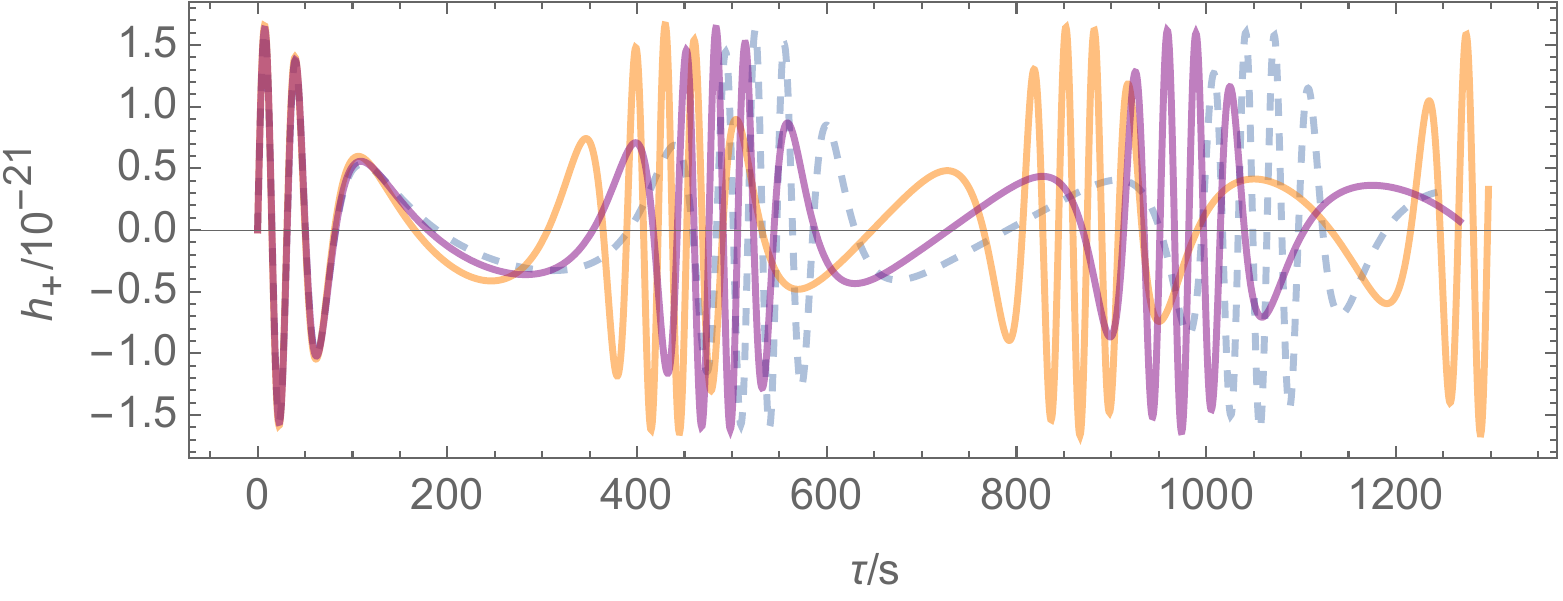} \\
    \includegraphics[width=1\textwidth]{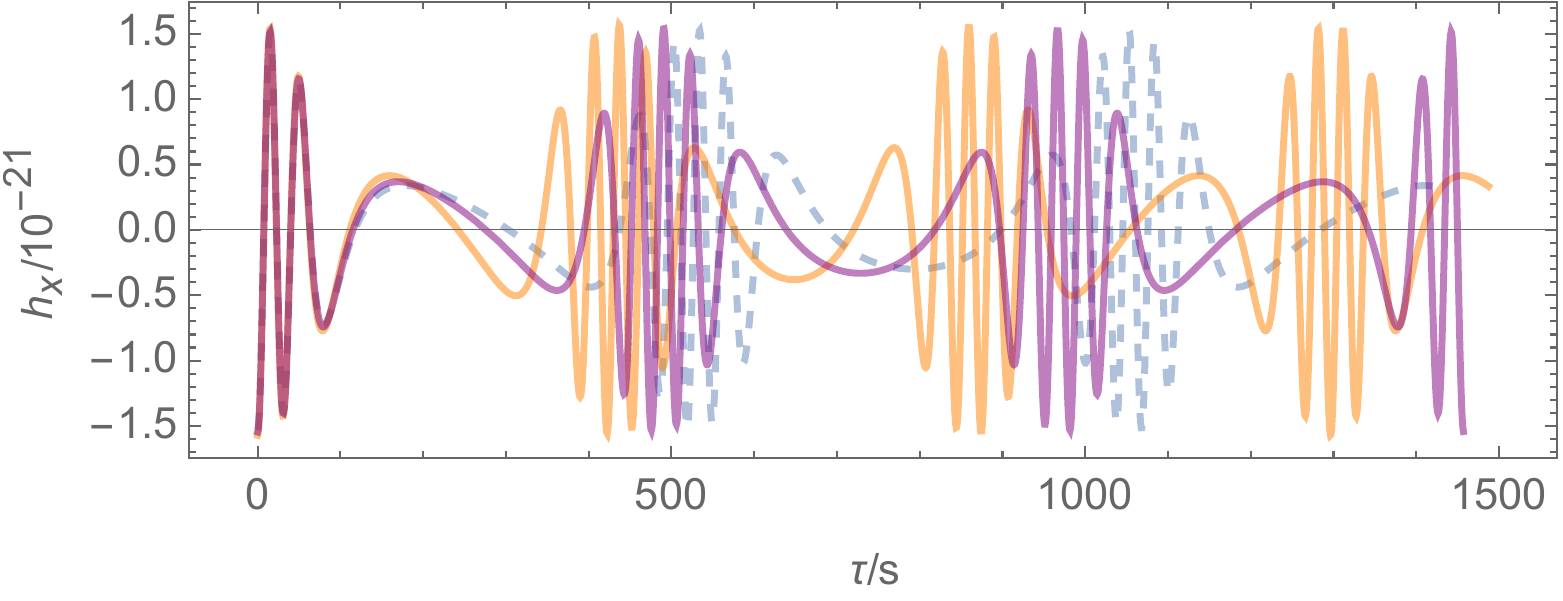} \\
  \end{minipage}
  \caption{The left figure is a sketch figure which shows a typical periodic orbit around a black hole with $(z, w, v)=(3, 1, 2)$. In the right figure, the dotted line represents the gravitational wave of the Schwarzschild black hole, while the orange solid line represents the gravitational wave of a polymer black hole in LQG with $A_\lambda=0.05$, $\epsilon=0.5$, and $q=1+2/3$. The purple solid line represents the gravitational wave of a polymer black hole in LQG with $A_\lambda=0.02$, $\epsilon=0.5$, and $q=1+2/3$.}
  \label{wave1}
\end{figure*}

\begin{figure*}
  \centering
  \begin{minipage}[b]{0.4\textwidth}
    \centering
    \includegraphics[width=\textwidth]{Orbitsa413.pdf}
  \end{minipage}
  \hfill
  \begin{minipage}[b]{0.5\textwidth}
    \centering
    \includegraphics[width=1\textwidth]{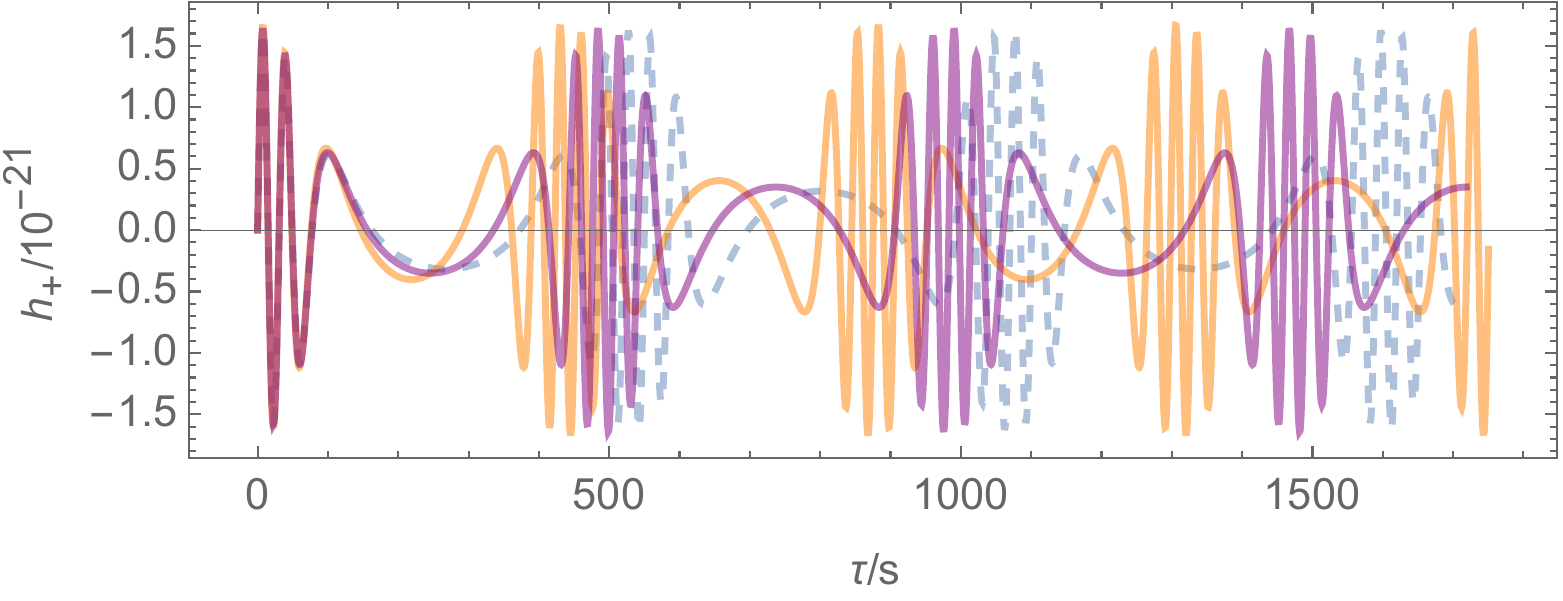} \\
    \includegraphics[width=1\textwidth]{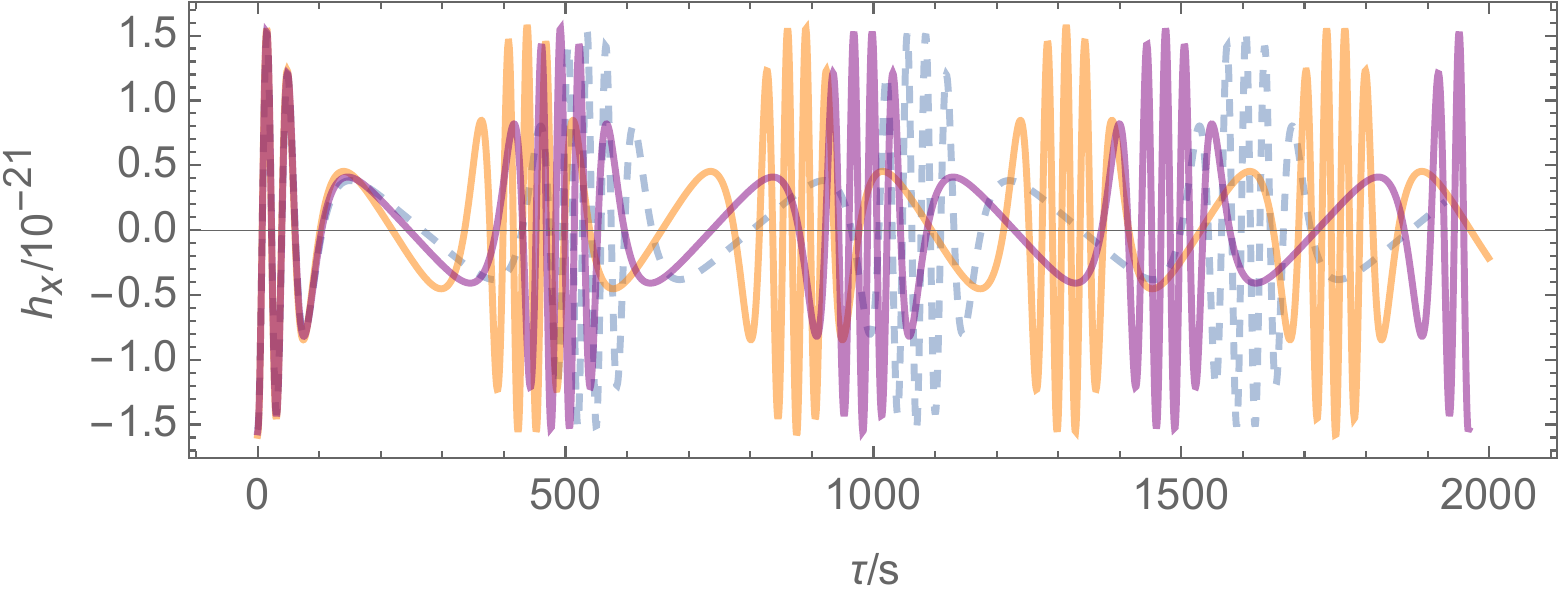} \\
  \end{minipage}
  \caption{The left figure is a sketch figure which shows a typical periodic orbit around a black hole with $(z, w, v)=(4, 1, 3)$. In the right figure, the dotted line represents the gravitational wave of the Schwarzschild black hole, while the orange solid line represents the gravitational wave of a polymer black hole in LQG with $A_\lambda=0.05$, $\epsilon=0.5$, and $q=1+3/4$. The purple solid line represents the gravitational wave of a polymer black hole in LQG with $A_\lambda=0.02$, $\epsilon=0.5$, and $q=1+3/4$. }
  \label{wave2}
\end{figure*}

In addition, we also take different values of energy to illustrate the rational number $q$ as a function of angular momentum $L$, as shown in Fig. \ref{veffb}. Among them, $E=0.95, 0.96, 0.97$. We found that the rational number $q$ decreases slowly with the increase of energy $L$. We magically found that $q$ became positive infinity when the angular momentum reached the minimum value. When $E$ is the same, the minimum angular momentum $L$ decreases with the increase of $A_\lambda$. By comparing different $E$, we can also find that the minimum angular momentum $L$ increases with the increase of $E$.

In Fig. \ref{orbitse}, we show periodic orbits with fixed energy $E=0.96$ and LQG parameter $A_\lambda=0.05$ with different integers ($z$,$w$,$v$). Obviously, $z$ describes the number of blade shapes for the orbit. As $z$ increases, the blade profile grows, and the trajectory becomes more complex. In Fig. \ref{orbitsl}, we show periodic orbits with fixed $A_\lambda=0.05$ and $\epsilon=0.5$ with different integers ($z$,$w$,$v$). It is evident that $z$ describes the number of blade shapes for the orbit. As $z$ increases, the blade profile grows, and the trajectory becomes more complex.

We also show the value of $q$ and energy $E$ in Tables \ref{tab1} and \ref{tab2} for different periodic orbits with $\epsilon=0.3$ and $\epsilon=0.5$, respectively. It can be seen from Tables \ref{tab1} and \ref{tab2} that the periodic orbit around the polymer black holes in LQG has lower energy than the classical Schwarzschild black hole. In addition, when we determine the values of $A_\lambda$ and $q$, the energy of the periodic orbit varies with $\epsilon$ and increases.


\section{Gravitational wave radiation from periodic orbits}
\label{gravitational}
\renewcommand{\theequation}{5.\arabic{equation}} \setcounter{equation}{0}

In this section, we provide a preliminary exploration of the gravitational radiation emitted by the periodic orbits of a test particle orbiting a supermassive polymer black hole. For this purpose, we consider an extreme-mass-ratio inspiral (EMRI) system, in which the smaller object has a mass much smaller than the supermassive black hole. In this way, it is convenient to treat this small object as a perturbation to the spacetime of the supermassive polymer black hole. With this treatment, when the change in the energy $E$ and angular momentum $L$ of the smaller object due to the gravitational radiation is sufficiently small over a few periods, one is able to adopt the adiabatic approximation, so that one can trace the periodic orbits which obey geodesic equation over a few orbital periods and calculate the corresponding gravitational radiation. 

We adopt the kludge waveform developed in \cite{Babak:2006uv} to calculate the gravitational wave emitted from the periodic orbits in the supermassive polymer black hole. The main strategy of the kludge waveform is as follows: treating the small object as a test particle, first, calculate the orbit of the particle (the periodic orbit in this paper) by solving the geodesic equation, and then use the quadrupole formula of gravitational radiation to get the corresponding gravitational waves. The periodic orbits of a test particle in the polymer black hole in LQG by solving the geodesic equation have been obtained in the above section. Then the gravitational waves emitted from these orbits can be calculated by using the following formula up to the quadratic order \cite{Liang:2022gdk, Maselli:2021men},
\bqn
h_{ij} = \frac{4 \eta M}{D_{\rm L}} \left(v_i v_j - \frac{m}{r} n_i n_j\right), \lb{quadratic}
\eqn
where $M$ is the mass of the polymer black hole, $m$ the mass of the test particle, $D_{\rm L}$ the luminosity distance of the EMRI system, $\eta=M m/(M+m)^2$ the symmetric mass ratio, $v_i$ the spatial velocity of the test particle, and $n_i$ is the unit vector which points to the radial direction associated to the motion of the test particle. 

Then one can project the above GW onto the detector-adapted coordinate system and in which the corresponding plus $h_+$ and cross $h_\times$, GW polarizations are given by \cite{Liang:2022gdk, Maselli:2021men}
\bqn
h_+ &=& - \frac{2 \eta}{D_{\rm L}} \frac{M^2}{r} (1+\cos^2\iota) \cos(2\phi+2 \zeta), \\
h_\times &=& - \frac{4 \eta}{D_{\rm L}} \frac{M^2}{r} \cos\iota \sin(2\phi+2 \zeta),
\eqn
where $\iota$ is the inclination angle between the EMRI's orbital angular momentum and the line of sight and $\zeta$ is the latitudinal angle. 

To illustrate the GW waveform of different periodic orbits and how the LQG effect can affect it, we consider an EMRI system that consists of a small component with mass $m=10 M_{\odot}$ and a supermassive black hole with mass $M = 10^7 M_{\odot}$ with $M_\odot$ being the solar mass. The inclination angle $\iota$ and the latitudinal angle $\zeta$ are set to be $\iota=\pi/4$ and $\zeta=\pi/4$ for simplicity, and we adopt the luminosity distance $D_{\rm L} = 200\;{\rm Mpc}$. In Figs.~\ref{wave1} and \ref{wave2}, as two examples, we show the GW waveforms emitted by two typical periodic orbits, with $(z, w, v) = (3, 1, 2)$ and  $(z, w, v) = (4, 1, 3)$ respectively. In both figures, we plot the plus $h_+$ and cross $h_\times$ of GW polarizations for GR, polymer black hole with $A_\lambda=0.05$, and $A_\lambda=0.02$, respectively. It is evident to see that the GW waveforms clearly exhibit the zoom-whirl behaviors of the periodic orbits. In Fig.~\ref{wave1}, for example, the periodic orbit (with $(z, w, v)=(3, 1, 2)$) has several zoom and whirl phases in one complete period. Correspondingly, the GW waveforms of the plus $h_+$ and $h_\times$ show distinctly quiet phases during the highly elliptical zooms followed by louder glitches during the nearly circular whirls. The number of the quiet phases is the same as the number of the leaves of the orbits, while the number of the louder glitches is the same as the number of whirls of the orbits. 
Compared to the Schwarzschild case (with $A_\lambda=0$), the LQG parameter $A_\lambda$ mainly changes the phases of the GW signals with a slight (almost negligible) change to the amplitudes. These properties suggest that the GW signals emitted by the periodic orbit can show the basic number of whirl and zoom phases and may be potentially useful for identifying the properties of the zoom-whirl orbits and constraining LQG effects in future GW detections. 

Here we would like to make a few remarks about the limitation of the calculations of the waveforms and the potential extensions of the current study. First, we use the adiabatic approximation with which we ignore the backreaction of the gravitational radiation to the periodic orbits. This approximation is sufficient if one only considers a few periods (in this section we only consider one complete period of the orbit). It is interesting to explore how gravitational radiation can affect the long-term evolution of periodic orbits. Second, by using the quadratic formula (\ref{quadratic}) to calculate the waveforms one in general ignores the contributions of multipoles higher than the quadratic order. The main purpose of this section is not to construct accurate waveforms for gravitational radiation, but to explore whether the GW signals emitted by the periodic orbits in polymer black holes can capture some basic orbital properties. It is quite important for future gw detections to construct more accurate waveforms by adding more multipole moments to the gravitational wave expansion formulae. And at last, the detection of GWs emitted by EMRI systems is one of the main targets of future space-based detectors, such as LISA, Taiji, Tianqin, etc. It is natural to ask how these future detectors can be used to constrain or test the effects of LQG in the periodic orbits. We expect to come back to these issues for a future study.

\section{Discussions and Conclusions}
\label{Conclusion}
\renewcommand{\theequation}{6.\arabic{equation}} \setcounter{equation}{0}

 In this study, we investigated the periodic orbit characteristics of polymer black holes in LQG. Firstly, we derived the geodesics of particles in the background of a polymer black hole in LQG, which significantly differs from the case of a Schwarzschild black hole but approaches the latter in the limit of $A_\lambda\rightarrow0$. Next, we numerically calculated the MBOs and ISCOs in the polymer black hole in LQG using its effective potential. The results showed that as $A_\lambda$ increases, the radius and angular momentum of both the MBOs and ISCOs decrease. Additionally, we analyzed the allowed parameter region $\Delta S$ in the $(L, E)$ plane and found that $\Delta S$ for the bound orbits around the polymer black hole in LQG decreases as $A_\lambda$ increases. Here $\Delta S$ denotes the area of the shadow region in the $(L, E)$ plane in Fig.~\ref{shadow}.

Based on the properties of the MBOs and ISCOs, we further investigated the periodic orbits of polymer black holes in LQG. We found that the rational number $q$ which characterizes the orbits increases with the particle's energy and decreases with its angular momentum. Specifically, for fixed energy $E$, $q$ increases with $A_\lambda$, while for fixed angular momentum $L$, $q$ decreases with $A_\lambda$. According to reference\cite{Levin:2008mq}, each periodic orbit is described by a set of parameters $(z, w, v)$, and we also extended the study to these orbits with the same $w$ and $v$. The results showed that energy decreases with decreasing $z$, for fixed angular momentum, higher $z$ orbits generally have lower energy. When $z\rightarrow \infty$, these orbits tend towards circles with the lowest energy. In the region $L_{\rm isco} < L < L_{\rm mbo}$, all eccentric periodic orbits around the polymer black hole in LQG exhibited some kind of scaled vortex behavior. These results may provide a way to distinguish between polymer black holes in LQG and Schwarzschild black holes by testing the periodic orbits around the central source. 

Furthermore, the gravitational wave radiation from the periodic orbits in polymer black holes in LQG is also preliminary explored. It is shown that the GW signals clearly exhibit the zoom-whirl behaviors of the periodic orbits. As shown in Figs.~\ref{wave1} and \ref{wave2}, the periodic orbits (with $(z, w, v)=(3, 1, 2)$ and $(z, w, v)=(4, 1, 3)$), the GW waveforms of the plus $h_+$ and $h_\times$ show distinctly quiet phases during the highly elliptical zooms followed by louder glitches during the nearly circular whirls. The number of the quiet phases is the same as the number of the leaves of the orbits, while the number of the louder glitches is the same as the number of whirls of the orbits. It is also shown that the LQG effects mainly affect the phases of GW rather than their amplitudes. These properties may be used for identifying the orbital structure of EMRI system and testing/constraining the polymer black hole in LQG with future GW detectors. 


\section*{Acknowledgements}

This work is supported by the National Key Research and Development Program of China under Grant No. 2020YFC2201503, the Zhejiang Provincial Natural Science Foundation of China under Grants No. LR21A050001 and No. LY20A050002, the National Natural Science Foundation of China under Grants No. 11675143 and No. 11975203,  and the Fundamental Research Funds for the Provincial Universities of Zhejiang in China under Grant No. RF-A2019015.


\begin{thebibliography}{399}

\bibitem{gws}
B. P. Abbott et al. [The LIGO Scientific Collaboration and the Virgo Collaboration],
Observation of Gravitational Waves from a Binary Black Hole Merger, 
Phys. Rev. Lett. {\bf 116}, 061102 (2016).

\bibitem{VLBI_deflection}
E. Fomalont, S. Kopeikin, G. Lanyi, and J. Benson, Progress in Measurements of the Gravitational Bending of Radio Waves Using the Vlba, Astrophys. J. {\bf 699}, 1395 (2009).

\bibitem{Akiyama:2019fyp}
K.~Akiyama {\it et al.} [Event Horizon Telescope Collaboration],
First M87 Event Horizon Telescope Results. V. Physical Origin of the Asymmetric Ring,
Astrophys. J. {\bf 875}, L5 (2019).

\bibitem{Akiyama:2019eap}
K.~Akiyama {\it et al.} [Event Horizon Telescope Collaboration],
First M87 Event Horizon Telescope Results. VI. The Shadow and Mass of the Central Black Hole,
Astrophys. J. {\bf 875}, L6 (2019).

\bibitem{Levin:2008mq} 
  J.~Levin and G.~Perez-Giz,
  A Periodic Table for Black Hole Orbits,
  Phys.\ Rev.\ D {\bf 77}, 103005 (2008)
  [arXiv:0802.0459 [gr-qc]].

\bibitem{Danzmann:1997hm}
K.~Danzmann,
LISA: An ESA cornerstone mission for a gravitational wave observatory,
Class. Quant. Grav. \textbf{14}, 1399-1404 (1997)

\bibitem{LISA:2017pwj}
P.~Amaro-Seoane \textit{et al.} [LISA],
Laser Interferometer Space Antenna,
[arXiv:1702.00786 [astro-ph.IM]].

\bibitem{CMS:2017asf}
A.~M.~Sirunyan \textit{et al.} [CMS],
Search for vectorlike light-flavor quark partners in proton-proton collisions at $\sqrt s$ =8  TeV,
Phys. Rev. D \textbf{97}, 072008 (2018)
[arXiv:1708.02510 [hep-ex]].

\bibitem{TianQin:2015yph}
J.~Luo \textit{et al.} [TianQin],
TianQin: a space-borne gravitational wave detector,
Class. Quant. Grav. \textbf{33}, 035010 (2016)
[arXiv:1512.02076 [astro-ph.IM]].

\bibitem{Gong:2021gvw}
Y.~Gong, J.~Luo and B.~Wang,
Concepts and status of Chinese space gravitational wave detection projects,
Nature Astron. \textbf{5}, 881-889 (2021)
[arXiv:2109.07442 [astro-ph.IM]].

\bibitem{LISA:2022kgy}
K.~G.~Arun \textit{et al.} [LISA],
Living Rev. Rel. \textbf{25}, no.1, 4 (2022)
doi:10.1007/s41114-022-00036-9
[arXiv:2205.01597 [gr-qc]].

\bibitem{LISACosmologyWorkingGroup:2022jok}
P.~Auclair \textit{et al.} [LISA Cosmology Working Group],
[arXiv:2204.05434 [astro-ph.CO]].


\bibitem{Glampedakis:2002ya}
K.~Glampedakis and D.~Kennefick,
Zoom and whirl: Eccentric equatorial orbits around spinning black holes and their evolution under gravitational radiation reaction,
Phys. Rev. D \textbf{66}, 044002 (2002)
[arXiv:gr-qc/0203086 [gr-qc]].

\bibitem{Levin:2008ci}
J.~Levin and B.~Grossman,
Dynamics of Black Hole Pairs. I. Periodic Tables,
Phys. Rev. D \textbf{79}, 043016 (2009)
[arXiv:0809.3838 [gr-qc]].
  
\bibitem{Levin:2009sk}
J.~Levin,
Energy Level Diagrams for Black Hole Orbits,
Class. Quant. Grav. \textbf{26}, 235010 (2009)
[arXiv:0907.5195 [gr-qc]].
  
\bibitem{Misra:2010pu}
V.~Misra and J.~Levin,
Rational Orbits around Charged Black Holes,
Phys. Rev. D \textbf{82}, 083001 (2010)
[arXiv:1007.2699 [gr-qc]].
  
\bibitem{Babar:2017gsg}
G.~Z.~Babar, A.~Z.~Babar and Y.~K.~Lim,
Periodic orbits around a spherically symmetric naked singularity,
Phys. Rev. D \textbf{96}, 084052 (2017)
[arXiv:1710.09581 [gr-qc]].
  
\bibitem{Liu:2018vea}
C.~Liu, C.~Ding and J.~Jing,
Periodic orbits around Kerr Sen black holes,
Commun. Theor. Phys. \textbf{71}, 1461 (2019)
[arXiv:1804.05883 [gr-qc]].

\bibitem{Lin:2023rmo}
H.~Y.~Lin and X.~M.~Deng,
Precessing and periodic orbits around hairy black holes in Horndeski\textquoteright{}s Theory,
Eur. Phys. J. C \textbf{83}, 311 (2023).

\bibitem{Wang:2022tfo}
R.~Wang, F.~Gao and H.~Chen,
Periodic orbits around a static spherically symmetric black hole surrounded by quintessence,
Annals Phys. \textbf{447}, 169167 (2022).


\bibitem{Mummery:2022ana}
A.~Mummery and S.~Balbus,
Inspirals from the Innermost Stable Circular Orbit of Kerr Black Holes: Exact Solutions and Universal Radial Flow,
Phys. Rev. Lett. \textbf{129}, 161101 (2022)
[arXiv:2209.03579 [gr-qc]].

\bibitem{Habibina:2022ztd}
A.~S.~Habibina and H.~S.~Ramadhan,
Bound orbits around charged black strings,
Annals Phys. \textbf{448}, 169169 (2023)
[arXiv:2205.14635 [gr-qc]].

\bibitem{Zhang:2022psr}
J.~Zhang and Y.~Xie,
Probing a self-complete and Generalized-Uncertainty-Principle black hole with precessing and periodic motion,
Astrophys. Space Sci. \textbf{367}, 17 (2022).

\bibitem{Lin:2022wda}
H.~Y.~Lin and X.~M.~Deng,
Precessing and periodic orbits around Lee\textendash{}Wick black holes,
Eur. Phys. J. Plus \textbf{137}, 176 (2022).

\bibitem{Gao:2021arw}
B.~Gao and X.~M.~Deng,
Bound orbits around modified Hayward black holes,
Mod. Phys. Lett. A \textbf{36}, 2150237 (2021).

\bibitem{Lin:2021noq}
H.~Y.~Lin and X.~M.~Deng,
Rational orbits around 4 $\mathcal D$ Einstein\textendash{}Lovelock black holes,
Phys. Dark Univ. \textbf{31}, 100745 (2021).

\bibitem{Deng:2020yfm}
X.~M.~Deng,
Geodesics and periodic orbits around quantum-corrected black holes,
Phys. Dark Univ. \textbf{30}, 100629 (2020).

\bibitem{Zhou:2020zys}
T.~Y.~Zhou and Y.~Xie,
Precessing and periodic motions around a black-bounce/traversable wormhole,
Eur. Phys. J. C \textbf{80}, 1070 (2020).

\bibitem{Gao:2020wjz}
B.~Gao and X.~M.~Deng,
Bound orbits around Bardeen black holes,
Annals Phys. \textbf{418}, 168194 (2020).

\bibitem{Deng:2020hxw}
X.~M.~Deng,
Periodic orbits around brane-world black holes,
Eur. Phys. J. C \textbf{80}, 489 (2020).

\bibitem{Azreg-Ainou:2020bfl}
M.~Azreg-A\"\i{}nou, Z.~Chen, B.~Deng, M.~Jamil, T.~Zhu, Q.~Wu and Y.~K.~Lim,
Orbital mechanics and quasiperiodic oscillation resonances of black holes in Einstein-\AE{}ther theory,
Phys. Rev. D \textbf{102}, 044028 (2020)
[arXiv:2004.02602 [gr-qc]].

\bibitem{Wei:2019zdf}
S.~W.~Wei, J.~Yang and Y.~X.~Liu,
Geodesics and periodic orbits in Kehagias-Sfetsos black holes in deformed Ho\v{r}ava-Lifshitz gravity,
Phys. Rev. D \textbf{99}, 104016 (2019)
[arXiv:1904.03129 [gr-qc]].


\bibitem{Pugliese:2013xfa}
D.~Pugliese, H.~Quevedo and R.~Ruffini,
General classification of charged test particle circular orbits in Reissner--Nordstr\"om spacetime,
Eur. Phys. J. C \textbf{77}, 206 (2017)
[arXiv:1304.2940 [gr-qc]].

\bibitem{Healy:2009zm}
J.~Healy, J.~Levin and D.~Shoemaker,
Zoom-Whirl Orbits in Black Hole Binaries,
Phys. Rev. Lett. \textbf{103}, 131101 (2009)
[arXiv:0907.0671 [gr-qc]].

\bibitem{Zhang:2022zox}
J.~Zhang and Y.~Xie,
Probing a black-bounce-Reissner\textendash{}Nordstr\"om spacetime with precessing and periodic motion,
Eur. Phys. J. C \textbf{82}, 854 (2022).

\bibitem{Lin:2022llz}
H.~Y.~Lin and X.~M.~Deng,
Bound Orbits and Epicyclic Motions around Renormalization Group Improved Schwarzschild Black Holes,
Universe \textbf{8}, 278 (2022).


\bibitem{Bambhaniya:2020zno}
P.~Bambhaniya, D.~N.~Solanki, D.~Dey, A.~B.~Joshi, P.~S.~Joshi and V.~Patel,
Precession of timelike bound orbits in Kerr spacetime,
Eur. Phys. J. C \textbf{81}, 205 (2021)
[arXiv:2007.12086 [gr-qc]].


\bibitem{Rana:2019bsn}
P.~Rana and A.~Mangalam,
Astrophysically relevant bound trajectories around a Kerr black hole,
Class. Quant. Grav. \textbf{36}, 045009 (2019)
[arXiv:1901.02730 [gr-qc]].

\bibitem{LQG_BH1}
N. Bodendorfer, F.M. Mele, and J. Münch, 
(b,v)-type variables for black to white hole transitions in effective loop quantum gravity,
Phys. Lett. B {\bf 819}, 136390 (2021).


\bibitem{LQG_BH2}
N. Bodendorfer, F.M. Mele, and J. Münch,
Effective quantum-extended spacetime of polymer Schwarzschild black hole,
Class. Quant. Grav. {\bf38}, 095002 (2021).

\bibitem{Ashtekar20} 
A. Ashtekar, 
Black Hole evaporation: A Perspective from Loop Quantum Gravity, 
[arXiv:2001.08833].

\bibitem{Brahma:2020eos}
S.~Brahma, C.~Y.~Chen and D.~h.~Yeom,
Testing Loop Quantum Gravity from Observational Consequences of Nonsingular Rotating Black Holes,
Phys. Rev. Lett. \textbf{126}, 181301 (2021)
[arXiv:2012.08785 [gr-qc]].

\bibitem{Papanikolaou:2023crz}
T.~Papanikolaou,
Primordial black holes in loop quantum gravity: The effect on the threshold,
[arXiv:2301.11439 [gr-qc]].

\bibitem{Islam:2022wck}
S.~U.~Islam, J.~Kumar, R.~Kumar Walia and S.~G.~Ghosh,
Investigating Loop Quantum Gravity with Event Horizon Telescope Observations of the Effects of Rotating Black Holes,
Astrophys. J. \textbf{943}, 22 (2023).

\bibitem{Afrin:2022ztr}
M.~Afrin, S.~Vagnozzi and S.~G.~Ghosh,
Tests of Loop Quantum Gravity from the Event Horizon Telescope Results of Sgr A$^*$,
[arXiv:2209.12584 [gr-qc]].

\bibitem{Vagnozzi:2022moj}
S.~Vagnozzi, R.~Roy, Y.~D.~Tsai, L.~Visinelli, M.~Afrin, A.~Allahyari, P.~Bambhaniya, D.~Dey, S.~G.~Ghosh and P.~S.~Joshi, \textit{et al.}
Horizon-scale tests of gravity theories and fundamental physics from the Event Horizon Telescope image of Sagittarius A$^*$,
[arXiv:2205.07787 [gr-qc]].

\bibitem{KumarWalia:2022ddq}
R.~Kumar Walia,
Observational Predictions of LQG Motivated Polymerized Black Holes and Constraints From Sgr A$^*$ and M87$^*$,
[arXiv:2207.02106 [gr-qc]].

\bibitem{Yan:2022fkr}
J.~M.~Yan, Q.~Wu, C.~Liu, T.~Zhu and A.~Wang,
Constraints on self-dual black hole in loop quantum gravity with S0-2 star in the galactic center,
JCAP \textbf{09}, 008 (2022).

\bibitem{Alesci:2011wn}
E.~Alesci and L.~Modesto,
Particle Creation by Loop Black Holes,
Gen. Rel. Grav. \textbf{46}, 1656 (2014)
[arXiv:1101.5792 [gr-qc]].

\bibitem{Chen:2011zzi}
J.~H.~Chen and Y.~J.~Wang,
Complex frequencies of a massless scalar field in loop quantum black hole spacetime,
Chin. Phys. B \textbf{20}, 030401 (2011)

\bibitem{Dasgupta:2012nk}
A.~Dasgupta,
Entropy Production and Semiclassical Gravity,
SIGMA \textbf{9}, 013 (2013)
[arXiv:1203.5119 [gr-qc]].

\bibitem{Barrau:2014yka}
A.~Barrau, C.~Rovelli and F.~Vidotto,
Fast Radio Bursts and White Hole Signals,
Phys. Rev. D \textbf{90}, 127503 (2014)
[arXiv:1409.4031 [gr-qc]].

\bibitem{Hossenfelder:2012tc}
S.~Hossenfelder, L.~Modesto and I.~Premont-Schwarz,
Emission spectra of self-dual black holes,
[arXiv:1202.0412 [gr-qc]].

\bibitem{Sahu:2015dea}
S.~Sahu, K.~Lochan and D.~Narasimha,
Gravitational lensing by self-dual black holes in loop quantum gravity,
Phys. Rev. D \textbf{91}, 063001 (2015)
[arXiv:1502.05619 [gr-qc]].

\bibitem{Cruz:2015bcj}
M.~B.~Cruz, C.~A.~S.~Silva and F.~A.~Brito,
Gravitational axial perturbations and quasinormal modes of loop quantum black holes,
Eur. Phys. J. C \textbf{79}, 157 (2019)
[arXiv:1511.08263 [gr-qc]].

\bibitem{Pugliese:2020ivz}
D.~Pugliese and G.~Montani,
Constraining LQG Graph with Light Surfaces: Properties of BH Thermodynamics for Mini-Super-Space, Semi-Classical Polymeric BH,
Entropy \textbf{22}, 402 (2020).


\bibitem{Pawlowski:2014nfa}
T.~Paw\l{}owski,
Observations on interfacing loop quantum gravity with cosmology,
Phys. Rev. D \textbf{92}, 124020 (2015).

\bibitem{Vaid:2012pr}
D.~Vaid,
Quantum Hall Effect and Black Hole Entropy in Loop Quantum Gravity,
[arXiv:1208.3335 [gr-qc]].

\bibitem{Barrau:2011md}
A.~Barrau, T.~Cailleteau, X.~Cao, J.~Diaz-Polo and J.~Grain,
Probing Loop Quantum Gravity with Evaporating Black Holes,
Phys. Rev. Lett. \textbf{107}, 251301 (2011).

\bibitem{Modesto:2009ve}
L.~Modesto and I.~Premont-Schwarz,
Self-dual Black Holes in LQG: Theory and Phenomenology,
Phys. Rev. D \textbf{80}, 064041 (2009).

\bibitem{Yang:2023gas}
S.~Yang, W.~D.~Guo, Q.~Tan and Y.~X.~Liu,
Axial gravitational quasinormal modes of a self-dual black hole in loop quantum gravity,
[arXiv:2304.06895 [gr-qc]].

\bibitem{Fu:2023drp}
G.~Fu, D.~Zhang, P.~Liu, X.~M.~Kuang and J.~P.~Wu,
Peculiar properties in quasi-normal spectra from loop quantum gravity effect,
[arXiv:2301.08421 [gr-qc]].

\bibitem{Addazi:2021xuf}
A.~Addazi, J.~Alvarez-Muniz, R.~Alves Batista, G.~Amelino-Camelia, V.~Antonelli, M.~Arzano, M.~Asorey, J.~L.~Atteia, S.~Bahamonde and F.~Bajardi, \textit{et al.}
Prog. Part. Nucl. Phys. \textbf{125}, 103948 (2022)
doi:10.1016/j.ppnp.2022.103948
[arXiv:2111.05659 [hep-ph]].

\bibitem{Garcia-Chung:2020zyq}
A.~Garcia-Chung, J.~B.~Mertens, S.~Rastgoo, Y.~Tavakoli and P.~Vargas Moniz,
Phys. Rev. D \textbf{103}, no.8, 084053 (2021)
doi:10.1103/PhysRevD.103.084053
[arXiv:2012.09366 [gr-qc]].

\bibitem{Garcia-Chung:2022pdy}
A.~Garcia-Chung, M.~F.~Carney, J.~B.~Mertens, A.~Parvizi, S.~Rastgoo and Y.~Tavakoli,
JCAP \textbf{11}, 054 (2022)
doi:10.1088/1475-7516/2022/11/054
[arXiv:2208.09739 [gr-qc]].

\bibitem{Garcia-Chung:2023oul}
A.~Garcia-Chung, M.~F.~Carney, J.~B.~Mertens, A.~Parvizi, S.~Rastgoo and Y.~Tavakoli,
[arXiv:2305.18192 [gr-qc]].

\bibitem{Cutler:1994pb}
C.~Cutler, D.~Kennefick and E.~Poisson,
Gravitational radiation reaction for bound motion around a Schwarzschild black hole,
Phys. Rev. D \textbf{50}, 3816-3835 (1994)


\bibitem{Babak:2006uv}
S.~Babak, H.~Fang, J.~R.~Gair, K.~Glampedakis and S.~A.~Hughes,
``Kludge" gravitational waveforms for a test-body orbiting a Kerr black hole,
Phys. Rev. D \textbf{75}, 024005 (2007)
[arXiv:gr-qc/0607007 [gr-qc]].

\bibitem{Maselli:2021men}
A.~Maselli, N.~Franchini, L.~Gualtieri, T.~P.~Sotiriou, S.~Barsanti and P.~Pani,
Detecting fundamental fields with LISA observations of gravitational waves from extreme mass-ratio inspirals,
Nature Astron. \textbf{6}, 464-470 (2022)
[arXiv:2106.11325 [gr-qc]].



\bibitem{Liang:2022gdk}
D.~Liang, R.~Xu, Z.~F.~Mai and L.~Shao,
Probing vector hair of black holes with extreme-mass-ratio inspirals,
Phys. Rev. D \textbf{107}, no.4, 044053 (2023)
[arXiv:2212.09346 [gr-qc]].


\end{thebibliography}
\end{document}